\documentclass[11pt]{article}

\usepackage{preamb}

\title{Min(d)ing the President:\\A text analytic approach to measuring tax news}
\author[1]{Lenard Lieb%
\thanks{Corresponding author. Department of Macro, International and Labour Economics, Maastricht University, P.O. Box 616, 6200 MD Maastricht, The Netherlands. Email: \href{mailto:l.lieb@maastrichtuniversity.nl}{l.lieb@maastrichtuniversity.nl}}}
\author[2]{Adam Jassem}
\author[2,3]{Rui Jorge Almeida}
\author[2]{Nalan Ba\c{s}t\"{u}rk}
\author[2]{Stephan Smeekes}
\affil[1]{Department of Macro, International and Labour Economics, Maastricht University}
\affil[2]{Department of Quantitative Economics, Maastricht University}
\affil[3]{Department of Data Analytics and Digitilisation, Maastricht University}
\date{\today}

\begin{document}

\maketitle

\begin{abstract}
Economic agents react to signals about future tax policy changes. Consequently, estimating their macroeconomic effects requires identification of such signals. We propose a novel text analytic approach for transforming textual information into an economically meaningful time series. Using this method, we create a tax news measure from all publicly available post-war communications of U.S. presidents. Our measure predicts the direction and size of future tax changes and contains signals not present in previously considered (narrative) measures of tax changes. We investigate the effects of tax news and find that, for long anticipation horizons, pre-implementation effects lead initially to contractions in output.

\end{abstract}

\onehalfspacing

\section{Introduction}
In this paper we propose a novel approach for incorporating textual information into the empirical analysis of causal relations in economics. We complement the literature with three main contributions. First, we develop a two-step semi-supervised topic model that automatically extracts specific information about future policy changes from textual data. Second, we provide methodology for transforming such textual information into economically meaningful time series to be included in causal econometric models as variable of interest or instrument. Third, we study fiscal foresight by extracting information about future tax reforms from speeches of U.S.~presidents. We find that economic agents react to these news signals.

News about future policy changes is likely to have effects today. When receiving new information, economic agents’ forward-looking behavior implies that economic decisions are also contemporaneously affected. There is a growing empirical literature supporting these findings.\footnote{For example \citep{jaimovich2009can,blanchard2013news,beaudry2014news,forni2017noisy} investigate news shocks related to future productivity, \citet{Ramey11} considers fiscal news, monetary policy news shocks are analysed in \citet{NakamuraSteinsson18}; see also \citet{Ramey16} for a comprehensive overview of the recent literature.} 
Taking foresight into account is particularly important when analyzing effects of fiscal policy: economic agents typically receive clear signals about future tax reforms long before a particular bill is enacted. \citet{yang2007chronology} shows that, in the U.S., almost all implemented tax changes were preceded by legislative lags ranging between a quarter and three years. The Tax Reform Act of 1986 provides an illustrative example of how fiscal foresight impacts economic behavior. This Act stipulated an increase of the effective maximum tax rate on capital gains from 20\% to 28\%, to be implemented in the following year. As a result, tax revenues from capital gains jumped by 90\% before the act came into force \citep{auerbach1997economic}. 
Assessing this particular tax reform's effect is flawed if fiscal foresight is not accounted for, and one may easily confuse directions of causality. 

As a general critique, \citet{leeper2013fiscal} argue that standard empirical approaches such as structural vector autoregressions (SVARs) often fail to properly account for fiscal foresight, as the variables typically included do not span the full information set available to economic agents. Estimated tax change effects using such models may as a result be biased and misleading; tax multipliers might even be of the wrong sign. 
Several studies therefore use instead a narrative approach to trace the arrival of information on future tax policy changes from the legislative process. \cite{romer2009narrative,RomerRomer10} (henceforth RR) identify the key motivation behind all legislated post-war tax changes in the U.S.~and determine their impact on government revenues. Using various sources of narrative records, such as presidential speeches and Congressional records, they identify those tax changes that are not systematically related to changes in output and classify them as \emph{exogenous}.\footnote{The exogeneity of RR's tax narratives is a matter of debate \citep{brueckner2021fiscal}. Nevertheless, it is widely used in the literature as measure of exogenous legislated tax changes in the U.S.} RR additionally define a series of tax \textit{news} as the present value of tax changes discounted back to the date of enactment (in contrast to the exogenous tax changes which are assigned to the implementation date), with the aim of capturing anticipation effects. \citet{mertens2012empirical,mertens2014reconciliation} further split the RR series of exogenous tax changes into two components and use these to identify anticipated and unanticipated tax shocks. 
 
We find that neither RR’s tax news, nor derivatives of RR's series as for example \citet{mertens2012empirical,mertens2014reconciliation}, fully get the timing of arrival of new information right. In addition, narrative approaches that identify tax shocks solely from tax changes that were eventually implemented do not take into account that a news signal might be noisy and that policy plans, and alongside them economic agents’ expectations, are subject to revision over time. Before a tax bill is signed into law, information on the exact design of the tax reform is imprecise. Moreover, some proposals of tax changes initially put forward by the administration never come to fruition at all, while nonetheless influencing expectations.

While similar to RR, our approach is conceptually different in two important aspects. First, while analyzing similar auxiliary sources of data, our goal is not to search for the motivation behind each tax change but rather to trace the arrival of information regarding future tax reforms. In the U.S., the president is the main driving force behind tax policy legislation \citep{yang2007chronology,RomerRomer10}. Presidential speeches are therefore a highly informative source for signals about future tax reforms. We review the U.S. president’s speeches and communications and quantify how prominently tax reforms, as well as their direction (cuts vs.~hikes), are featured on the political agenda at a given point in time. To implement this idea, we build on the following hypothesis: if a policy maker repeatedly emphasises the importance of future tax changes, the public should expect that these changes will likely be implemented in the future. As a result we obtain two measures of \emph{prevalence}: one capturing the importance of tax cuts on the administration's policy agenda and one capturing the importance of tax hikes.

Second, to construct these prevalence measures from a considerable number of documents (we consider \textit{all} of the President's public statements since 1949), we introduce a novel semi-supervised text-analytic approach. While a standard, unsupervised Latent Dirichlet Allocation (LDA) topic model would allow us to quantify the prevalence of specific political issues -- henceforth referred to as topics -- and distinguish speeches about tax reforms from speeches on other topics, further differentiating between tax hikes and tax cuts is not possible. Our proposed  novel strategy is to feed additional information into the LDA topic model in order to construct more informative priors for the tax hike/cut topics. To this aim, we combine lexical knowledge of a priori selected terms related to the direction of tax changes with the results from an unsupervised model. It is important to stress that our dictionary of selected terms only `nudges' the model towards the terms we deem to be important: the topic estimation remains data-driven and robust to misspecification.

We find that our tax prevalence series predict future federal tax reforms regardless of how we measure tax changes: both cyclically adjusted revenue changes as well as all RR narrative measures are Granger-caused by our prevalence series. In contrast, we do not find evidence for predictability in the opposite direction. Interestingly, our series also Granger-cause implicit tax rates, which are often used to proxy tax news in the U.S.\footnote{To construct implicit tax rates, i.e. a measure of expected future tax rates, \citet{leeper2012quantitative} use the yield spread between taxable treasury bonds and tax-exempt municipal bonds to identify economic agents' expectations about future personal tax changes.} Again, we do not find evidence for predictability in the opposite direction. These findings suggest that our tax prevalence series capture the timing of the arrival of tax news more accurately. Additionally, and importantly we do not find evidence that our tax prevalence series are driven by business cycle conditions in the (recent) past. 

We use our constructed measures to analyse the effects of tax news on economic activity. Along the lines suggested in \cite{SW2018}, we attach a scale, or monetary value, to the news by using  the tax prevalence measures as instruments for future tax changes. We find that output expansion is driven by (post-) implementation effects of tax changes. However, allowing our measures to also instrument tax changes in the more distant future also reveals pre-implementation effects. More specifically, anticipation effects trigger initially a contraction in GDP if the anticipation horizon is long enough. We also find that the longer the anticipation horizon, the weaker the expansion due to implemented tax changes.

The remainder of this paper is organised as follows. Section \ref{sec:quant} illustrates the type of textual information we aim to quantify and gives an intuitive overview of our approach. In Section \ref{sec:topic_model} we outline the semi-supervised topic model in more detail and discuss how we estimate tax policy signals. The identified topics, including the tax prevalence measures, are presented in Section \ref{sec:res_topics}. In Section \ref{sec:struct} we use our prevalence measures to analyse the impact of tax news on economic activity. Section \ref{sec:conclusion} concludes. Additional methodological details and further empirical results are collected in several Appendices.

\section{Quantifying tax policy signals} \label{sec:quant}

In the U.S. the president is the main driving force behind tax policy legislation \citep{yang2007chronology,RomerRomer10}, making presidential speeches a highly informative source for signals about future tax reforms. Below we illustrate the information flow and the various types of signals we aim to quantify. All four statements were made by Ronald Reagan in the months leading up to enactment of the Economic Recovery Act of 1981.

Throughout his term, the president outlines tax reforms he wants to pursue, albeit in general terms:
\begin{quote}
``It is time to reawaken this industrial giant, to get government back within its means, and to lighten our punitive tax burden.''
\end{quote}
\begin{flushright} - \textit{Inaugural Address} January 20, 1981 \end{flushright}

When the president believes the tax law should be changed he recommends that to the House of Representatives:
\begin{quote}
``At the same time, however, we cannot delay in implementing an economic program aimed at both reducing tax rates to stimulate productivity and reducing the growth in government spending to reduce unemployment and inflation. On February 18th, I will present in detail an economic program to Congress embodying the features I've just stated.''
\end{quote} \begin{flushright} - \textit{Address to the Nation on the Economy}, February 05, 1981 \end{flushright}

In that announcement, as well as in the following months, the president may offer details about particular measures included in the upcoming tax law changes:
\begin{quote}
``I shall ask for a 10-percent reduction across the board in personal income tax rates for each of the next 3 years. Proposals will also be submitted for accelerated depreciation allowances for business to provide necessary capital so as to create jobs.''
\end{quote} \begin{flushright} - \textit{Address to the Nation on the Economy}, February 05, 1981 \end{flushright}

Once the bill is passed through the Congress, the president provides final remarks as he signs it and ends the legislative process:
\begin{quote}
``These bills that I'm about to sign—not every page—this is the budget bill, and this is the tax program—but I think they represent a turnaround of almost a half a century of a course this country's been on and mark an end to the excessive growth in government bureaucracy, government spending, government taxing.''
\end{quote} \begin{flushright} - \textit{Remarks on Signing the Economic Recovery Tax Act of 1981 and the Omnibus Budget Reconciliation Act of 1981}, August 13, 1981 \end{flushright}

Quantifying individual statements in terms of monetary value is difficult as they often lack sufficient details about the proposed reform. While the Congressional Budget Office (CBO) might publish an estimate of its tax revenue impact, this is done only in the last stages of the legislative process, once the law is drafted. Moreover, the president’s statements are often subject to revisions: statements might be retracted, announced changes might not come to fruition or they might include different measures than initially planned. The goal of our approach is to capture those signals as well.

Our solution is to instead quantify how prominently each direction of tax reforms (cuts vs.~hikes) is featured on the political agenda of the president. We do that by measuring the proportion of presidential speeches in a given quarter that is devoted to tax cuts and tax hikes respectively.

The idea of measuring how often a particular issue is mentioned in a body of texts is not new. Most prominently, \citet{baker2016measuring} develop an Economic Policy Uncertainty (EPU) index by quantifying the proportion of news article referencing various types of uncertainty. The main challenge of such an approach is determining which texts, or which parts of a specific text, are relevant. The simplest approach is to check whether (or how many of) the words in a document belong to a pre-defined set of terms (a so-called lexicon). For example, \citet{baker2016measuring} use a rule-based extension of this approach, which consists of checking if a combination of certain terms appears in a given text document. Lexicon-based approaches are particularly useful in sentiment analysis, since readily available sentiment lexicons can often be used in a variety of applications. \citet{shapiro2020measuring} combine multiple lexicons to produce a robust analysis of economic news sentiment and its effects.
In the case of our application however a lexicon-based approach would require that we exactly specify the words and phrases which the president uses to reference tax reforms. This introduces a risk of misspecification, especially if the relevant terms can appear in different contexts.

Topic modelling solves this problem by jointly considering the whole text instead of looking for particular terms. We can compare the terms used in a document with the term distributions used to discuss a certain topic to determine to what degree the text is about it. Crucially, in contrast to lexicon- or rule-based approaches, a topic model ``learns'' those distributions from the data in an unsupervised fashion. Given a collection of documents and a specified number of topics, we find topics which best explain the way the words are used in the texts. The method we chose is the Latent Dirichlet Allocation (LDA) model proposed by \citet{blei2003latent} which is discussed further in the following section. It allows us to identify the way in which the president talks about tax policy changes, expressed as a probability distribution over the vocabulary, and determine which texts discuss them.

A similar approach has been used for example by \citet{larsen2019value}. Their LDA-based analysis of news outlets shows the impact that various types of news have on financial markets. We are also not the first ones in the economic literature who use topic models to analyse the statements of policymakers. LDA has been widely used to analyse the economic impact of the communication by central banks. For instance \citet{hansen2016shocking,10.1093/qje/qjx045} analyse the minutes of the deliberations of the \textit{Federal Open Market Committee} to identify topics of discussion and measure their impact on the economy.
In a study similar to ours, \citet{dybowski2018economic} also use a LDA model to identify the part of presidential speeches devoted to taxation in general. Using a lexicon-based approach they further measure how optimistic or pessimistic the identified tax communication is and investigate whether the effect of the communicated tax changes on economic activity depends on the tone. Crucially, the effect they capture is by design that of changes in perceived uncertainty. Since those are only very loosely connected with the communicated tax changes, their approach is unfortunately not well suited for analyzing the effects of tax news. The above examples are part of a fast growing body of literature using text-mining in economic research; for a recent survey see \citet{gentzkow2019text}.

Our approach is meant to explicitly distinguish between signals about tax hikes and tax cuts. Because standard LDA estimates the topics in an unsupervised fashion, there is little control over their composition. Intuitively, the discussion about both tax increases and decreases relies on a relatively similar, tax-oriented subset of vocabulary. As a result, when using the standard approach, the two are ``grouped'' together into a single topic, preventing us from determining the direction of the discussed changes. By reading through (some of) the speeches that we know are tax related we can gather precise information about terms which differentiate the two types of signals. Including this additional (prior) information in the model requires however a departure from the conventional LDA approach.
In our two-step approach we combine the so obtained information with the results from the unsupervised approach to construct informed priors for the topics. By using  those priors in an LDA model we are able to differentiate between the content devoted to discussing tax hikes and tax cuts. We aggregate the per-document results for each quarter to obtain a measure of the relative prominence of each political issue on the presidents agenda which we refer to as prevalence. In Section \ref{sec:pred} we show that the prevalence measures of the two tax topics contain information about future tax changes. 

\section{A topic model for measuring tax news} \label{sec:topic_model}

In this section we outline our methodology to extract information about tax policy from presidential statements using topic modelling.
Intuitively, our model assumes that when the president discusses different topics, he uses a different distribution over words (his vocabulary). Identifying these distributions, and their occurrence in each statement, therefore allows us to identify the topics the president talks about. We hypothesise that one or more of these topics can be linked to instances when the president talks about tax policy.
Indeed, as we show in Section \ref{sec:res_topics}, after estimating an unsupervised LDA model, one of the estimated topics can be labeled as the \textit{tax} topic.
However, as shown in Section \ref{sec:pred}, such a tax topic contains information too imprecise to be useful for predictive or causal analysis. We therefore aim to explicitly differentiate the discussion about tax policy into \textit{tax increase} and \textit{tax decrease} topics. For this purpose we introduce a two-step topic modelling approach. Before we discuss the topic model in detail, we first describe the data and the steps we take in pre-processing.

\subsection{Text data and pre-processing}
Our analysis is based on \textit{the Public Papers of the Presidents}, a compilation of all documents originating from the president. We obtain the raw texts from the \textit{American Presidency Project} (APP).\footnote{\href{http://www.presidency.ucsb.edu}{www.presidency.ucsb.edu}, retrieved on 2019-03-25.} We analyse 59,214 texts spanning from 1949-01-20 to 2017-01-19. 

The raw text of the speeches needs to be pre-processed to quantify relevant features of the text data, facilitating further statistical analysis. This process is described in detail in Appendix \ref{sec:text_data}. 

First, the documents in the dataset range from short remarks consisting of several sentences to long speeches, such as the \textit{State of the Union Address}, which cover a variety of otherwise unrelated issues. We therefore split the texts into a total of 1,119,200 individual paragraphs of roughly the same length. For the remainder of the analysis we treat those as separate \emph{documents}.

Next, we prepare a matrix of \textit{word counts} which is used as the input to our algorithm. Informally, for each text we count how often every word is used in it. To define the possible `words' we take two steps. First, we clean the text and exclude function words such as ``a'' or ``and'' and rare words. In this step we also transform words to their `root' form, e.g.~``taxes'' and ``tax'' are both counted as ``tax'' and ``implements'' and ``implemented'' both count as ``implement''. In the second step we identify combinations of two subsequent words that occur frequently together (so-called \emph{bigrams}), which are then also counted as a `word'. This is important for our analysis as combinations of words such as ``tax cut'' may contain very different information than either of the words would contain separately. To avoid confusion with the actual words used in the speeches, henceforth we refer to each of the `words' we count as \emph{terms}, which then refers to either individual (pre-processed) words or the identified bigrams.

\subsection{LDA topic model}\label{subsec:LDA}
In this subsection we present the Latent Dirichlet Allocation (LDA) topic model. For this purpose we first define the following elements of the pre-processed data:
\begin{itemize}
    \item The \textbf{vocabulary} $\mathcal{V}= \{v_1, \dots, v_{V} \}$ is the set of all \textbf{terms} $v_i$ that appear in the data, where $V = \abs{\mathcal{V}}$ is the total number of terms in our dataset. The vocabulary consists of $V=50,851$ individual words and bigrams.
    \item The \textbf{corpus} $\mathcal{W}$ is the collection of all \textbf{documents} $\boldsymbol{w}_d$, where $d=1,\ldots, D$.
    In our analysis the corpus consists of $D=1,119,200$ individual paragraphs.
    \item Each \textbf{document} $\boldsymbol{w}_d$ is defined as a vector of $N_d$ \textbf{tokens}, which are random variables each taking one value from the vocabulary $\mathcal{V}$. 
    $$\boldsymbol{w}_d = (w_{d,1}, \dots, w_{d,N_d})^T, \qquad w_{d,n} \in \mathcal{V}, \quad n=1, \ldots, N_d, \quad d=1,\ldots,D.$$
    Intuitively, instead of thinking about a document as a continuous string of text, we consider it as a sequence of $N_d$ slots, for which a particular realization is chosen from the set of possible values $\mathcal{V}$ when the document is created.
\end{itemize}

Our approach is based on the LDA topic model proposed by \citet{blei2003latent}. LDA is a relatively straightforward, unsupervised approach that often leads to easily interpretable results, and thus has become one of the most popular choices for topic modelling \citep{jelodar2019latent}. In LDA the process of creating a document is modelled as a series of independent draws from a particular distribution over the terms in the vocabulary. The key idea behind topic modelling is that the particular distribution changes depending on the \emph{topic}, i.e.,~what the text is about. For example, we expect the president to use certain terms with different frequency when talking about education versus for example military build-up. As such, with $K$ topics,\footnote{In the model that follows, the number of topics $K$ is assumed to be known. Of course, in practice this has to be estimated. The particular choice of $K$ for our dataset is described later.} the distribution with which terms are used overall, is a mixture of the $K$ distributions per topic, which we label as the \emph{topic-term distributions}. Furthermore, each document can consist of multiple topics. The proportions of each topic's occurrence in a single document is labelled the \emph{mixing proportion} of that document, which is of key interest for our analysis, as informally it addresses how much each topic (such as tax) is discussed in the document. The mixing proportions therefore allow us to identify documents which relate predominantly to tax policy. It is crucial to allow documents to discuss multiple topics in varying proportions. For instance, the president will generally not discuss tax in isolation, but in conjunction with other topics such as the need to balance the budget, stimulating the economy or creating the funding for particular investments such as military expenditures in times of war.

Hence, the creation of a document -- and the tokens inside that document -- can be seen as a two-step process. First, the creator (in our case the president), decides on the proportions of each topic to be discussed in that document (the mixing proportion). Then, given that proportion, each token in the document is randomly assigned to one topic. Next, given this \emph{topic assignment}, a term from the vocabulary is drawn from the distribution corresponding to the assigned topic. We now formalise this as follows.
\begin{itemize}
    \item The $n$-th token in document $d$, $w_{d,n}$, has a \textbf{topic assignment}, denoted as $z_{d,n}\in \{1,\dots,K\}$. We collect these 
    for the $d$-th document in the vector $\boldsymbol{z}_d=\left(z_{d,1},\ldots,z_{d,N_d}\right)$, and for the whole corpus in the matrix $\mathcal{Z}=\{\boldsymbol{z}_1,\ldots,\boldsymbol{z}_D \}$. 
    \item Within one document, all tokens have the same probability of `belonging' to a certain topic,
    specified by that \textbf{document's mixing proportions}. For document $\bm{w}_d$, those proportions are defined as a $K$-dimensional 
    vector $\boldsymbol{\theta}_d = \left( \theta_{d,1},\dots,\theta_{d,K} \right)^T$ such that
    \begin{align}
        \mathbb{P}(z_{d,n}=k|\boldsymbol{\theta}_d)=\theta_{d, k}, \quad n=1,\ldots, N_d,
     \label{eq_gen_z}
    \end{align}
    where $\theta_{d,k} \geq 0$ and $\sum_{k=1}^K\theta_{d,k} = 1$ for all $d=1, \ldots, D$. We denote those proportions jointly as $\boldsymbol{\Theta}=\left(\boldsymbol{\theta}_1,\dots,\boldsymbol{\theta}_D \right)^T$, a matrix of size $D\times K$ where the $d$-th row is the vector of the $d$-th document.
    
    \item All tokens assigned to topic $k$ share the same \textbf{topic-term distribution}. This distribution is parametrised by a $V$-dimensional vector $\boldsymbol{\phi}_k = \left( \phi_{k,1},\dots,\phi_{k,V} \right)^T$ such that
    \begin{align*}
        \mathbb{P}(w_{d,n}=v_i|z_{d,n}=k)=\phi_{k, i}, \quad n = 1, \ldots, N_d,\;  d=1,\ldots,D,\; i = 1, \ldots, V.
    \end{align*}
    where $\phi_{k,i} \geq 0$ and $\sum_{i=1}^V\phi_{k,i} = 1$ for all $k=1,\ldots,K$. We denote those distributions jointly as $\boldsymbol{\Phi}= \left(\boldsymbol{\phi}_1,\dots,\boldsymbol{\phi}_K \right)^T$, a matrix of size $K\times V$ where the $k$-th row is the vector of topic $k$.
\end{itemize}
We estimate the topic-term distributions and the mixing proportions in a Bayesian procedure by means of the posterior parameter distributions using Gibbs sampling. We denote them as $\hat{\bm{\Phi}}$ and $\hat{\bm{\Theta}}$ respectively. In order to do so, LDA puts Dirichlet priors on both the mixing proportions $\boldsymbol{\Theta}$ and the topics' term distributions $\boldsymbol{\Phi}$. Appendix \ref{sec:dirichlet} provides an intuitive description of the properties of the Dirichlet distribution $\bm{\theta}_d \sim \text{Dir}(\bm{\alpha}), \quad d= 1,\ldots,D$ and the implications of its use on estimation. In particular, we assume that $\boldsymbol{\theta}_1, \dots, \boldsymbol{\theta}_D$ share the same Dirichlet distribution with $K$-dimensional parameter vector $\boldsymbol{\alpha} = \left( \alpha_{1},\dots,\alpha_{K} \right)^T$. Following the recommendations of \citet{wallach2009rethinking}, we do not treat $\bm\alpha$ as a fixed hyper-parameter but place another level of (Gamma) priors on its elements, such that $\bm\alpha$ is estimated along with the lower-level parameters. 

Dirichlet priors are also placed on the topic-term distributions. How we do that exactly depends on the first or second step of our model, and is the source of our methodological innovation in the second step. 
Generally, each topic-term distribution has a Dirichlet prior $\bm{\phi}_k \sim \text{Dir}(\bm{\eta}_k), \quad k= 1,\ldots,K$ with $V$-dimensional parameter vector $\boldsymbol{\eta}_k=\left( \eta_{k,1},\dots,\eta_{k,V} \right)^T$. In the first step we use the conventional LDA setup where no `expert knowledge' about topic composition is formulated. In particular, all topics share the same symmetric Dirichlet prior, such that $\eta_{k,i} = \eta$ for all $k=1,\ldots,K$ and $i=1,\ldots,V$ - the prior probability of every term in every topic is equal. This parameter $\eta$ is then also estimated by assuming a Gamma prior \citep{wallach2009rethinking}. These uninformative priors imply that the topic-term distributions is fully driven by the data.

Crucially however, and as discussed before, a pure data-driven, unsupervised approach is unsuited to differentiate between \textit{tax increase} and \textit{tax decrease} topics, for which purpose we add a novel second step LDA estimation. Based on our first-step estimates $\hat{\bm{\Phi}}$, we first identify a \textit{general tax} topic and use it to develop informed priors for the two topics of interest for the second step of LDA estimation. The exact process is described in the following section. The result is a set of parameter vectors $\bm{\eta}_k, \quad k=1,\ldots,K$ which specify separate, informed, prior for each topic, where we postulate that particular terms occur either more frequently, or less frequently, in specific topics.  
These priors are then used in another Gibbs sampling to estimate the posterior means of the parameters, which are denoted as $\doublehat{\bm{\Phi}}$ and $\doublehat{\bm{\Theta}}$. The full details of the estimation for both steps are provided in Appendix \ref{sec:estimation}.

The approach described above relies on the assumption that the number of topics $K$ is known, but in practice this number needs to be estimated. An incorrect specification of the number of topics decreases the overall quality of the model and the interpretability of the topics. While the choice of $K$ can be guided by metrics based on the likelihood of observing the corpus given the model, we base this choice on the interpretability of our results instead. Specifically, our approach requires that discussion about tax policy forms a distinct topic. Intuitively, specifying too few topics is likely to lead to a topic that includes also other issues, not necessarily related to tax policy. On the other hand, setting the number of topics too high introduces unnecessary computational complexity and can result in the tax topic being split, for example into personal and corporate tax policy discussion. As a result we choose the number of topics to be $K=25$ in the first step. In the second step, because we split the \textit{general tax} policy topic into the \textit{tax increase} and \textit{tax decrease} topics, we use a total of $26$ topics. In section \ref{sec:res_topics} we substantiate our choice by inspecting the estimated topic-term distributions and mixing proportions.

\subsection{Two-step `semi-supervised' prior construction}\label{subsec:twostepLDA}

By using different priors for the topic-term distributions we can include `expert knowledge' as a priori beliefs about topic composition to `steer' the topics into the direction we want them to go, in our case topics about tax increase and tax decrease. Here we describe how we achieve this in our two-step procedure. 
The main challenge is now to correctly translate a priori beliefs about occurrence of particular terms into meaningful prior parameters $\bm\eta_i$ of the Dirichlet distribution. To illustrate, note that for any topic-term distribution $\bm{\phi}_k \sim \text{Dir}(\bm{\eta}_k)$ 
the Dirichlet prior implies that
$\mathbb{E}[\phi_{k,i}|\boldsymbol{\eta}_k] = \frac{\eta_{k,i}}{\sum_{j=1}^V \eta_{k,j}}$. Hence, in order to obtain sensible priors we need to formulate beliefs about the relative occurrence of \emph{all} terms. For example, we may believe that terms belonging to a certain set $L$, a so-called \emph{lexicon}, are likely to be mostly associated with a particular topic of interest. Such a belief could be expressed by choosing a topic $k$ and setting $\eta_{k,i}=a \mu_L$ if $v_i \in L$ and $\eta_{k,i}=a \mu_{\neg L}$ if $v_i \notin L$ where $\mu_L > \mu_{\neg L}$. However, the shape of the resulting a priori distribution is not in line with the way terms occur in text in general, as empirical term distributions tend to roughly follow a power law \citep{sato2010topic}. Setting such a prior would therefore lead to a highly distorted topic distribution; instead we must make sure the prior parameters are in line with the empirical properties of the text.

To address this issue, we first learn about the general shape of the topic-term distributions from the data in the first-step unsupervised estimation and then \emph{modify} these distributions using lexicons of terms relevant to the topics of interest. From the first-step unsupervised estimation we are able to identify a single topic that encompasses all the discussion about tax changes, which we refer to as the \textit{general tax} topic,\footnote{This interpretation is based on the fact that this distribution assigns uniquely high probabilities to terms related to tax policy (e.g. ``tax''), which is discussed in detail in Section \ref{sec:res_topics}.}
and denote the corresponding estimated topic-term distribution as $\hat{\bm{\phi}}_{k^*}$. 

We then construct the priors for the \emph{tax increase} and \emph{tax decrease} topics based on the assumption that both topics of interest have a similar distribution over the vocabulary, except for some key differentiating terms. Based on reading of documents related to tax policy we manually identify the terms which are predominantly used when discussing changes in a particular direction.\footnote{For this purpose we use all off the speeches identified by \citet{romer2009narrative} and \citet{yang2007chronology} as announcements of tax changes.} When then group them into two lexicons - $L_\text{inc}$ for terms related to tax increases, and $L_\text{dec}$ for terms related to tax decreases. The composition of the lexicons and details concerning their creation are presented in Appendix \ref{sec:lexicons}. 

We then `guide' the algorithm towards the two tax topics by modifying the prior probabilities of terms which are contained in either of the lexicons. For the tax increase prior $\boldsymbol{\eta}_\text{inc}$, we modify $\hat{\bm{\phi}}_{k^*}$ by multiplying probabilities of terms in $L_\text{inc}$ by a constant $m_1>1$ to `up-weight' them, and simultaneously multiplying probabilities of terms in $L_\text{dec}$ by $m_2<1$ to down-weight those. For the prior of the tax decrease topic we do the reverse. For the priors of the other topics we use their respective distributions estimated in the first step, without modifying their shape. This allows us to `fix' the other topics while splitting up the \textit{general tax} topic. The last step in the modification of those priors is to choose the strength of the effect of our prior, which is done by multiplying the vectors by a scalar $m_{3,k}$. Hence, we construct our prior parameters for all $i=1, \ldots,V$ as 
\vspace{-0.4cm}
\begin{equation}
\begin{split}
\eta_{\text{inc}, i} &= m_{3,\text{inc}} \left[\hat{\phi}_{k^*, i} + (m_1 - 1) \hat{\phi}_{k^*, i} \mathbbm{1}\left(v_i \in L_\text{inc} \right) + (m_2 - 1) \hat{\phi}_{k^*, i} \mathbbm{1}\left(v_i \in L_\text{dec} \right)\right],\\
\eta_{\text{dec}, i} &= m_{3,\text{dec}} \left[\hat\phi_{k^*, i} + (m_1 -1)  \hat\phi_{k^*, i} \mathbbm{1}\left(v_i \in L_\text{dec} \right) + (m_2 - 1) \hat\phi_{k^*, i} \mathbbm{1}\left(v_i \in L_\text{inc} \right)\right], \\
\eta_{k,i} &= m_{3,k} \hat\phi_{k-1,i}, \qquad k=3,\ldots,K+1
\end{split}
\label{eq:eta_values_step2} \nonumber
\end{equation}
We set $m_1=100$ and $m_2=100^{-1}$, while $m_{\text{inc}},m_{\text{dec}}, m_{k}$ are chosen such that the sum of the vectors is $\sum_i\eta_{\text{inc}, i} = \sum_i\eta_{\text{dec}, i} = \sum_i\eta_{k,i} =10,000$. This prior can be interpreted as postulating a prior belief equivalent to an additional observation of 10,000 tokens assigned to a given topic from the corresponding topic-term distribution; a relatively weak prior given the size of our dataset. In practice, the modified $\boldsymbol{\eta}_\text{inc}$ and $\boldsymbol{\eta}_\text{dec}$ act as `seeds', during each iteration of the estimation `nudging' the distributions of the relevant topics through the mechanism described in \ref{sec:dirichlet}. 

While the priors help us discern between the two topics of interest, the distributions are still predominantly determined through learning from the data. In particular, we see that a \textit{tax increase (decrease)} topic does not necessarily assign high probabilities to all the terms included in its lexicon. Moreover, each tax topic features terms that are not included in its lexicon, and can even feature terms from the other topic's lexicon. As a result, our approach is more robust to semantic misspecification than typical lexicon-based approaches as their impact is relatively small compared to that of the observed dataset. 

The additional benefit of using an LDA model is that the topic assignment on any given token is based not only on the topic distributions, but also on the other tokens in the document. In Appendix \ref{sec:pred_app} we compare the two approaches, and show that prevalence measures derived from a two-step LDA approach have much higher predictive power for future tax changes.

\subsection{Constructing measure of tax policy signals}

The final step of our text-analytic approach is to transform the estimation results from the topic model into numerical measures that reflect the prominence of tax-related discussions by the president. That is, we aim to construct a quantitative measure of the prominence of signals about tax policy over time, which does not follow directly from our topic model. The estimated topic model allows us to estimate the mixing proportion $\doublehat{\bm{\theta}}_d$ for each document $d=1, \ldots,D$ as the posterior means of the second-step LDA. These estimates signify what proportion of its words is associated with a given topic: documents for which $\doublehat{\theta}_{d,k}$ is estimated to be high discuss topic $k$ for a significant portion relative to documents for which $\doublehat{\theta}_{d,k}$ is low. We now use this property to aggregate estimation results for individual documents to create a measure for the topics' \emph{prevalence}, or popularity, and its evolution over documents registered over time. Because we are using our measure together with macroeconomic data in the empirical models discussed later, we opt for aggregating the documents to quarterly frequency. 

Formally, let $T$ denote the total number of quarters, and let $T_d \in (0, T]$ denote the normalised date corresponding to the publication of document $d$. For the measure in quarter $t$, we then average over all documents published in the period $(t-1,t]$ to obtain the measure
\begin{equation*}
    \text{Prevalence}_{t,k} = \frac{\sum_{d=1}^{D}\doublehat{\theta}_{d,k} \mathbbm{1}\left(t-1 < T_d \leq t \right)}{\sum_{d=1}^{D} \mathbbm{1}\left(t-1 < T_d \leq t \right)}, \qquad t = 1, \ldots, T, \quad k = 1, \ldots, K.
\end{equation*}
To illustrate the proposed prevalence measure, consider two extreme cases:
\begin{itemize}
    \item If all tokens in all documents in the time period belong to topic $k$, we have $\doublehat{\theta}_{d,k}=1$, $\doublehat{\theta}_{d,l}=0$ for $l\neq k$ for all documents $d$. In that case $\text{prevalence}_{t,k} = 1$, while $\text{prevalence}_{t,l} = 0$ for all other topics $l\neq k$.
    \item If all tokens in all documents are uninformative, so equally belonging to $k=1,\ldots,K$ topics, that is $\doublehat{\theta}_{d,k} = 1/K$ for all $k=1,\ldots,K$, then the prevalence of all topics is $1/K$, implying that all topics are appearing equally likely for the time period considered.
\end{itemize}
Arguably, the myriad of choices made in pre-processing, topic model specification and measure creation may make our final measures seem arbitrary. Indeed, while we made those decisions generally in accordance with standards used in the literature, alternative choices appear equally plausible to justify. Therefore, rather than arguing that our choices are the optimal ones, we empirically investigate if our resulting estimates have the properties we attribute to a measure of tax policy signals. In the next section we assess the topics estimated through our two-step LDA approach and evaluate our model's ability to properly classify the tax content of the documents. 

\section{Identified policy topics} \label{sec:res_topics}
In this section we investigate in how far the fitted topic model captures our concepts of policy topics, with special attention to the tax (increase and decrease) topics. We first evaluate the two steps of constructing tax topics in detail. Next, we also briefly consider the other topics in order to understand how well the topic model captures the general essence of the speeches.

\subsection{Tax topics}\label{sec:tax_topics}

\begin{figure}[t]
\centering
\begin{subfigure}{0.49\textwidth}
\caption{Tax increase topic}
\includegraphics[width=0.95\linewidth]{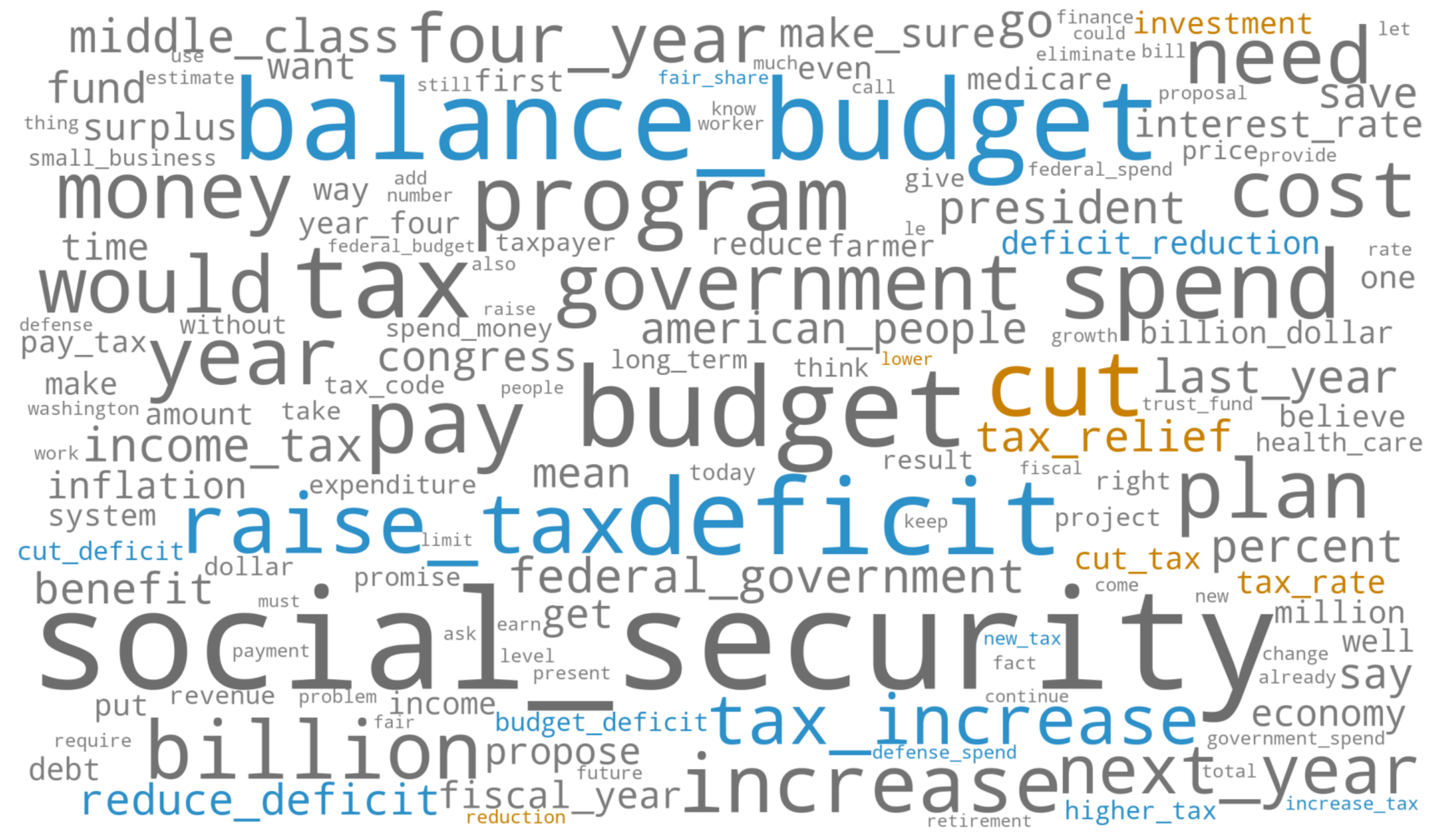}
\label{fig:wc_inc}
\end{subfigure}
\begin{subfigure}{0.49\textwidth}
\caption{Tax decrease topic}
\includegraphics[width=0.95\linewidth]{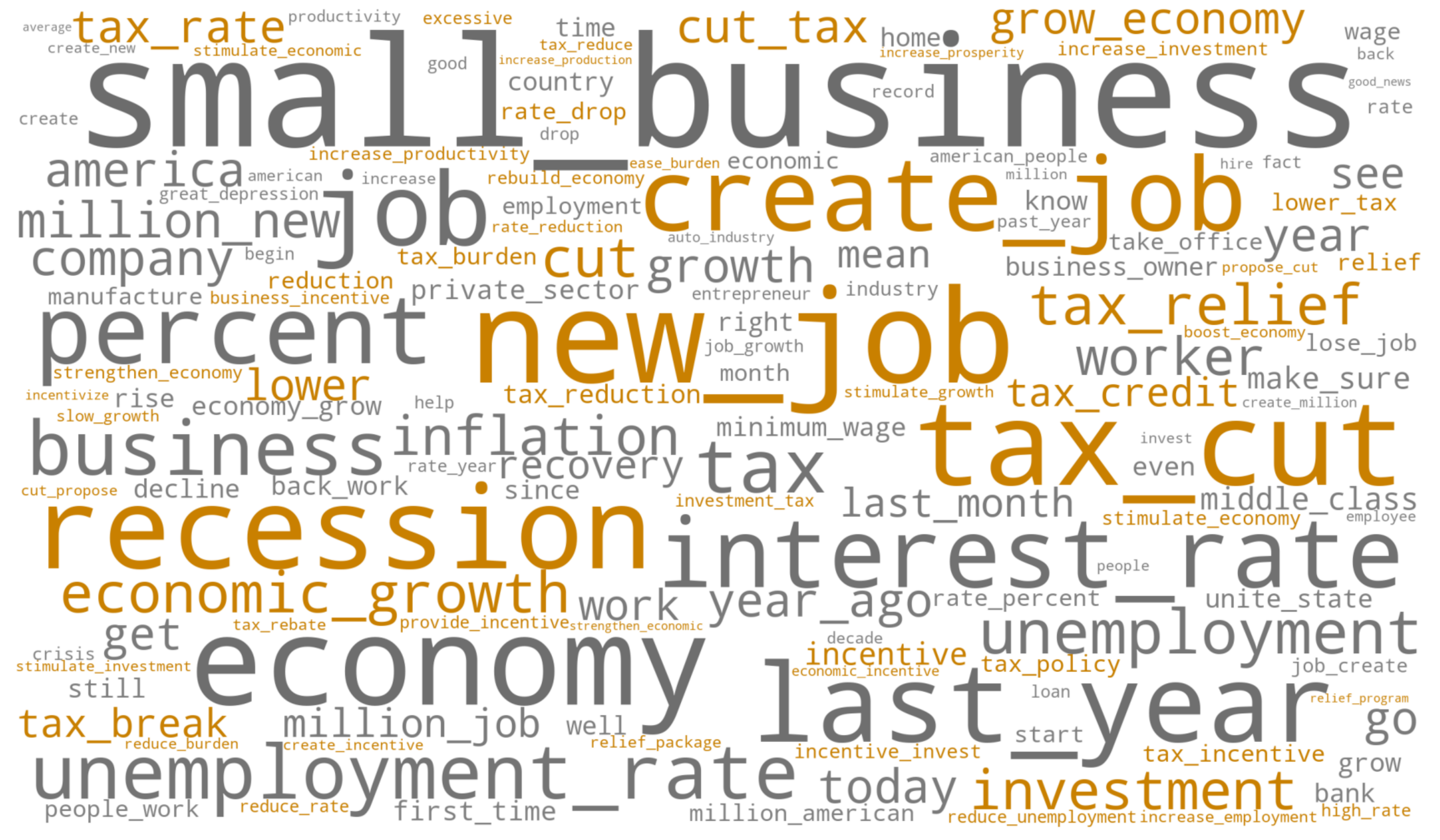}
\label{fig:wc_dec}
\end{subfigure}
\caption{Wordclouds of the estimated tax topics. The size of a term represents the probability attributed to it in topic's distribution. The terms included in the lexicon for tax increase (decrease) are highlighted in blue (orange).}
\end{figure}

The general tax topic from the first-stage, unsupervised LDA, combined with our lexicons, is used in the second-stage guided LDA to obtain the \emph{tax increase} and \emph{tax decrease} topics presented in Figures \ref{fig:wc_inc} and \ref{fig:wc_dec}, respectively. The terms in our lexicons driving the prior for tax increase (decrease) are highlighted in blue (orange). Although both distributions have many terms in common with the general tax topic estimated in the first step,\footnote{The wordcloud for the first-stage general tax topic is shown in Appendix \ref{sec:wordclouds}.} there are crucial differences extending beyond the terms whose prior probability was modified. Those include such terms as `social security' which is used predominantly when discussing tax increases, and `small business' which is referenced often when discussing tax decreases. Lastly, we can see that the topics still feature, to some extent, terms which we determined relate to tax changes in the other direction (e.g. `cut tax' still appears in the `tax increase' topic). This is an example of how the data overrides the priors. Apparently, even when discussing increasing taxes, the presidents sometimes make a reference to tax cuts.

In both steps we verify that the identified tax topics capture all of the content devoted to tax policy changes. We do that by inspecting the probabilities assigned to tax-related terms by the non-tax topics. For example, in the first step we find that the topic devoted to state and local policies has the second highest probability of using the term ``tax'', which is still over 100 times smaller than the probability assigned to that term by the tax topics. Overall, in both steps the identified tax topics account for over 99\% of the usage of the term ```tax''.

\begin{figure}[t]
    \centering
    \includegraphics[width=\columnwidth]{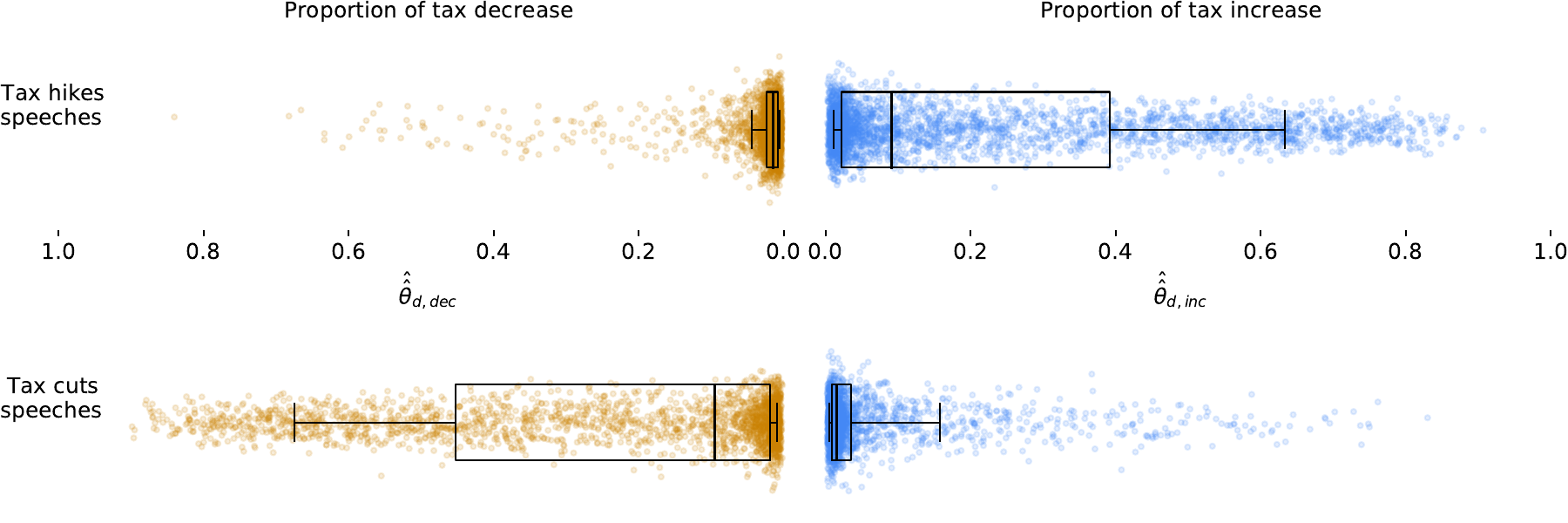}
    \caption{Mixing proportions of \textit{tax increase} and \textit{tax decrease} topics for paragraphs in speeches announcing tax changes. Upper (lower) row shows documents coming from speeches that announce tax hikes (cuts). Boxes show the 1st, 2nd and 3rd quartile. The whiskers show 1st and 9th decile}
    \label{fig:theta_boxplots}
\end{figure}

For our approach, the estimated mixing proportions of each document are more important than the topic-term distributions. For our method to make sense, we specifically need to verify that the estimated mixing proportions for the two tax topics do indeed reflect our interpretations of the texts. Given the size of our dataset with 1,119,200 individual paragraphs, such analysis is possible only for a subset of documents. We first consider paragraphs from speeches that \citet{RomerRomer10} and \citet{yang2007chronology} list as announcements of tax policy changes. Figure \ref{fig:theta_boxplots} shows that, as expected, documents coming from speeches announcing tax hikes tend to have much higher mixing proportion for the \textit{tax increase} topic, and the same is true for those announcing tax cuts and the \textit{tax decrease} topic. In case of some documents we see that the opposite mixing proportion is high. This can be at least partially explained by the fact that some legislated changes include measures in both directions, lowering some taxes while raising others. Finally, in both groups a lot of documents do not relate to tax changes at all, since some of the speeches considered (e.g. \textit{State of the Union Addresses}) concern more issues than just taxation.

Out of those documents we randomly select 100 for which either $\doublehat{\theta}_{d,inc}>0.3$ or $\doublehat{\theta}_{d,dec}>0.3$ and compare the estimated mixing proportions with our interpretation of their content. Additionally, we consider the estimated maximum a posteriori (MAP) topic assignment labels for particular tokens.\footnote{For a token $w_{d,n}=v_i$ the maximum a posteriori (MAP) topic assignment estimate is $z_{d,n}^{*} = \argmax_{k} \doublehat{\theta}_{d,k} \doublehat{\phi}_{k,i}$} While those labels are not directly necessary for our analysis, they help visualise the clustering property of LDA. In Tables \ref{tb:tax_inc} and \ref{tb:tax_dec} we present an example for each direction of change, where tokens assigned to tax increase (tax decrease) topic are highlighted in blue (orange).\footnote{Words not in bold are removed during pre-processing.} We present details of this part of analysis and summarise our findings for the rest of the selected speeches in Appendix \ref{sec:speeches}. We find that overall the estimated mixing proportions correctly capture the direction of the implied changes.

\begin{table}
\caption{Example of a ``tax increase'' speech} \label{tb:tax_inc}
\fbox{
  \parbox{0.9\textwidth}{\small
\begin{quote}
    That's my \clt{t_inc}{commitment}. That's what's at \clt{t_inc}{stake in this election}. \clt{t_inc}{Change} is a \clt{t_inc}{future} where we have to \clt{t_inc}{reduce our deficit}, but do it in a \clt{t_inc}{balanced way}. And I've \clt{t_inc}{signed} a \clt{t_inc}{trillion dollars}' \clt{t_inc}{worth} of \clt{t_inc}{spending cuts}; I \clt{t_inc}{intend} to do more. But if we're \clt{t_inc}{serious} about \clt{t_inc}{reducing the deficit}, we've got to \clt{t_inc}{ask the wealthiest} \clt{t_inc}{Americans} to \clt{t_inc}{go back to the tax rates} they \clt{t_inc}{paid} when \clt{t_inc}{Bill Clinton} was in \clt{t_inc}{office}. Because, \clt{t_inc}{listen}, a \clt{t_inc}{budget} is about \clt{t_inc}{priorities}; it's about \clt{t_inc}{values}. And I'm not \clt{t_inc}{going} to \clt{t_inc}{kick} some \clt{t_inc}{kid} off of \clt{t_inc}{Head Start} so I can \clt{t_inc}{get} a \clt{t_inc}{tax break}. I'm not \clt{t_inc}{going} to \clt{t_inc}{turn} \clt{t_inc}{Medicare} into a \clt{t_inc}{voucher} just to \clt{t_inc}{pay} for \clt{t_inc}{another} \clt{t_inc}{millionaire}'s \clt{t_dec}{tax cut}. That's not who we are.
\end{quote} 
\textbf{President Barack Obama, 2012-11-05.}\\
\textit{Context}: One of President's Obama campaign speeches in which he advocated for an increase in taxation, in particular for the wealthiest Americans.\\
\textit{Future impact}: Raised the rate and introduced new surtax on capital and investment gains for households with income over \$250,000 as part of financing of Medicare.\\
$\doublehat{\theta}_{d,\text{inc}}=0.80, \quad \doublehat{\theta}_{d,\text{dec}}=0.03$
  }
}
\end{table}
\vspace{-0.2cm}
\begin{table}
\caption{Example of a ``tax decrease'' speech} \label{tb:tax_dec}
\fbox{
  \parbox{0.9\textwidth}{\small
\begin{quote}
    The \clt{t_dec}{tax reductions} I am \clt{t_dec}{recommending}, \textbf{together} with this \textbf{broad} \clt{t_dec}{upturn} of the \clt{t_dec}{economy} which has \clt{t_dec}{taken place} in the \clt{t_dec}{first half} of this \clt{t_dec}{year}, will \clt{t_dec}{move} us \clt{t_dec}{strongly} \clt{t_dec}{forward} \textbf{toward a goal} this \textbf{Nation} has not \clt{t_dec}{reached} \clt{t_dec}{since} 1956, 15 \clt{t_dec}{years} ago: \clt{t_dec}{prosperity} with \clt{t_dec}{full employment} in \textbf{peacetime}.
\end{quote}
\textbf{President Richard Nixon, 1971-08-15.}\\
\textit{Context}: Address to the Nation Outlining a New Economic Policy: ``The Challenge of Peace''. President Nixon announces a package of measures meant to stimulate the economy by decreasing taxation.\\
\textit{Future impact}: The package was voted in later that year. \\
$\doublehat{\theta}_{d,\text{inc}}=0.05, \doublehat{\theta}_{d,\text{dec}}=0.44$
}
}
\end{table}
\vspace{-0.5cm}
Looking at the estimated topic assignments for individual tokens we can see the clustering property of LDA. In many cases terms that are not clearly related to tax changes are classified as referring to one of the tax topics. This is the result of the context in which they appear - because so much of the rest of the speeches relate to tax increase or tax decrease the likelihood that they relate to one of those topics is higher. Even terms that intuitively refer to changes in one direction can be classified as referring to changes in the opposite direction (e.g.~`tax break' assigned to the tax increase topic in the 2012-11-05 speech). This property of the topic model is a crucial improvement over a simple lexicon-based approach, making it much less sensitive to misspecification.

It is important to note that despite the clustering property of LDA some paradoxical classifications can occur. This happens when the terms used in the text are not sufficient to capture the meaning contained in the syntax. For instance, speeches along the lines of ``we are not going to increase taxes'' would likely be classified as information about a future tax hike. This is inevitable given the \emph{bag-of-words} assumption underlying the LDA topic model. Addressing this would require methods able to recover the true intention of statements and their interpretation in a broader context. However, in our context, this is difficult and often impossible even for well-informed political observers. Thus, it cannot be reasonably expected to be achieved perfectly by any algorithm, and we believe it should therefore also not be held against our approach based on the bag-of-words assumption. Moreover, such mistakes are likely to average out over all speeches within each quarter.

\begin{figure}[t]
    \centering
    \includegraphics[width=0.99\columnwidth]{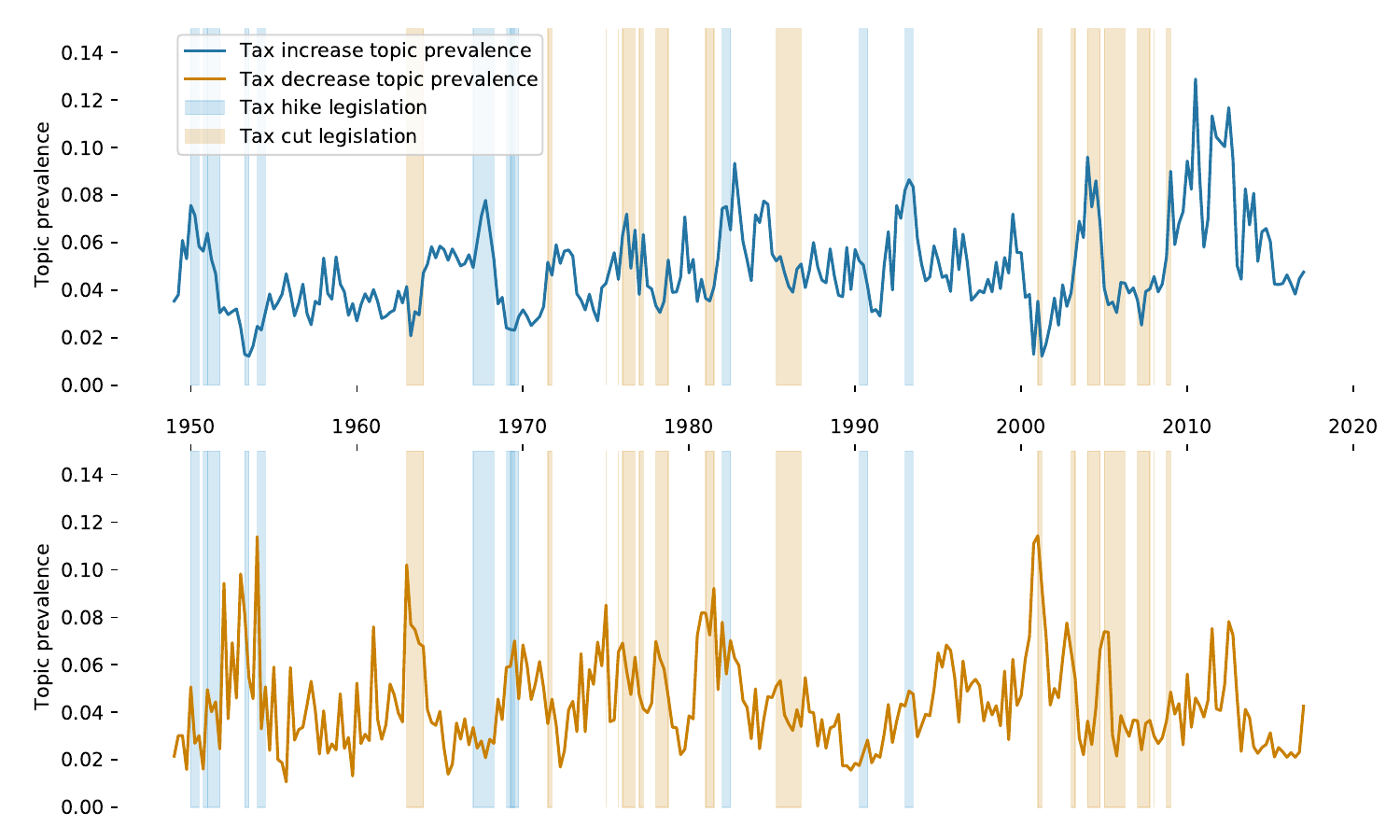}
    \caption{The prevalence of the \textit{tax increase} (upper) and \textit{tax decrease} (lower) topics. The shaded areas in blue (orange) indicate periods of legislative lag of tax hikes (cuts) as identified by \citet{yang2007chronology}.}
    \label{fig:prev_tax_topics}
\end{figure}

Our final diagnostic check in this section concerns the aggregated quarterly prevalence measures of tax increase and decrease, as presented in Figure \ref{fig:prev_tax_topics}, along with periods of legislative lag of tax hikes (blue) and cuts (orange) as identified by \citet{yang2007chronology}. Visual inspection shows that the topic prevalence of the relevant direction - but not the opposite one - generally increases leading up to an actual tax change in that same direction, with the peak around the enactment date. While by no means a formal analysis, this does seem to indicate that our identified tax increase and decrease topics have predictive power for actual tax changes. We investigate this more formally in the next section, but first we briefly investigate the other topics.

\subsection{Other identified topics}\label{sec:other_topics}
In addition to the two tax topics, we identify 24 other topics. While those topics are not the focus of our study, they can give an indication of the overall quality of the model. Their distributions remains relatively unchanged between the first and the second step of the topic model and the majority can clearly be attributed to specific policy issues such as public health, trade, and foreign policy. The distributions of all  the topics are presented in Appendix \ref{sec:speeches}.

\begin{figure}[t]
    \centering
    \begin{subfigure}[t]{0.49\textwidth}
        \centering
        \includegraphics[width=0.95\linewidth]{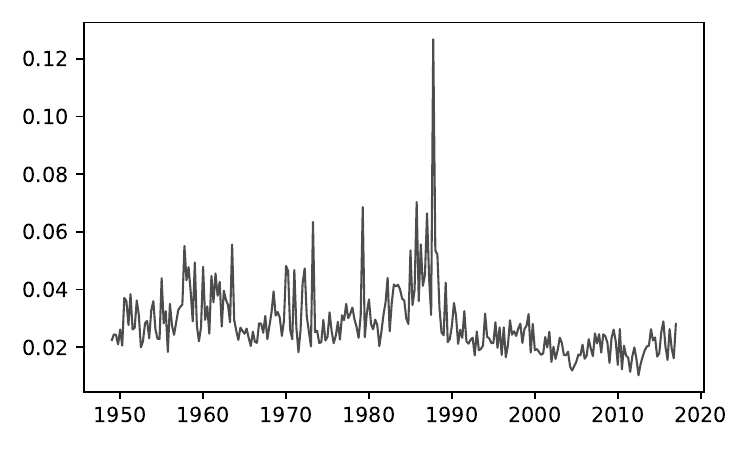}
        \caption{Cold War}
        \label{fig:prev_coldwar}
    \end{subfigure}%
    \hfill
    \begin{subfigure}[t]{0.49\textwidth}
        \centering
        \includegraphics[width=0.95\linewidth]{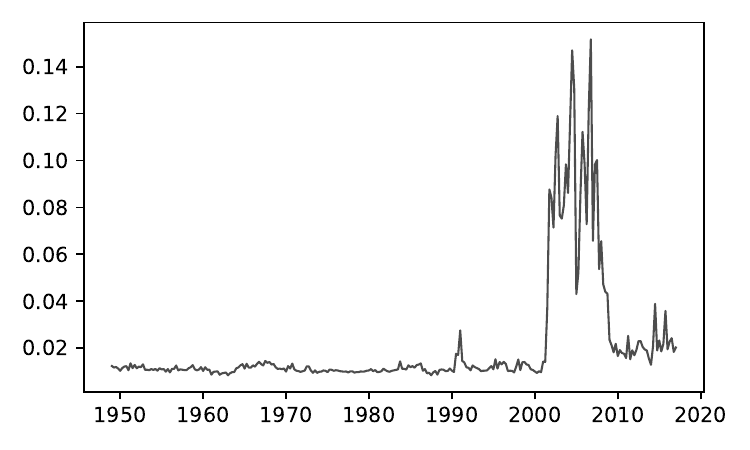}
        \caption{War on Terror}
        \label{fig:prev_warterror}
    \end{subfigure}
    \vspace{0.5cm}
    \caption{Prevalences of selected topics.}
    \label{fig:prev_other}
\end{figure}

In several cases we can quite intuitively see how the changing importance of certain issues is reflected in our prevalence measures. Perhaps the best examples are the \textit{Cold War} and \textit{War on Terror} topics displayed in Figure \ref{fig:prev_other}. The prevalence of the \textit{Cold War} topic has oscillated for decades, reaching a sharp peak in the late 1980s after which it declined considerably. In case of the \textit{War on Terror} topic we see a slight peak in the early 1990s - likely the discussion surrounding the First Gulf War, a steep rise after the 9/11 attacks and a considerable decline since Barack Obama entered office, reflecting a different approach of the new president. This also shows that our LDA approach is flexible enough to handle variation over time.

\subsection{Tax prevalence predictability} \label{sec:pred}
In this section we investigate how well the tax prevalence measures predict future tax changes. We initially estimate predictive regressions of various tax change series on our tax prevalence measures, assessing the strength of the correlation between tax changes and the lags of our measures by $F$-tests for the non-significance of the measures in the predictive regressions. We find that the general tax prevalence measure obtained from the first-stage unsupervised LDA is not a powerful predictor for any tax change measure considered. However, the second-step prevalence measures -- capturing tax cuts and hikes -- are strong predictors of future tax changes (regardless of how they are measured), with $F$-statistics well above the cut-off.
The superior predictive power of the second-step measures is also reflected in the much higher partial $R^2$ statistics. Details of this analysis can be found in Appendix \ref{sec:pred_app}.

To study (non-)predictive relationships in more detail, as well as to investigate the (conditional) exogeneity of our tax topics and to what extent they are driven by other factors such as macroeconomic conditions, we carry out Granger causality tests. We investigate contemporaneous correlations in Appendix A.1.

To test for Granger causality we consider a large VAR which includes  a large number of possibly relevant macroeconomic and financial variables. We also include variables that incorporate information on future policies concerning spending \citep{Ramey11,Ramey18} and taxation \citep{leeper2012quantitative}, as well as prevalence measures for the other topics.
This allows us to determine in how far our tax prevalence series contain unique predictive power for the various tax change measures that is not found in other series. Similarly, by testing for Granger causality from a variety of macroeconomic variables (including several tax change measures) to the two tax prevalence series we can determine what the causes of potential endogeneity are. To handle the high-dimensionality of the VAR, we use the post-selection test of \citet{hecq2023granger} which provides valid inference about Granger causality after selection of relevant covariates via the lasso.\footnote{All tests are based on a VAR with six lags and an intercept. Nonstationary macro variables are transformed to first differences of logs.}

\begin{figure}[t]
\begin{center}
\includegraphics[width=\textwidth]{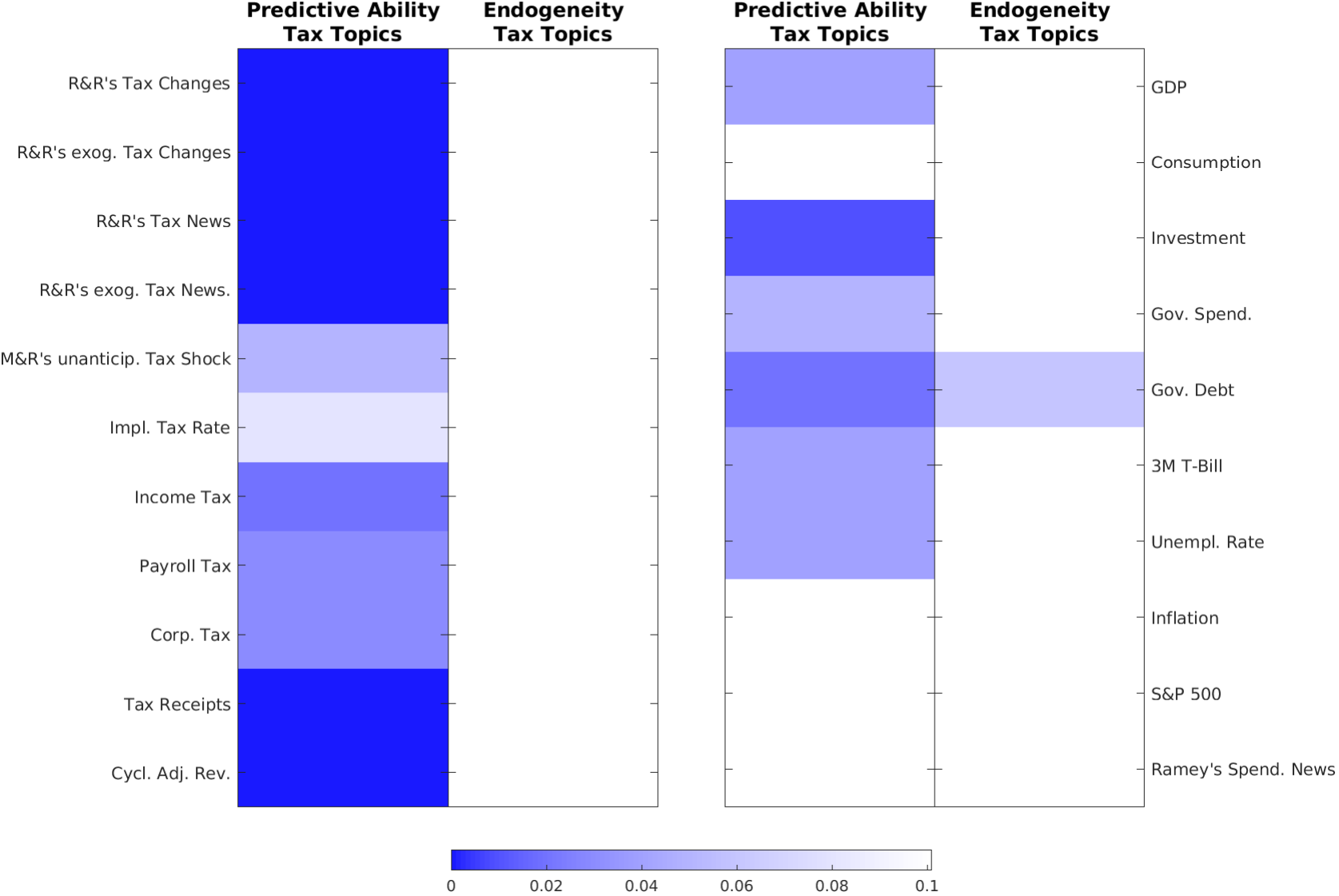}
\caption{$p$-Values from Granger causality tests. The left column of each panel (`\textit{Predictive Ability}') visualises to what extent the two tax prevalence measures jointly Granger-cause various measures of tax liability changes and macroeconomic variables. The right column of each panel (`\textit{Endogeneity}) depicts the extent to which the tax prevalence measures are Granger-caused by measures of tax liability changes and macroeconomic variables.}
\label{fig:GC1}
\end{center}
\end{figure}

The left column of Figure \ref{fig:GC1} illustrates which measures of tax changes are Granger-caused by our prevalence measures. It also shows whether any tax measure is predicting our tax prevalence measures. The tax prevalence topics Granger-cause all tax change measures. While predictive power is higher for aggregate measures of tax changes (federal revenue, or narratives of legislated tax changes), the tax topics also Granger-cause corporate, income, and payroll taxes. Most notably, we find that even  RR's exogenous news narrative, as well as \citepos{mertens2014reconciliation} unanticipated tax narrative, are Granger-caused by the tax prevalence measures. Those results support our hypothesis that tax policy changes are signalled by the president well ahead of time, even when the motivation is exogenous to recent economic conditions and the implementation lag is short. We further find evidence (p-value $=0.075$) that our prevalence measures Granger-cause implicit tax rates, which are considered as proxy for tax news in the literature.\footnote{We use the risk-adjusted implicit tax rate from \citet{leeper2012quantitative}. We use rates with maturity of one year which are considered by the authors to best predict tax changes in the near future.} Since investors are forward looking, the yield spread between taxable treasury bonds and tax-exempt municipal bonds likely mirrors anticipated federal tax changes. The results above seem to suggest that investors' expectations are driven - at least to some extent - by tax-related communications of the president. That is, if the president signals that federal taxes (on treasury bond income) may increase, investors will, as a response, demand higher yields on treasury bonds. Thus, in that case, our prevalence measure - and not the yield spread - would get the timing of the arrival of news regarding future tax changes right.  
In contrast, we do not find that any of the considered tax (news) measures contain information that help predict the tax prevalence measures.

We next investigate whether the prevalence measures predict macroeconomic and financial variables and vice versa. Results displayed in the right column of Figure \ref{fig:GC1} show that the tax prevalence topics Granger-cause the information typically included in a fiscal VAR. We find evidence for predictability in the other direction only for government debt. This suggests that the prevalence measures capture only information about tax changes that are not systematically correlated with short-run changes in economic activity, but are rather driven by other (long-run) policy goals, for example reducing the debt burden.

We also evaluate predictive relationships between the tax topics and other prevalence measures (results are shown in Figure \ref{fig:GC2} in Appendix \ref{sec:pred_app}). We find that some topics related to policies that affect the federal budget Granger-cause the tax topics. In particular, the \textit{War on Terror} topic Granger-causes the tax prevalence measures. We will use this insight when selecting control variables among the topics in the causal analysis in Section \ref{sec:struct} to eliminate possible endogeneity issues. 

\section{Output effects of tax news}\label{sec:struct}

\subsection{Empirical framework}
\label{sec:empirics}
Our tax prevalence measures do not have a direct, economically meaningful (monetary) quantitative interpretation. In order to enable a causal analysis using these measures, we link our tax prevalence measures to data that are informative about the monetary value of expected tax changes. We cast the analysis in the LP-IV framework of \cite{SW2018}, treating the prevalence series as imperfect measures of an unobservable tax news shock. That is, we assume that the prevalence measures (and their lags) are correlated with the unobserved tax news shock but contain measurement error. This setup  requires an observable, endogenous variable driven by the tax news shock, such as some monetary measure of future taxes, in order to treat the prevalence measures (and their lags) as instruments for this endogenous variable and estimate relative impulse responses by LP-IV.

As argued above, we believe that expectations and decisions of economic agents are influenced not only by news about implemented future tax liability changes but also by `noise'; that is, new information about possible future policy changes that are deemed plausible \textit{ex-ante},
even though such policy plans never come to fruition \textit{ex-post}. For LP-IV to work, the unobserved tax news shock must lead \textit{on average} to a change in the observable endogenous variable. Identifying the causal effect of a `noisy' tax news shock using our prevalence measures as instruments is possible as long as new \textit{ex-ante} information on future policies does not systematically differ from (\textit{ex-post}) implemented policy outcomes. Hence, we assume that economic agents' expectations regarding future tax changes may be noisy but are correct \textit{on average}.\footnote{We do not identify noise and news separately. The information we hope to capture relates to \emph{both} news about future fundamentals as well as changes influencing agents' beliefs only. One might argue that not being able to identify either component limits a (deep) structural interpretation. This criticism holds, however, for a variety of empirical papers. The previous investigation on the informational content of the tax prevalence measures indicates that the news signal about future tax changes, although not noise free, is strong enough allowing for a meaningful interpretation. Generally, noise and news are tightly related, further distinguishing both components and identifying them separately -- if this is at all possible -- is beyond the scope of this paper. We refer to \citet{chahrour2018news} for a detailed analysis of this relationship.}

Impulse responses are based on the following regressions
\begin{equation}
\label{eq:LPIV}
y_{t+h} = {\bm \gamma}_{h} + \beta_{1,h} x_{1,t} + {\bm x}_{2,t}' {\bm \beta}_{2,h} + u_{t,h} \qquad \text{ for } h=0,1,2,...
\end{equation}
where $x_{1,t}$ is an endogenous measure of future tax liability changes and ${\beta}_{1,h}$ is the the coefficient of interest, capturing the response to a tax news shock in the outcome variable $y_t$, $h$ periods after the shock occurred. Further, ${\bm x}_{2,t}$ gathers relevant control variables and ${\bm \gamma}_{h}$ captures deterministic components. Estimation and inference via two-stage least squares is straightforward.

We also need to choose the appropriate endogenous variable that carries the information about future tax changes. The construction of the tax prevalence measures does not rely on an accurate definition of the timing of future policy changes. Our prevalence measures predict tax hikes or cuts implemented as early as the following quarter, and are even stronger predictors for policy changes three or four quarters ahead. Thus, our measures contain information about (expected) tax changes implemented \textit{throughout} the following year(s) and hence do not allow us to pin down the precise timing of future tax liability changes. Indeed, economic agents are likely uncertain about the exact implementation timing of future tax changes, and this uncertainty is reflected in a predictive power over a longer time horizon. If we were to instrument (endogenous) tax changes in a given quarter only (vs. over a longer period of time), we would likely disregard meaningful information regarding (expected) tax changes at a different point in time and only partially capture expectations.

Bearing this in mind and to broadly capture `future tax changes' over a longer period of time, we propose to use the following rolling average as our endogenous measure of future tax changes:
\begin{equation}
\label{eq:AccTaxChanges}
    x_{1,t} = \sum_{j=M}^N \Delta T_{t+j},
\end{equation}
where $\Delta T_t$ is the (endogenous) tax liability change in period $t$. This simple aggregation scheme is  better suited for an analysis of expectations, considering the uncertainty of economic agents regarding the exact timing of expected future policy changes. In addition, this makes specifying the exact timing of when a tax change is expected no longer necessary. 

The choice of window start- and endpoints $M$ and $N$ is of crucial importance for identifying possible anticipation effects. For example, setting $M=1$ and $N=4$ aggregates all tax changes throughout the coming year, starting the following quarter. If the tax prevalence measures were most strongly correlated with tax changes implemented one quarter ahead, short-term (exogenous) variation in tax liabilities would drive the response of output to a tax news shock and we would inadvertently purge out information contained in the prevalence measures about expected tax changes in the more distant future. In addition, the resulting short anticipation horizon would likely not allow for large anticipation effects, since economic agents would have little time to react, and a potential impact from anticipation would likely by dominated by implementation effects. effects.\footnote{\cite{mertens2012empirical} also find that the shorter the anticipation horizon, the weaker the (negative) pre-implementation (or anticipation) effects on output.} In contrast, by increasing $M$ (and $N$) we would put less emphasis on the correlation between the prevalence measures and soon-to-be implemented tax changes and our prevalence measures would instead instrument tax changes that are expected in the more distant future only. This would allow for a longer anticipation horizon and potentially stronger anticipation effects.

\subsection{Empirical results}
\label{sec:results}

Based on the local projection regressions in (\ref{eq:LPIV}) we estimate responses of log real output and log real government tax receipts to (noisy) news of future tax cuts. To construct our measure of future tax changes as in (\ref{eq:AccTaxChanges}), we use RR's rich narrative account of all legislated federal tax liability changes in the US from 1947 to 2006. RR link tax changes directly to the legislative process and, thus, their narrative record contains precise information about the timing of tax changes.  We use our tax prevalence measures (including eight lags) as instruments. The deterministic components ${\bm \gamma}_{h}$ consist of an intercept, a linear, and a quadratic trend. 12 lags of the dependent variable are always included in the set of regressors. This is a rather conservative choice and possibly more than necessary to capture the temporal dependency in $y_t$. However, it is in line with the suggestions in \citet{MOPM20} regarding robust inference in the presence of persistent data.

Furthermore, ${\bm x}_{2,t}$ includes 12 lags of the following macroeconomic controls: log real GDP, log real government spending, log real government debt, and the 3-month Treasury Bill. Based on the analysis in section \ref{sec:pred}, it may be necessary to include lags of other, non-tax topics as well. In particular, the two prevalence measures for \textit{War on Terror} and \textit{Natural Resources, Energy \& Technology} Granger-cause the tax topics and are related to political decisions likely to affect the federal budget. 

In addition, we take into account other policy changes that may coincide with tax changes. In Appendix \ref{sec:pred_app}, we investigate the contemporaneous relationships between the tax topics and the other identified topics. This analysis reveals that the topic related to regulatory policies (\textit{Public Administration}) as well as the one we associate with polices aiming at improving long-run economic conditions (\textit{Economic Development}) correlate significantly with the two tax prevalence measures. Thus, we add contemporaneous values of these two series to the set of regressors in ${\bm x}_{2,t}$. Finally, the president may at times refer to past tax changes in his speeches. This does not add new information about future tax changes and can therefore not be considered as news.  12 lags of RR's series of all tax changes are therefore included as additional controls in ${\bm x}_{2,t}$.\footnote{Our baseline specification does not include contemporaneous values of most controls; in Section \ref{sec:robustness_main} we investigate the sensitivity to this choice. In Appendix \ref{sec:robustness} we further investigate whether results are sensitive to the choice of control variables. We find that using various subsets of the regressors discussed above, does not affect the shape and magnitude of responses very much.}

Before discussing the response of economic activity to news about future tax changes, we investigate the strength of the correlation between the tax prevalence measures and our endogenous measure of future tax changes $x_{1,t}$. Table \ref{tab:FStatsMain} shows $F$-statistics on the joint exclusion of our tax prevalence measures from the first-stage of the LP-IV regressions, as specified above, for various choices of $M$ and $N$. The tax prevalence measures are fairly strong predictors of future tax changes implemented throughout the coming two years. Predictability is weaker for soon-to-be implemented tax changes. Predictive power also rapidly declines for tax changes implemented in the more distant future (after six quarters).   \\

\begin{table}
\caption{$F$-statistics on the joint exclusion of the tax prevalence measures from the first-stage regression associated with (\ref{eq:LPIV}) for different choices of $M$ and $N$.} \label{tab:FStatsMain}
\begin{center}
\scalebox{0.95}{
\begin{tabular}{r c c c c c c c c}
\toprule
$M$ & $0$ & $1$ & $2$ & $3$ & $4$ & $5$ & $6$ & $7$ \\
$N $ & $ 3 $ & $ 4 $ & $ 5 $ & $ 6 $ & $ 7 $ & $ 8 $ & $ 9 $ & $10$ \\
\midrule
$F$-statistics & 1.6 &  16.7 &   21 &   30.1 &   17.5  &  3.4  &  0.5 & 0.05 \\
\bottomrule
\end{tabular}
}
\end{center}
\end{table}

Estimated impulse responses of output and tax receipts to news about a future tax cut for different values of $M$ and $N$ are displayed in Figure \ref{fig:IRsMain}. To make results quantitatively comparable across different specifications, responses are normalised such that the log change of government tax receipts is minus unity at its trough. Point estimates are shown together with 68\% and 90\% confidence intervals.

We start by investigating responses of output and government receipts for a shorter anticipation horizon. Setting $M=1$ and $N=4$ means that we use our prevalence measures to instrument tax news about policy changes that will have been implemented throughout the following four quarters. Results are displayed in Figures \ref{fig:IRsMain}(a) and \ref{fig:IRsMain}(b). Tax receipts react almost instantaneously, declining for several quarters and reaching the trough response one year after the arrival of the news. The associated response of output is insignificant and close to zero for about four quarters. After that, output increases significantly and steadily, peaking after 10 quarters. Comparing the timing of the decline in tax receipts and the increase in output indicates that the expansion in GDP is primarily driven by (post-)implementation effects. The results do not provide evidence for significant anticipation effects. Instead, the slightly delayed but large positive response in GDP is qualitatively and quantitatively comparable with findings in the literature  on effects of aggregate tax changes \citep{RomerRomer10,mertens2012empirical}, providing additional evidence for large output effects due to tax cuts. 

As alluded to above, it could be that pre-implementation effects depend on the anticipation horizon and, thus,  on the specification of the aggregation window in (\ref{eq:AccTaxChanges}). \cite{mertens2012empirical} indeed find that the longer the anticipation horizon, the stronger the pre-implementation contraction in output. By setting $M=5$ and $N=8$ we use the tax prevalence measure to instrument news about tax changes in the more distant future, i.e.~policy changes that will have been implemented throughout four consecutive quarters in one year from now.\footnote{This specification is not chosen randomly. \cite{mertens2012empirical} find that the median anticipation horizon (the time frame from signing of the bill to implementation) of anticipated tax changes (i.e. tax changes that are not implemented in the quarter they are enacted) is six quarters.} Output and tax receipts responses for this specification are displayed in Figures \ref{fig:IRsMain}(c) and \ref{fig:IRsMain}(d). 

The results are strikingly different compared to the findings discussed before. First, the contraction in tax receipts is delayed, with a trough response eight quarters after the arrival of the news (instead of four as in the previous experiment). The response of aggregate tax revenue is initially slightly positive (although barely significant) and only turns negative after a year. Overall, the response of tax receipts confirms that by setting $M=5$ and $N=8$ we indeed seem to successfully instrument policy changes in the more distant future, which means a longer anticipation horizon.
Second, and most importantly, the delayed tax cut does not simply trigger a delayed output expansion with an otherwise similar pattern as observed in Figure \ref{fig:IRsMain}(a). While GDP  still does not react at impact, output starts to decline after two quarters for about a year and only starts to recover simultaneously with the implementation of the tax cut. We thus conclude that the initial contraction of output is due to anticipation (resp. pre-implementation) effects. Once the policy is implemented and taxes are cut, economic recovery follows a similar path as in the previous specification, but with a smaller peak expansion of output.

Although we use a conceptually fundamentally different approach, our results confirm the evidence about contractionary effects of anticipated tax cuts in \cite{mertens2012empirical}. We investigate next whether we also find that the longer tax cuts are anticipated the more severe the decline in economic activity. To extend the anticipation horizon, we move the aggregation window (by increasing $M$ and $N$) and instrument tax changes that will have been implemented in an increasingly distant future.  The results displayed in Figure \ref{fig:IRsMain}(e) and \ref{fig:IRsMain}(f) confirm indeed that: (i) the further in the future the tax cut is implemented the stronger the contractionary pre-implementation effects on output; and (ii) the smaller the post-implementation effect triggering economic expansion.\footnote{In Appendix \ref{sec:robustness} we investigate to what extend the window size in the construction of $x_{1,t}$ (i.e. $N-M$) affects the results. We find that allowing for a larger window over which future tax changes are aggregated does not qualitatively alter the results. In particular we find that fixing $M=1$ and extending the time window over which we aggregate tax policy changes (i.e. increasing $N$) does not lead to results indicating any anticipation effects. It seems indeed that the stronger correlation of the tax prevalence measures with soon-to-be implemented policy changes dominates.}

\subsection{Robustness of results}\label{sec:robustness_main}
\begin{figure}
\begin{center}
\includegraphics[width=\textwidth]{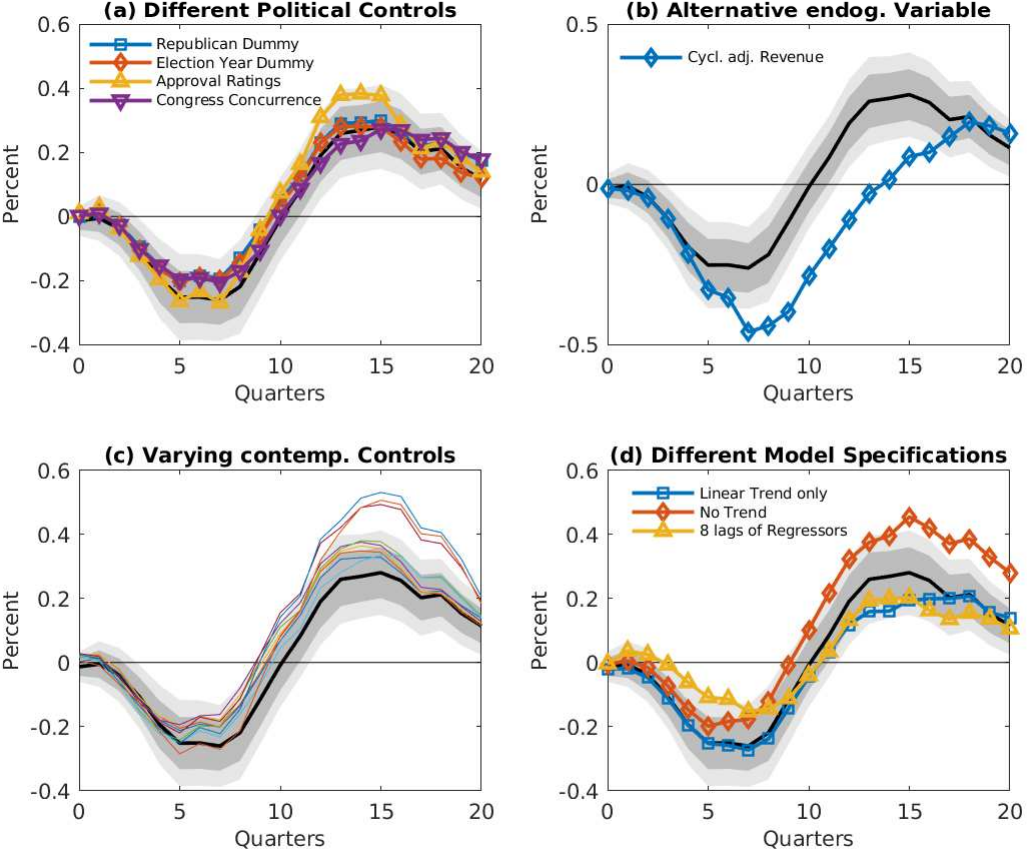}
\caption{Output responses for various specifications of (\ref{eq:LPIV}). For comparison, point estimates and confidence intervals of output responses based on the benchmark specification are plotted as well (solid black lines and grey shaded area, respectively). The upper left \textbf{quadrant (a)} shows output responses when contemporaneous values of specific political controls are added to the set of benchmark regressors. The upper right \textbf{quadrant (b)} shows the reaction of GDP when cyclically adjusted revenue is used  to construct the endogenous measure of future tax changes in (\ref{eq:AccTaxChanges}). The lower left \textbf{quadrant (c)} shows output responses using the benchmark set of regressors and ``\textit{all possible}'' recursive identification assumptions (``orderings'') in the LP-IV regressions. The lower right \textbf{quadrant (d)} shows output responses for different specification of deterministic terms in the benchmark LP-IV regression as.}
\label{fig:IRsRobustness}
\end{center}
\end{figure}

We investigate next whether the findings above are sensitive to specific modelling choices. We present results of this robustness analysis for the arguably more interesting case of a longer anticipation horizon ($M=5$ and $N=8$) for which we find contractionary pre-implementation effects on output. For an easier comparison, output responses to news about a tax cut, as displayed in Figure \ref{fig:IRsRobustness}, always contain estimates (and confidence intervals) of the benchmark specification presented in Section \ref{sec:results}. As before, output responses are normalised, such that the trough response of tax receipts is minus one. Robustness checks for a shorter anticipation horizon ($M=1$ and $N=4$) can be found in Appendix \ref{sec:robustness}.\footnote{Results for a shorter anticipation horizon are qualitatively and quantitatively very similar to the findings discussed in Section \ref{sec:results}.}

It is plausible that political considerations drive the president's communication. One could imagine that Republican presidents would conduct different economic policies and that people are more likely to expect tax cuts from them than from Democratic presidents (or would deem initial announcements more credible). Similarly, upcoming presidential elections may affect our results if the president is campaigning for re-election with potentially less credible announcements. Moreover, presidents with low approval ratings may be inclined to propose policy reforms that increase popularity. And finally, the president's relation with Congress may influence the tone in his public communications, as well as the kind of policy reforms brought forward. Figure \ref{fig:IRsRobustness}(a) shows output responses incorporating these different political controls and indicates that none of these aspects seem to matter.\footnote{To control for the president's popularity in public, we consider quarterly averages of Gallup's job approval ratings. The House and Senate concurrence is the percentage of members of congress who agree with the president's position on a roll call vote. Both series are obtained from UC Santa-Barbara's American Presidency Project \url{https://www.presidency.ucsb.edu/}. We add current observations and 12 lags of these additional controls to the benchmark set of regressors in the local projection regressions.}

As endogenous measure of future taxes we have used RR's narrative records which quantify, in monetary terms, all post-war legislated tax policy changes. RR's narrative series is an obvious choice due to the similarity of the underlying auxiliary data (presidential speeches and addresses). However, alternative measures of tax liability changes may be considered. A standard measure to proxy tax changes is cyclically adjusted revenue.\footnote{We use the revenue measure constructed by RR which is expressed in terms of change in revenues as a percent of GDP. Thus RR's narrative series as well as their measure of cyclically adjusted revenue show tax changes in the same unit and are hence comparable.} Figure \ref{fig:IRsRobustness}(b) shows the response of output when using this endogenous series to construct the measure of future tax changes in (\ref{eq:AccTaxChanges}). Our previous finding of an initial pre-implementation decline in economic activity is robust to this alternative choice. Using cyclically adjusted revenue as endogenous variable results in an even more severe pre-implementation contraction. In addition, we observe a slightly smaller (and delayed) expansion in output after the tax cut is implemented.
 
The underlying identification assumption of the LP-IV approach is that our tax prevalence series are exogenous conditionally on the control variables. As motivated in \ref{sec:results}, we include two prevalence measures as contemporaneous controls; one series related to regulatory policies and one associated with polices aimed at improving long-run economic conditions. All other control variables in ${\bm x}_{2,t}$ are included as lagged values. This is akin to a recursive identification structure in a VAR where tax news is ordered after the prevalence measures associated with regulatory and long-term economic policies. 
Although we find that our tax prevalence measures cannot be predicted by economic conditions (see Section \ref{sec:pred}), it may still be the case that, for example, current recessionary shocks trigger an instantaneous discussion about tax cuts. In this case, the considered identification scheme would be flawed. To mitigate these concerns to some extent, we estimate the output response using the set of regressors of our benchmark specification but varying the variables which we include as contemporaneous regressors (thus essentially considering all possible ordering in the recursive identification scheme). The results, displayed in Figure \ref{fig:IRsRobustness}(c), indicate that the shape of output responses is little affected by different orderings; at worst, our benchmark specification may somewhat underestimate expansionary effects due to implemented tax cuts.

Finally, Figure \ref{fig:IRsRobustness}(d) shows the effect of using different model specifications. Including eight instead of 12 lags of all past regressors has little quantitative impact on the response of output. Excluding any trend leads - unsurprisingly - to a more persistent response.
 
\begin{figure}
\begin{center}
\includegraphics[width=\textwidth]{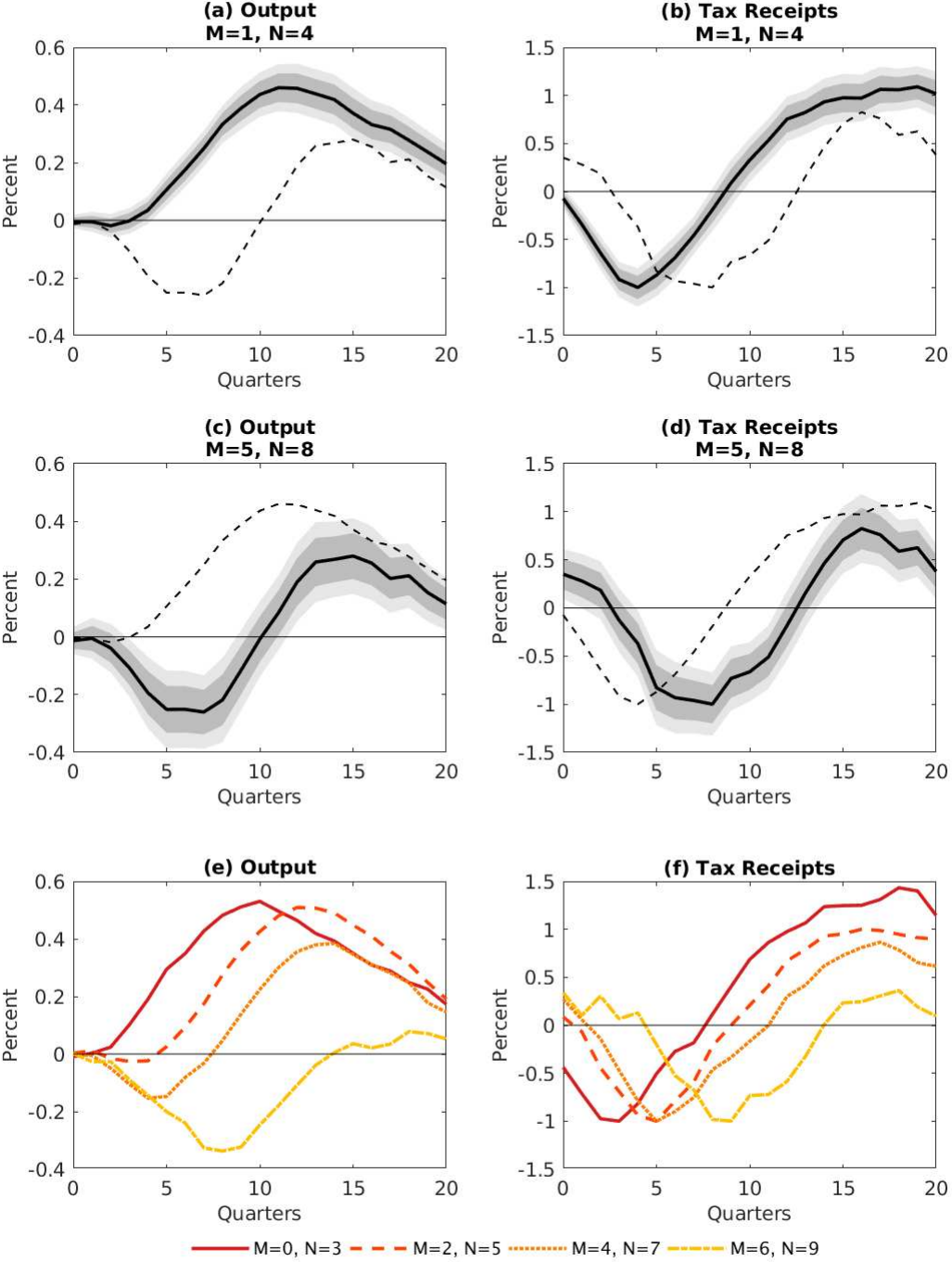}
\caption{The first two rows, i.e. Figures \textbf{(a)-(d)}, show output and government revenue responses to news about a future tax cut. Displayed are impulse responses (solid black lines) as well as 68\% and 90\% confidence intervals (gray shaded area). For ease of comparison, impulse responses with longer, resp. shorter, anticipation horizon are displayed in black dashed lines. The last row, i.e. Figures \textbf{(e)-(f)}, show impulse responses for different values of $M$ and $N$.}
\label{fig:IRsMain}
\end{center}
\end{figure}

Appendix \ref{sec:robustness} gathers additional robustness checks. We investigate the role of our instruments by comparing the benchmark results presented in Figure \ref{fig:IRsMain} with impulse responses based on the regression specified in equation \eqref{eq:LPIV} estimated by OLS, using RR's exogenous tax changes to construct $x_{1,t}$. We find that anticipated future tax cuts do not lead to an initial contraction of economic activity when RR's narrative is used to construct future tax changes. There is no significant response of GDP prior to implementation of the tax change, suggesting that our tax prevalence measures contain important information regarding anticipation of future tax changes. 
 
\section{Conclusion}\label{sec:conclusion}
In this paper we analyse the public communications of U.S.~presidents to identify information regarding planned tax reforms. Our semi-supervised topic modelling approach allows us to automatically determine what issues are mentioned in the texts, and in particular when the president discusses tax changes. Based on those results we create a measure of prevalence for the \textit{tax increase} and \textit{tax decrease} topics, which reflects the relative prominence across time of those policy issues on the president's agenda. We show that they are strong predictors for a variety of measures of tax changes, including those usually considered as unanticipated. This predictive power is not connected to other macroeconomic conditions, but is rather the effect of capturing the legislative process behind those changes.

To analyse the effects of tax news, we use our identified topics to instrument future changes in tax liabilities and estimate the reaction of output to news about a future tax cut. We find that output when tax cuts are implemented. However, a longer anticipation horizon leads to initial contractionary effects in output and also dampens (post-)implementation expansion.

There are several methodological implications of our findings. First, we show that presidential speeches contain signals about future tax changes, and as such might be crucial for solving the problem of fiscal foresight. Moreover, it is possible to meaningfully quantify those signals using an automated text-analytic approach. In particular, we propose a two-step estimation procedure for the LDA model, in which informative lexical priors are constructed to differentiate between similar topics. Our approach, while relatively simple, proves crucial for determining the direction of the discussed changes.

As an extension, our two-step approach could prove helpful in identifying news regarding different types of taxes (cf. \citet{mertens2013dynamic}), however the identification of relevant lexicons would be more challenging than in this study. Our approach could in principle be used to identify news about government spending. Because text-analytic methods scale well, in both cases the analysis can be augmented by considering other channels of communication such as congressional records, news outlets or even Twitter. 

From an econometric perspective an important extension would be to integrate estimation of the text model and the causal econometric model into a single procedure. While considering two separate steps has the advantage that one can build on established methods for both the text mining and the causal modeling, a single unified framework could potentially better exploit the causality that runs in both directions. By directly extracting only the exogenous components of the president's speeches, one could measure the (noisy) news content of speeches more accurately and thereby mimic economic agents' reception of potential news more closely. Such an approach, however, would require the development of new models and estimation methods combining text and economic data in a single approach. Given the current abundance of textual data sources, such an approach would be greatly beneficial to economic analysis, and therefore seems a promising future research agenda.

\newpage

\bibliographystyle{chicago}
\bibliography{refs}

\begin{appendices}

\counterwithin{figure}{section}
\counterwithin{table}{section}
\counterwithin{equation}{section}

\section{Additional results for the prevalence and tax news measures}\label{sec:results_app}
\subsection{Results about predictability} \label{sec:pred_app}

In Tables \ref{tab:LDA1}-\ref{tab:RsqITR} we investigate the predictive ability of the tax prevalence measures for a variety of tax change measures by means of $F$-tests and $R^2$ statistics. We regress various measures of tax changes on the tax topics and a set of control variables. More specifically, we estimate the following predictive regressions
\begin{equation}
    \label{eq:1stStage}
    \Delta T_t = \gamma_0 + \bm{\gamma}_1^\prime \bm{z}_{t-j} + \sum_{i=1}^{12} \bm{\beta}_i^\prime \bm{x}_{t-i} + u_t \quad \text{ for } j=0,1,2,\ldots,6,
\end{equation}
where $\Delta T_t$ is a tax change series, $\bm{z}_t$ contains tax topic prevalence measures, and $\bm{x}_t$ contains control variables. $\gamma_0$ is an intercept. We estimate (\ref{eq:1stStage}) for several endogenous tax change series. Controls always include lags of the dependent variable, first differences of log GDP and first differences of log government spending, as well as interest rates.\footnote{Our analysis is build on time series sampled at quarterly frequency from 1949Q1-2007Q4. Details about the economic data used in this paper can be found in Appendix \ref{sec:other_data}.} The strength of the correlation between the tax topics contained in $\bm{z}_{t-j}$ and $\Delta T_t$ $j$ quarters ahead can be assessed by comparing the (marginal) $F$-statistic to their $\chi_p^2/p$ critical values, with the degrees of freedom $p$ equal to the number of variables in $\bm{z}_t$.\footnote{At 95\%, these values are equal to 3.84 and 3.00 for $p=1$ and $p=2$, respectively.} As measures of tax changes we consider changes in (cyclically adjusted) tax receipts and two of RR's narrative measures of tax changes.\footnote{RR link tax changes directly to the legislative process and, thus, their narrative contains more precise information about the timing of tax changes. We consider RR's series quantifying expected changes in current tax liabilities at time of implementation (i.e. for the first fiscal year the law was scheduled to be in effect) and the associated discounted present value at time of enactment. The latter measure is interpreted by RR as a signal containing tax news (in the strict sense of providing (almost) perfect foresight, i.e. there is no uncertainty of whether the tax change will happen or not). Further, RR construct  sub-components of those measures capturing ``exogenous'' tax changes only (i.e. changes not driven by recent economic conditions). All narratives taken from RR are expressed as ratio to nominal GDP.}

Table \ref{tab:LDA1} reveals that the prevalence measure obtained from the first-stage unsupervised LDA (i.e. the general tax topic) is not a powerful predictor for any tax change measure considered. This finding serves as a note of caution against a too confident use of unsupervised text analytics. 
Table \ref{tab:LDA2} shows the (marginal) $F$-statistic on the (joint) exclusion of the tax cut and tax hike topics from various predictive regressions. Regardless of how we measure future tax changes, the $F$-statistic is, for several forecast horizons, well above the cut-off for any reasonable significance level.

The above conclusions are also supported when considering the predictive power of the various measures in terms of change in $R^2$. Table \ref{tab:RsqLDA1} shows that the general tax topic from first-stage LDA offers little improvement over the controls in predicting the measures of tax changes. The tax cut and tax hike topics obtained from the second-stage LDA jointly offer far greater improvement in predictive power across all measures of tax changes, as seen in Table \ref{tab:RsqLDA2}. Additionally, Tables \ref{tab:Lexicon} and \ref{tab:RsqLexicon} suggest that those topics tend to be stronger predictors than analogous measures constructed using a simple lexicon-based approach.

One may also use the same predictive regressions and $F$-statistics for testing whether these measures can be used as strong instruments of the regressand using the procedure laid out in \cite{stock2005testing}. Comparing to the rule of thumb cut-off point of 10 \citep{staiger1997instrumental}, but also to the stricter critical values tabled in \cite{stock2005testing} that guarantee small size distortion,\footnote{As argued by \cite{stock2005testing}, while the cut-off of 10 is reasonable to control bias, for having low size distortion we need to consider larger cut-off points. To control size distortion at maximum 5\% (10\%), the 95\% critical values are 16.38 (8.96) and 19.93 (11.59) for 1 and 2 instruments respectively. We see that the $F$-statistics for the second-stage tax topics regularly exceed the cut-offs, unlike the first stage tax topic or the lexicon-based series, and we may therefore reject the null hypothesis that the second-stage tax topics are weak instruments. This suggests that the second-stage tax topics series can not only be considered strong predictors of future tax changes, but could also be used as strong instruments for tax changes.}

One may argue that, due to their construction, our prevalence series may pick up news signals about other policy plans that would affect the federal budget in the future, notably government spending. However, as indicated in the last rows of Tables \ref{tab:LDA2} and \ref{tab:RsqLDA2}, our measures do not possess much power in predicting federal spending (news).\footnote{Overall, the above findings are robust across different specifications of the regression in (\ref{eq:1stStage}). Unreported results, considering other control variables such as the party of the sitting president, election year, etc., do not alter the results.}

For comparison, we additionally summarise $F$-tests and changes in $R^2$ indicating the predictive strength of the implicit tax rates from \citet{leeper2012quantitative} (see Tables \ref{tab:ITR} and \ref{tab:RsqITR}). 

\begin{figure}
\begin{center}
\includegraphics[width=\textwidth]{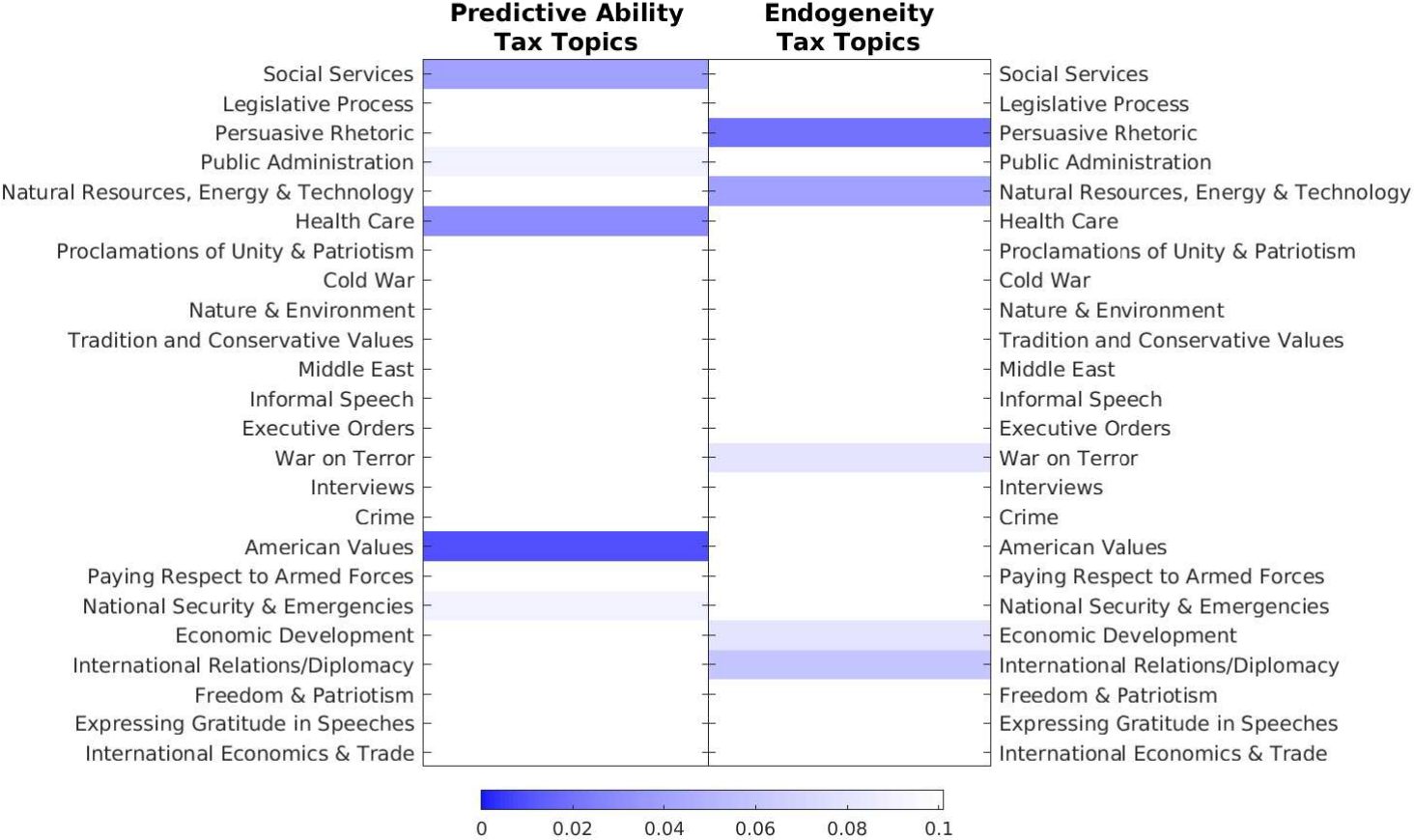}
\caption{$p$-Values from Granger causality tests. The left column (`\textit{Predictive Ability}') visualises to what extend the two tax prevalence measures jointly Granger-cause other identified topics. The right column (`\textit{Endogeneity}) depicts the extend to which the tax prevalence measures are Granger-caused by other identified topics.}
\label{fig:GC2}
\end{center}
\end{figure}

Finally, In Figure \ref{fig:GC2} we report the results of the Granger causality tests of \citet{hecq2023granger} described in Section \ref{sec:pred}. 

As mentioned in Section \ref{sec:results}, we investigate next the contemporaneous relationship between our tax prevalence measures and other identified topics. Figure \ref{fig:Correlations} displays estimates of the correlation coefficient as well as 95\% confidence intervals. In particular, we find that the topic related to regulatory policies (\textit{Public Administration}) as well as the one we associate with polices aiming at improving long-run economic conditions (\textit{Economic Development}) correlate significantly with the the two tax prevalence measures. 

\begin{table}
\caption{$F$-statistics on the exclusion of the prevalence measure obtained from the first-stage unsupervised LDA.} \label{tab:LDA1}
\begin{small}
\begin{center}
\begin{tabular}{p{4.2cm} c c c c c c c c c}
\toprule
Forecast Horizon & 0 & 1 & 2 & 3 & 4 & 5 & 6 & $H_{max}$ & $F_{max}$ \\
\midrule
\textbf{Tax change measures}: & & & & & & & & & \\[0.1cm]
Change in aggregate tax receipts  & 0.03 & 0.00   & 0.11 & 0.30 & 0.63 & 0.04 & 0.31 & 4 & 0.63  \\
\rowcolor{gray!20} Change in cycl. adj. revenue & 0.06 & 2.44 & 3.40  & 2.10 & 3.18 & 0.00    & 2.90  & 2 & 3.4  \\
R\&R (2010) -- Expected change in tax liabilities at time of implementation
& 0.73 & 5.38 & 1.59 & 1.04 & 0.83 & 0.58 & 6.37 & 6 & 6.37 \\
\rowcolor{gray!20} R\&R (2010) --
Expected change in tax liabilities at time of enactment
& 11.31 & 1.57 & 0.33 & 2.19 & 0.07 & 0.4 & 0.27 & 0 & 11.31 \\[0.2cm]
\midrule
\textbf{Spending change measures}: & & & & & & & & & \\[0.2cm]
Change in government 
consumption expenditures & 1.73 &   0.00  &   3.60  &   1.30  &  3.38 & 0.01  &  0.13 & 2 & 3.6  \\[0.1cm]
\rowcolor{gray!20} Ramey's (2011) spending news  
& 0.11  &  0.66 &   0.18  &  1.17  &  0.65 & 0.12 &  0.03  &  3  & 1.17\\
\bottomrule
\end{tabular}
\end{center}
\end{small}
\end{table}
\begin{table}
\caption{Percentage point change in $R^2$ from including tax topic prevalence measures obtained from the first-stage unsupervised
LDA.} \label{tab:RsqLDA1}
\begin{small}
\begin{center}
\scalebox{0.9}{
\begin{tabular}{p{4.2cm} c c c c c c c c c c}
\toprule
Forecast Horizon &  0 & 1 & 2 & 3 & 4 & 5 & 6 & $H_{max}$ & $\Delta R^2_{max}$ \\
\midrule
\textbf{Tax change measures}: & & & & & & & & \\[0.1cm]
Change in aggregate tax receipts  
& 0.01 & 0 & 0.04 & 0.1 & 0.2 & 0.01 & 0.1 & 4 & 0.2 \\
\rowcolor{gray!20} Change in cycl. adj. revenue 
& 0.02 & 0.83 & 1.16 & 0.71 & 1.07 & 0 & 0.98 & 2 & 1.16 \\
R\&R (2010) -- Expected change in tax liabilities at time of implementation
& 0.23 & 1.67 & 0.5 & 0.32 & 0.25 & 0.18 & 1.9 & 6 & 1.9 \\
\rowcolor{gray!20} R\&R (2010) --
Expected change in tax liabilities at time of enactment
& 2.99 & 0.43 & 0.09 & 0.6 & 0.02 & 0.11 & 0.08 & 0 &2.99 \\[0.2cm]
\midrule
\textbf{Spending change measures}: & & & & & & & & & \\[0.2cm]
Change in government 
consumption expenditures & 0.54 & 0 & 1.18 & 0.43 & 1.11 & 0 & 0.04 & 2 & 1.18 \\[0.1cm]
\rowcolor{gray!20} Ramey's (2011) spending news  
& 0.04 & 0.23 & 0.06 & 0.4 & 0.23 & 0.04 & 0.01 & 3 & 0.4\\
\bottomrule
\end{tabular}}
\end{center}
\end{small}
\end{table}

\begin{table}
\caption{$F$-statistics on the joint exclusion of the tax cut and tax hike topics from a simple Lexicon approach.} \label{tab:Lexicon}
\begin{small}
\begin{center}
\scalebox{0.95}{
\begin{tabular}{p{3.4cm} c c c c c c c c c}
\toprule
Forecast Horizon & 0 & 1 & 2 & 3 & 4 & 5 & 6 & $H_{max}$ & $F_{max}$ \\
\midrule
\textbf{Tax change measures}: & & & & & & & & & \\[0.1cm]
Change in aggregate tax receipts & 1.31 & 4.57 & 2.17  & 0.93 & 4.91 & 1.59 & 4.64 & 4 & 4.91 \\
\rowcolor{gray!20} Change in cycl. adj. revenue & 0.39 & 5.24 & 2.07 & 0.63 & 6.98 & 2.14 & 6.08 & 4 & 6.98 \\ 
R\&R (2010) -- Expected change in tax liabilities at time of implementation
& 0.35 & 5.05 & 5.08 & 3.13 & 4.72 & 4.13 & 2.43 & 2 & 5.08  \\
\rowcolor{gray!20} R\&R (2010) -- Expected change in tax liabilities at time of enactment &  12.37 & 7.35 & 4.88 & 3.73 & 2.92 & 0.48 & 1.14 & 0 & 12.37  \\
\midrule
\textbf{Spending change measures}: & & & & & & & & & \\[0.1cm]
Change in government consumption expenditures & 1.18 & 0.88 & 6.24 & 3.23 & 3.86 & 1.27 & 2.55 & 2 & 6.24 \\
\rowcolor{gray!20} Ramey's (2011) spending news & 0.28 & 0.29 & 0.67 & 1.74 & 2.03 & 0.71 & 0.06 & 4 & 2.03\\
\bottomrule
\end{tabular}
}
\end{center}
\end{small}
\end{table}
\begin{table}
\caption{Percentage point change in $R^2$ from jointly including the tax cut and tax hike topics from a simple Lexicon approach.} \label{tab:RsqLexicon}
\begin{small}
\begin{center}
\scalebox{0.9}{
\begin{tabular}{p{4.2cm} c c c c c c c c c}
\toprule
Forecast Horizon &  0 & 1 & 2 & 3 & 4 & 5 & 6 & $H_{max}$ & $\Delta R^2_{max}$ \\
\midrule
\textbf{Tax change measures}: & & & & & & & & & \\[0.1cm]
Change in aggregate tax receipts & 1.09 & 5.94 & 6.7 & 10.7 & 3.9 & 3.51 & 0.45 & 3 & 10.7\\
\rowcolor{gray!20} Change in cycl. adj. revenue &0.06 & 4.05 & 6.83 & 7.26 & 4.22 & 4.7 & 1.97 & 3 & 7.26\\
R\&R (2010) -- Expected change in tax liabilities at time of implementation
& 1.1 & 2.72 & 5.54 & 8.91 & 10.07 & 6.71 & 7.18 & 4 & 10.07\\
\rowcolor{gray!20} R\&R (2010) -- Expected change in tax liabilities at time of enactment & 6.07 & 3.65 & 11.57 & 7.44 & 6.58 & 0.41 & 1.7 & 2 & 11.57\\
\midrule
\textbf{Spending change measures}: & & & & & & & & & \\[0.1cm]
Change in government consumption expenditures & 1.41 & 0.61 & 1.43 & 0.36 & 0.65 & 0.83 & 0.56 & 2 & 1.43\\
\rowcolor{gray!20} Ramey's (2011) spending news & 0.27 & 0.56 & 0.5 & 0.37 & 0.38 & 0.03 & 0.01 & 1  & 0.56  \\
\bottomrule
\end{tabular}
}
\end{center}
\end{small}
\end{table}

\begin{table}
\caption{$F$-statistics on the joint exclusion of the tax cut and tax hike topics from the second-stage semi-supervised LDA.} \label{tab:LDA2}
\begin{small}
\begin{center}
\scalebox{0.95}{
\begin{tabular}{p{3.4cm} c c c c c c c c c}
\toprule
Forecast Horizon & 0 & 1 & 2 & 3 & 4 & 5 & 6 & $H_{max}$ & $F_{max}$ \\
\midrule
\textbf{Tax change measures}: & & & & & & & & & \\[0.1cm]
Change in aggregate tax receipts & 1.76 & 10.25 & 11.65  & 19.87 & 6.48 & 5.77 & 0.70 & 3 & 19.87 \\
\rowcolor{gray!20} Change in cycl. adj. revenue & 0.09 & 6.20 & 10.87 & 11.85 & 6.56 & 7.34 & 2.95 & 3 & 11.85 \\ 
R\&R (2010) -- Expected change in tax liabilities at time of implementation
&  1.76 & 4.44 & 9.56 & 16.76 & 19.23 & 12.05 & 13.09 & 4 & 19.23 \\
\rowcolor{gray!20} R\&R (2010) -- Expected change in tax liabilities at time of enactment & 12.11 & 6.96 & 25.5 & 15.23 & 13.24 & 0.73 & 3.11 & 2 & 25.5 \\
\midrule
\textbf{Spending change measures}: & & & & & & & & & \\[0.1cm]
Change in government consumption expenditures &  2.27 & 0.93 & 2.20 & 0.55 & 0.99 &  1.24 & 0.82 & 0 & 2.27 \\
\rowcolor{gray!20} Ramey's (2011) spending news & 0.38 & 0.80 & 0.73 & 0.54 & 0.54 & 0.04 &    0.01 & 1 & 0.8\\
\bottomrule
\end{tabular}
}
\end{center}
\end{small}
\end{table}
\begin{table}
\caption{Percentage point change in $R^2$ from jointly including the tax cut and tax hike topics from the second-stage semi-supervised LDA.} \label{tab:RsqLDA2}
\begin{small}
\begin{center}
\scalebox{0.9}{
\begin{tabular}{p{4.2cm} c c c c c c c c c}
\toprule
Forecast Horizon & 0 & 1 & 2 & 3 & 4 & 5 & 6 & $H_{max}$ & $\Delta R^2_{max}$ \\
\midrule
\textbf{Tax change measures}: & & & & & & & & & \\[0.1cm]
Change in aggregate tax receipts &0.82 & 2.78 & 1.35 & 0.58 & 2.99 & 1 & 2.86 & 4 & 2.99\\
\rowcolor{gray!20} Change in cycl. adj. revenue &0.27 & 3.44 & 1.4 & 0.42 & 4.47 & 1.43 & 3.96 & 4 & 4.47\\
R\&R (2010) -- Expected change in tax liabilities at time of implementation
& 0.22 & 3.07 & 3.06 & 1.86 & 2.79 & 2.46 & 1.46 & 1 & 3.07\\
\rowcolor{gray!20} R\&R (2010) -- Expected change in tax liabilities at time of enactment &6.19 & 3.84 & 2.61 & 2.01 & 1.59 & 0.27 & 0.63 & 0 & 6.19\\
\midrule
\textbf{Spending change measures}: & & & & & & & & & \\[0.1cm]
Change in government consumption expenditures & 0.74 & 0.58 & 3.92 & 2.08 & 2.49 & 0.85 & 1.73 & 2 & 3.92 \\
\rowcolor{gray!20} Ramey's (2011) spending news & 0.2 & 0.2 & 0.45 & 1.18 & 1.39 & 0.5 & 0.04 & 4 & 1.39\\
\bottomrule
\end{tabular}
}
\end{center}
\end{small}
\end{table}

\begin{table}
\caption{$F$-statistics on the exclusion of implicit tax rates from \citet{leeper2012quantitative}.}\label{tab:ITR}
\begin{small}
 \begin{center}
\begin{tabular}{p{4.2cm} c c c c c c c c c}
\toprule
Forecast Horizon & 0 & 1 & 2 & 3 & 4 & 5 & 6 & $H_{max}$ & $F_{max}$ \\
\midrule
\textbf{Tax change measures}: & & & & & & & & & \\[0.1cm]
Change in aggregate tax receipts & 2.15 & 3.62 & 0.06  & 0.27 & 0.57 & 0.34 & 0.17 & 1 & 3.62 \\
\rowcolor{gray!20} Change in cycl. adj. revenue & 0.42 & 2.50 & 0.27 & 0.75 & 0.28 & 0.44 & 0.01 & 1 & 2.50 \\ 
R\&R (2010) -- Expected change in tax liabilities at time of implementation &  1.51 & 0.36 & 0.81 & 0.05 & 1.57 & 2.69 & 3.80 & 6 & 3.80 \\
\rowcolor{gray!20} R\&R (2010) -- Expected change in tax liabilities at time of enactment & 0.64 & 0.64 & 0.28 & 1.49 & 0.07 & 0.71 & 1.43 & 3 & 1.49\\
\midrule
\textbf{Spending change measures}: & & & & & & & & & \\[0.1cm]
Change in government consumption expenditures &  0.05 & 0.06 & 0.00 & 0.59 & 0.66 &  1.68 & 0.32 & 5 & 1.68 \\
Ramey's (2011) spending news & 2.39 & 0.55 & 1.74 & 1.43 & 0.08 & 0.34 &    1.42 & 0 & 2.39\\
\bottomrule
\end{tabular}
\end{center}
\end{small}
\end{table}
\begin{table}
\caption{Percentage point change in $R^2$ from including implicit tax rates from \citet{leeper2012quantitative}.} \label{tab:RsqITR}
\begin{small}
\begin{center}
\scalebox{0.9}{
\begin{tabular}{p{4.2cm} c c c c c c c c c}
\toprule
Forecast Horizon & 0 & 1 & 2 & 3 & 4 & 5 & 6 & $H_{max}$ & $\Delta R^2_{max}$ \\
\midrule
\textbf{Tax change measures}: & & & & & & & & & \\[0.1cm]
Change in aggregate tax receipts &0.73 & 1.23 & 0.02 & 0.09 & 0.2 & 0.12 & 0.06 & 1 & 1.23 \\
\rowcolor{gray!20} Change in cycl. adj. revenue & 0.15 & 0.91 & 0.1 & 0.28 & 0.1 & 0.16 & 0 & 1 & 0.91\\
R\&R (2010) -- Expected change in tax liabilities at time of implementation
& 0.49 & 0.12 & 0.27 & 0.02 & 0.52 & 0.89 & 1.25 & 6 & 1.25\\
\rowcolor{gray!20} R\&R (2010) -- Expected change in tax liabilities at time of enactment & 0.19 & 0.19 & 0.08 & 0.45 & 0.02 & 0.22 & 0.43 & 3 & 0.45\\
\midrule
\textbf{Spending change measures}: & & & & & & & & & \\[0.1cm]
Change in government consumption expenditures & 0.02 & 0.02 & 0 & 0.23 & 0.26 & 0.67 & 0.13 & 5 & 0.67\\
\rowcolor{gray!20} Ramey's (2011) spending news & 0.88 & 0.21 & 0.65 & 0.54 & 0.03 & 0.13 & 0.54 & 0 & 0.88 \\
\bottomrule
\end{tabular}
}
\end{center}
\end{small}
\end{table}

\begin{figure}
\begin{center}
\includegraphics[width=\textwidth]{img/FigA2_Correlations.pdf}
\caption{Point estimates as well as 95\% confidence intervals of the correlation coefficient between each of the tax prevalence measures (cuts and hikes) and other identified topics.}
\label{fig:Correlations}
\end{center}
\end{figure}

\clearpage

\subsection{Effects of tax news for various model specifications}\label{sec:robustness}

\begin{figure}
\begin{center}
\includegraphics[width=0.95\textwidth]{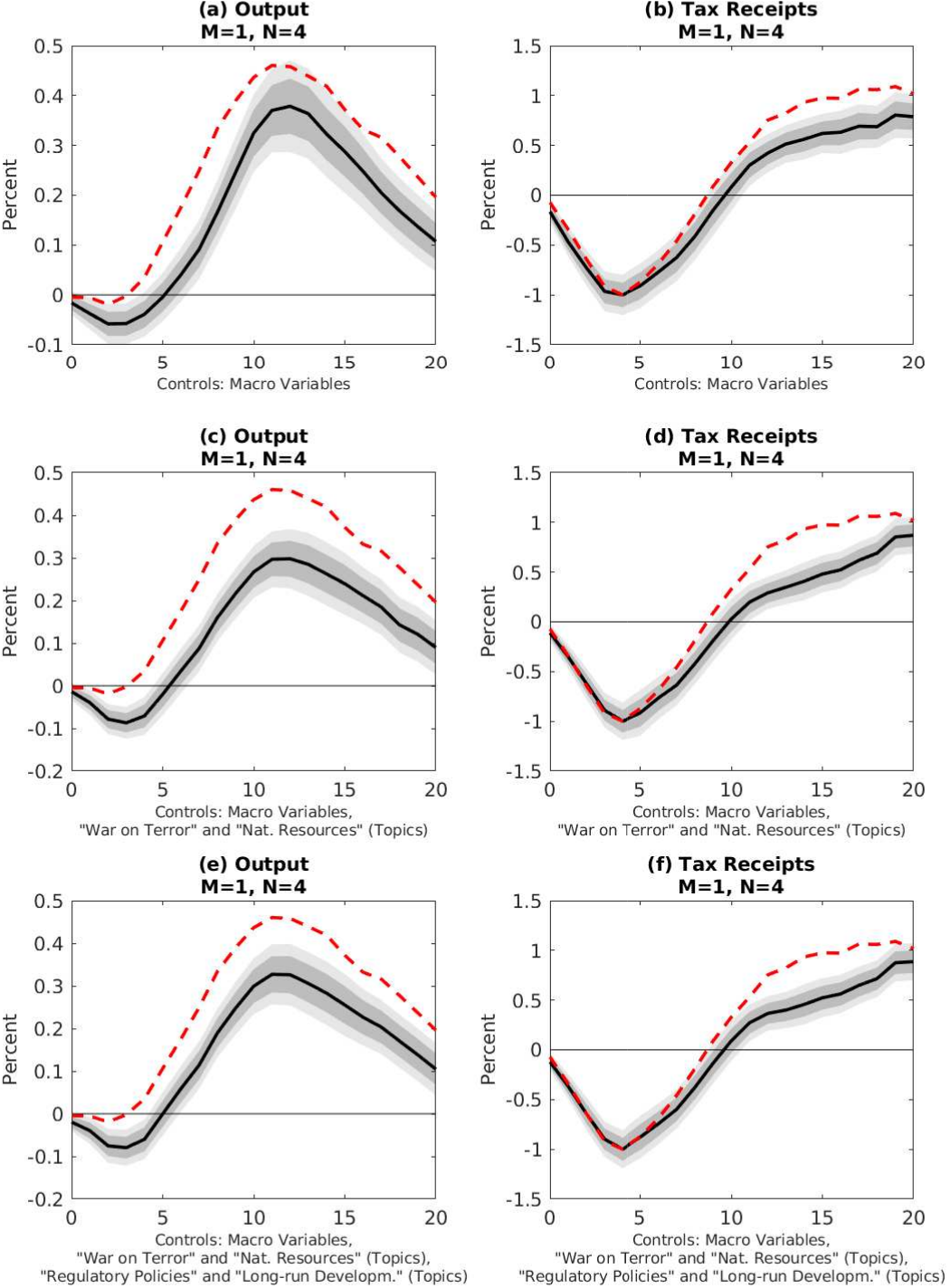}
\caption{Output and government revenue responses to news about a future tax cut with a short(er) anticipation horizon ($M=1$ and $N=4$) for various sets of control variables. Displayed are impulse responses (solid black lines) as well as 68\% and 90\% confidence intervals (gray shaded area). For comparison, impulse responses based on our benchmark set of regressors (cf. Figure \ref{fig:IRsMain}) are also plotted (red dashed lines).}
\label{fig:Controls1}
\end{center}
\end{figure}

\begin{figure}
\begin{center}
\includegraphics[width=0.95\textwidth]{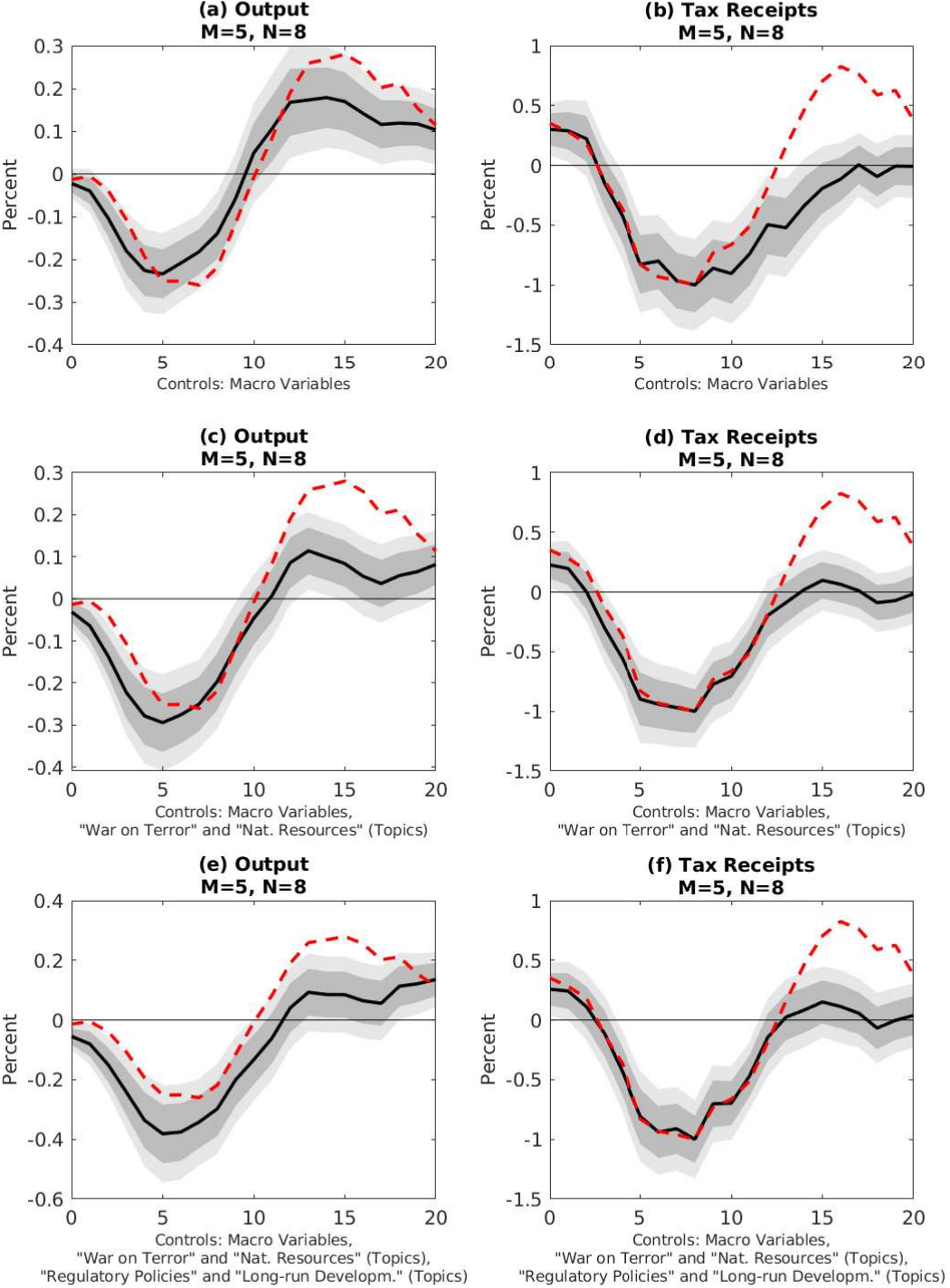}
\caption{Output and government revenue responses to news about a future tax cut with a long(er) anticipation horizon ($M=5$ and $N=4$) for various sets of control variables. Displayed are impulse responses (solid black lines) as well as 68\% and 90\% confidence intervals (gray shaded area). For comparison, impulse responses based on our benchmark set of regressors (cf. Figure \ref{fig:IRsMain}) are also plotted (red dashed lines).}
\label{fig:Controls2}
\end{center}
\end{figure}

In this section we further investigate the robustness of the results presented in Section \ref{sec:results}. We start with checking the effect of modifying the set of controls in the LP-IV regression. Figures \ref{fig:Controls1} and \ref{fig:Controls2} show impulse responses of output and tax receipts to news about a future tax cut for a short(er) and long(er) anticipation horizon, respectively, and for different sets of control variables. Apart from different regressors the model specification corresponds to the benchmark model discussed in Section \ref{sec:results} (i.e. trends, lags, etc.). Responses of both variables don't differ much across the three different sets of controls considered, neither for short(er) nor for long(er) term anticipation. However, and in particular for $M=1$ and $N=4$, it is interesting to note that including in addition past (RR) tax changes (i.e. extending the set of regressors in Figures \ref{fig:Controls1}(e) resp. \ref{fig:Controls1}(f) by past tax changes) leads to a stronger post-implementation response of output, and to a weaker pre-implementation contraction. In contrast, the response of tax receipts is largely unaffected. This is possibly suggesting that tax changes with long phase-ins  have indeed less strong expansionary effects on economic activity.

\begin{figure}
\begin{center}
\includegraphics[width=\textwidth]{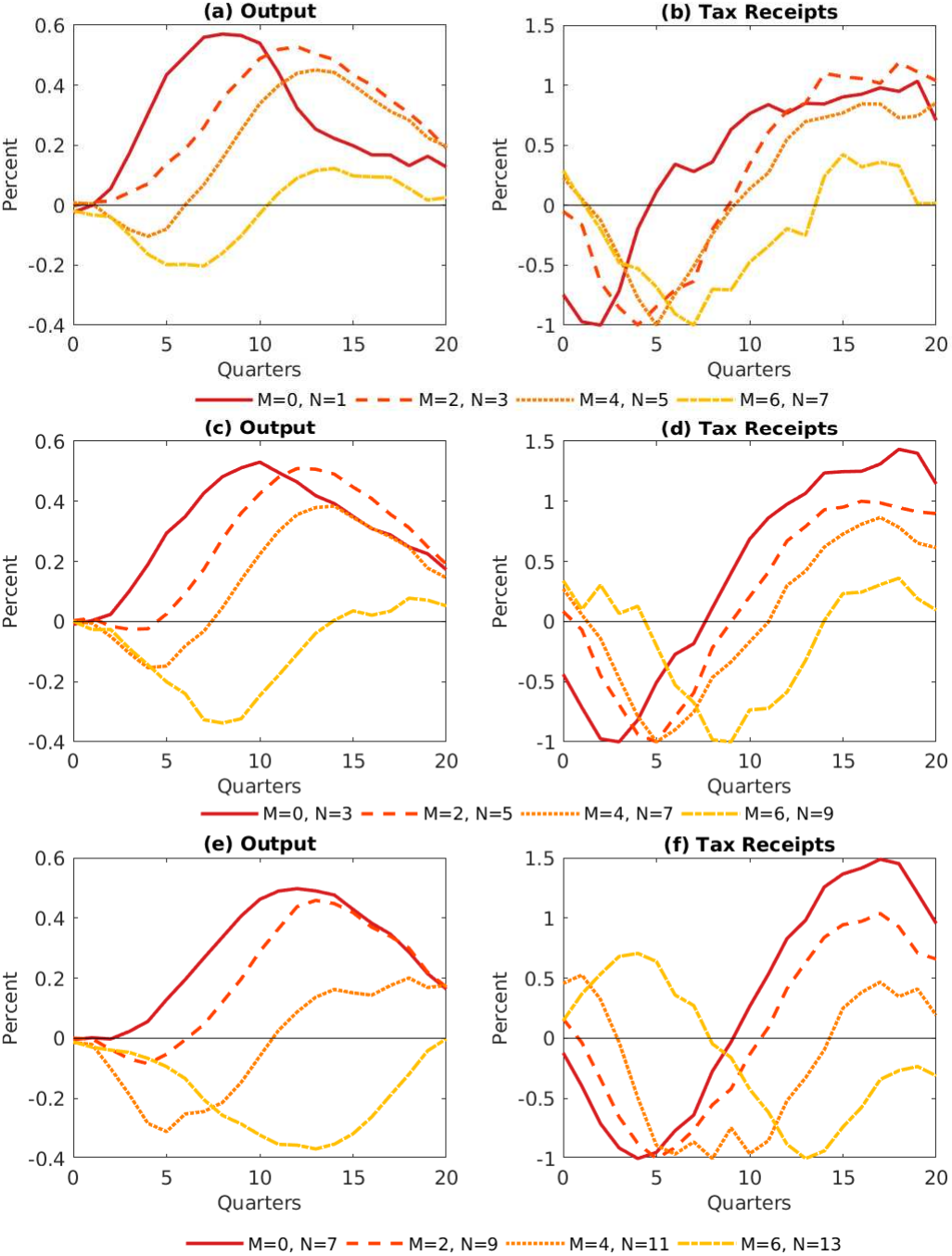}
\caption{Sensitivity of responses to varying anticipation horizon. Apart from different values of $M$ and $N$ the model specification (deterministic components, regressors, lags) correspond to the benchmark model discussed in Section \ref{sec:results}. }
\label{fig:Robustness2}
\end{center}
\end{figure}

We next turn to the question whether the specification of the anticipation horizon, i.e. the choice of $M$ and $N$ affects the results. Investigating Figure \ref{fig:Robustness2} leads to two conclusions. First, (i) the larger $M$, the larger the contractionary pre-implementation effect on output and the later the (trough) response in tax receipts. That is, the more we ``force'' the tax prevalence measures to instrument tax news relating to policy changes farther in the future, the stronger the (negative) anticipation effect on output. Second, comparing now responses across the rows in Figure \ref{fig:Robustness2}, (ii) the length of the ``aggregation window'' ($N-M$) does not alter the results as drastically. One can only observes a slightly lagged and more persistent response in both variables.

\begin{figure}
\begin{center}
\includegraphics[width=\textwidth]{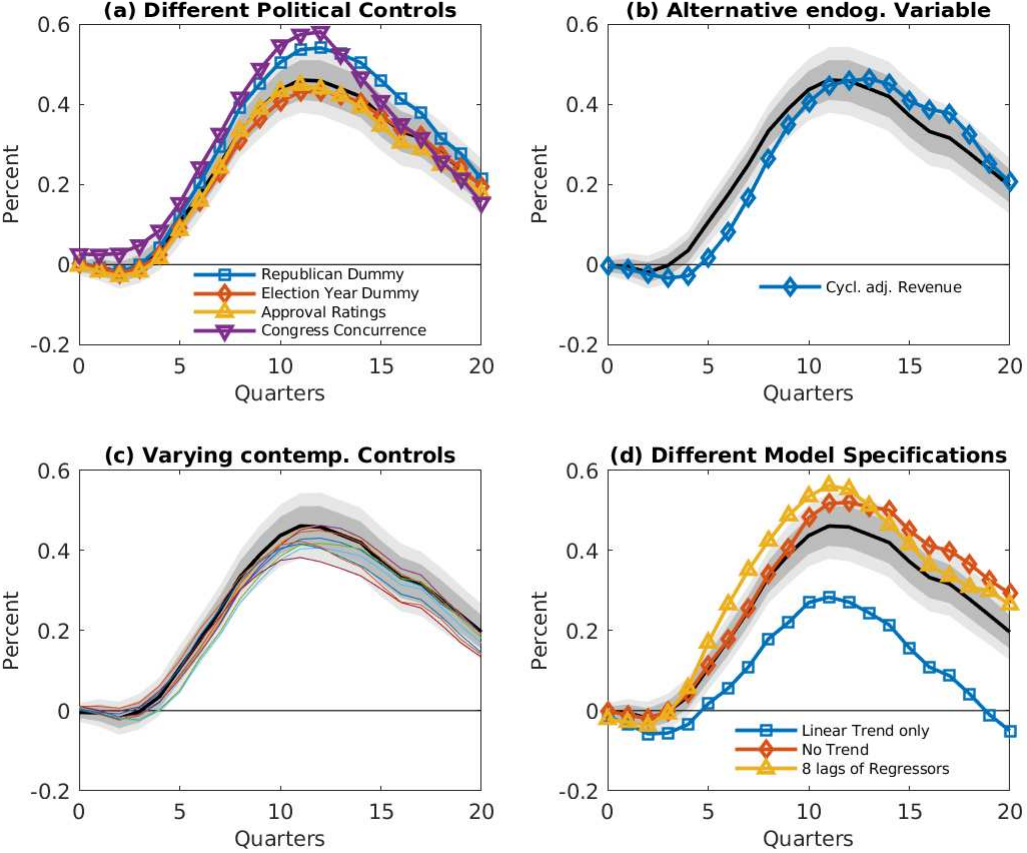}
\caption{Short(er) anticipation horizon (i.e. $M=1$ and $N=4$.)}
\label{fig:Robustness3}
\end{center}
\end{figure}

Figure \ref{fig:Robustness3} complements Figure \ref{fig:IRsRobustness} in the text. That is a variety of changes in the model specification are investigated for the case of a short(er) anticipation horizon ($M=1$ and $N=4$). As can be seen from Figure \ref{fig:Robustness3} no modification leads to a significantly different output response. Only changing the trend specification, that is considering only a linear trend, instead of a linear and quadratic trend, has a dampening effect on the response of GDP. 

\begin{figure}
\begin{center}
\includegraphics[width=\textwidth]{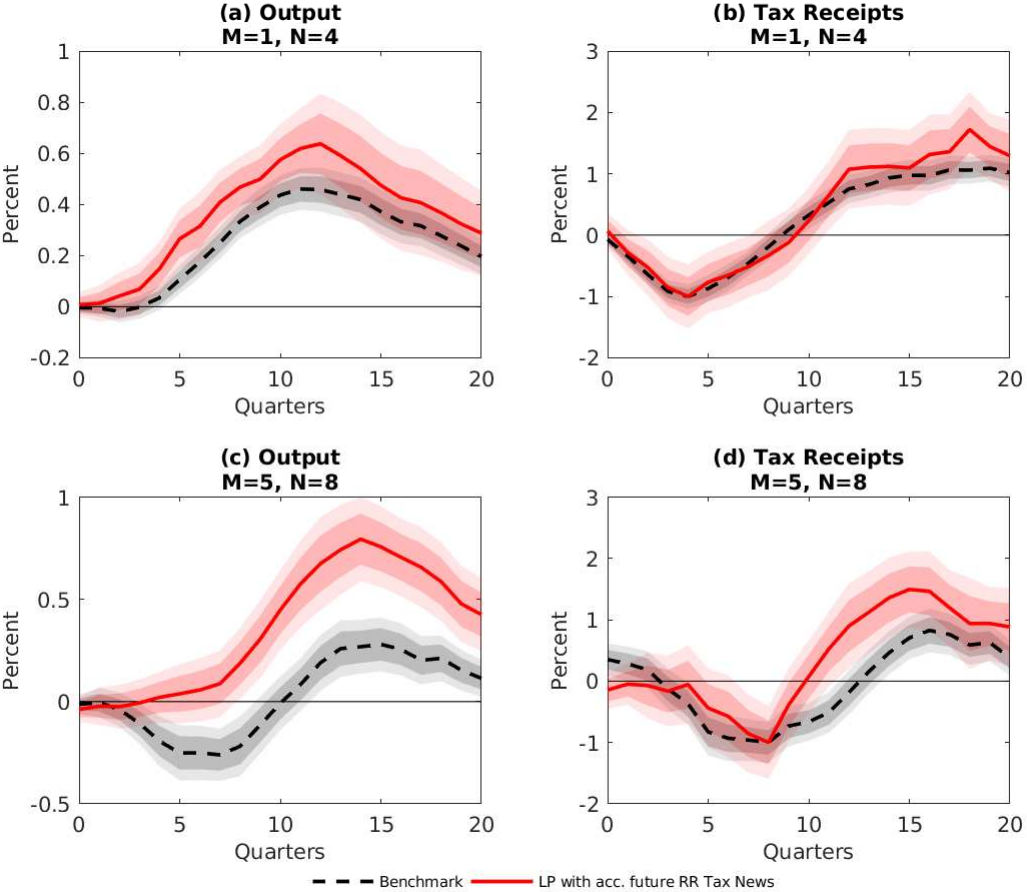}
\caption{Impulse responses estimated using local projections and a measure of future tax changes as constructed in (\ref{eq:AccTaxChanges}) as shock variable of interest. The measure of future tax changes is constructed from RR's exogenous tax changes.}
\label{fig:RobustnessRR}
\end{center}
\end{figure}

Next, we investigate how our approach of constructing tax news complements and compares to existing narrative approaches. With this aim, we analyse impulse responses of output and tax receipts to a change in $x_{1,t}$, which is constructed as in (\ref{eq:AccTaxChanges}) using RR's exogenous tax changes as a measure for $\Delta T_t$. Since we do not instrument $x_{1,t}$ with the tax prevalence measures, estimation is done by OLS using the benchmark specification of the local projection regression discussed in Section \ref{sec:results}. To make results comparable, the same normalization of impulse responses is applied. Results are displayed in Figure \ref{fig:RobustnessRR}. For soon-to-be implemented tax changes output responses are (qualitatively) similar for both tax news measures. However, when constructed using RR's narrative, news about a tax cut implemented in the more distant future does not lead to initial contractionary effects. Output does not react before tax changes are implemented. This suggests that RR's (exogenous) tax change series -- if not further decomposed as in \cite{mertens2012empirical} -- does not contain enough information to clearly identify anticipation effects.

\begin{figure}
\begin{center}
\includegraphics[width=\textwidth]{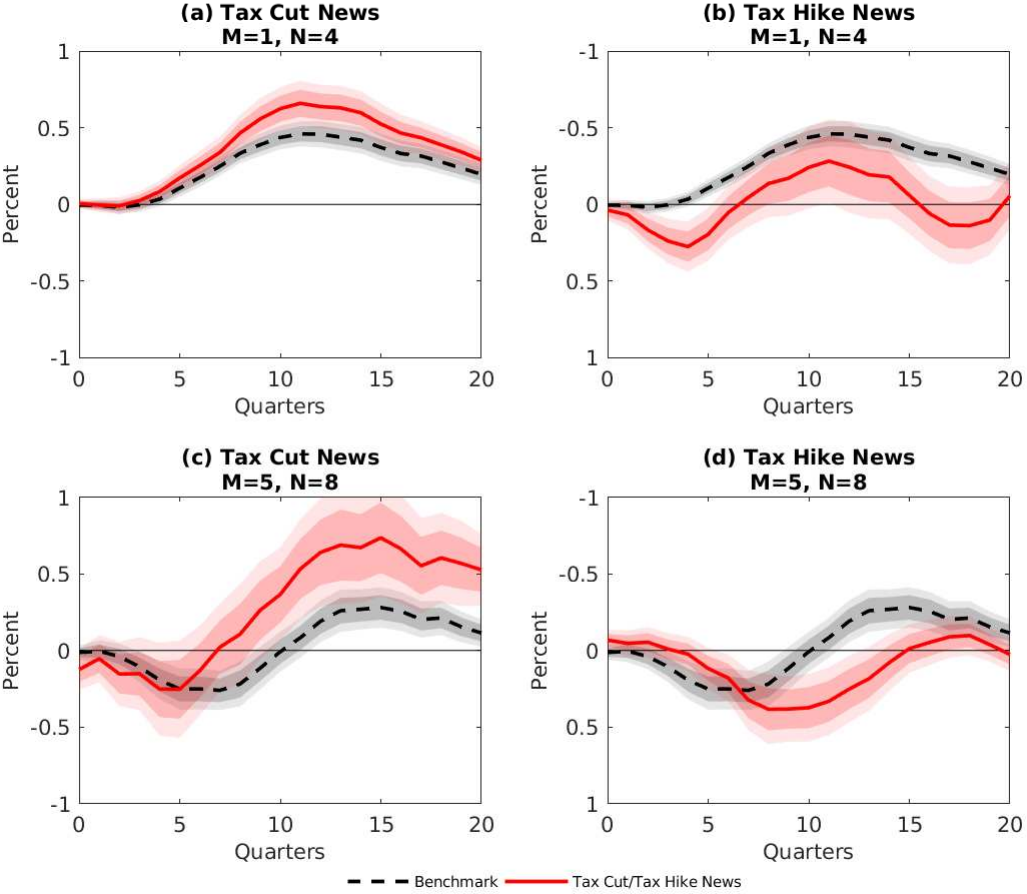}
\caption{Output responses to news about future tax cuts and to news about future tax hikes, respectively. For comparison, impulse responses based on the benchmark specification are plotted as well.}
\label{fig:Robustness4}
\end{center}
\end{figure}

Finally, we take a first step in analyzing whether tax cuts more effectively impact GDP compared to tax hikes. To investigate this we use the empirical framework as outlined in Section \ref{sec:empirics}, however, we use our tax cut prevalence measure (and not our tax hike measure) to instrument tax cuts (thus only negative tax changes) in the RR series. We estimate responses to news about a tax hike similarly. The results in Figure \ref{fig:Robustness4} suggest that regardless of the anticipation horizon, tax cuts have a stronger effect on output than tax hikes. We want to emphasise however, that we don't interpret these results as conclusive evidence of non-linear effects of tax changes - a much more careful investigation would be needed. Rather, we see these initial findings as motivation to possibly analyse this question more carefully in future research.

\clearpage

\section{Details for the two-step LDA model}

\subsection{Dirichlet distribution} \label{sec:dirichlet}

\begin{figure}[ht]
    \centering
    \includegraphics[width=\textwidth]{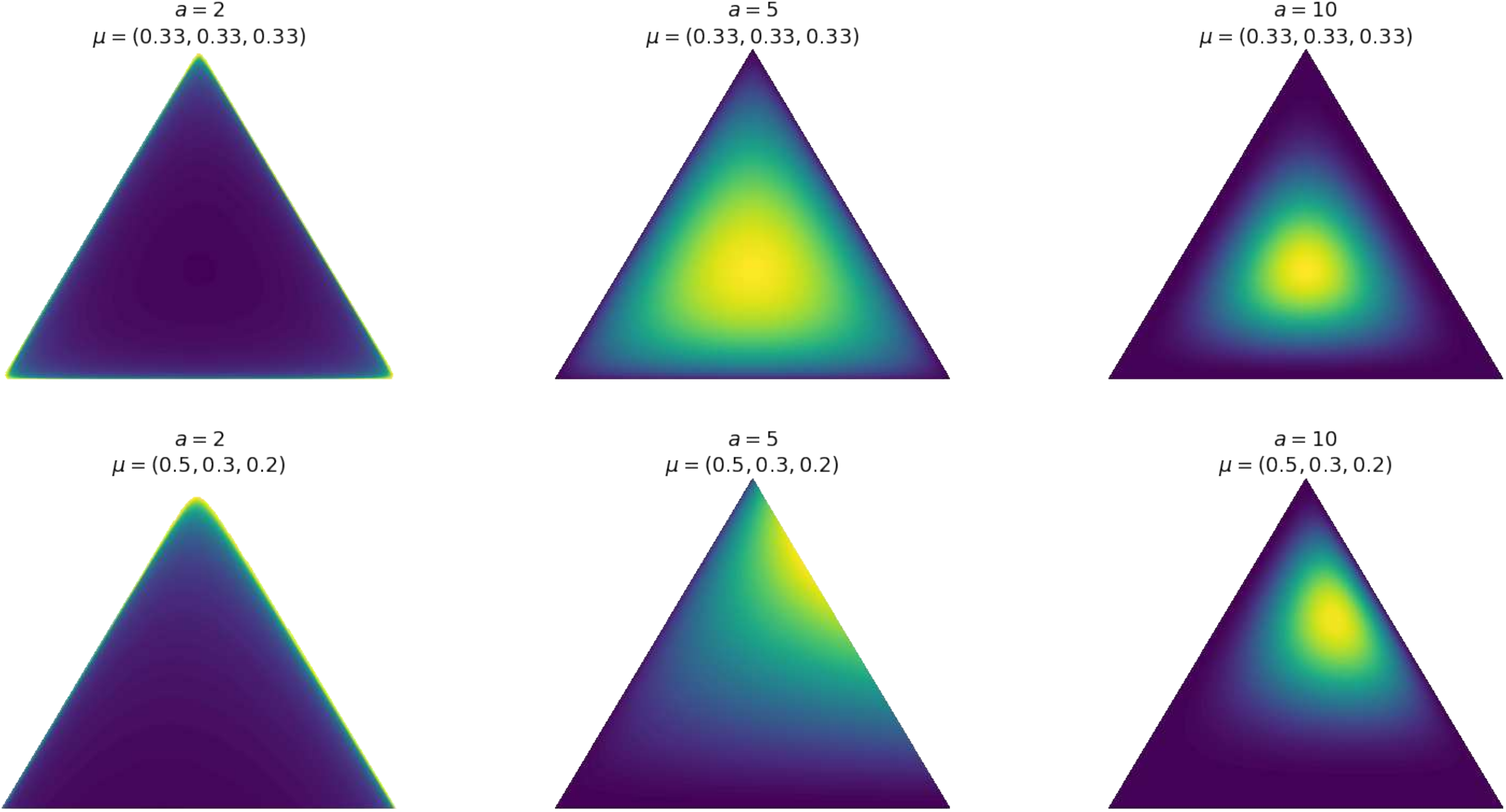}
    \caption{Probability density graphs for $K=3$-dimensional Dirichlet distributions with mean $\boldsymbol{\mu}$ and concentration parameter $a$. The lighter the colour, the higher the density.}
    \label{fig:dirichlet}
\end{figure}

A $K$-dimensional Dirichlet distribution is a continuous probability distribution, where a draw from the distribution is a vector of $K$ non-negative numbers summing up to $1$. For $\boldsymbol{\theta} \sim \text{Dir}(\boldsymbol{\alpha})$ the parameter vector $\boldsymbol{\alpha} = a\boldsymbol{\mu}$ is  decomposed into two parts - a~scalar concentration parameter $a = \sum_{k=1}^K\alpha_k$ and the mean
vector $\boldsymbol{\mu} = \mathbb{E}[\boldsymbol{\theta}] = \frac{1}{a} \boldsymbol{\alpha}$. The higher the $a$ the more concentrated the draws will be around the mean, following $\text{Var}[\theta_k] = \frac{\mu_k (1 - \mu_k)}{a + 1}$. This is visualised in Figure \ref{fig:dirichlet}.

A specific advantage of the Dirichlet distribution for estimating LDA models is the conjugacy property of this distribution in Bayesian inference. As an example, consider a 3-topic model with $\theta_d\sim \text{Dir}(2, 1, 0.5)$. If based on the data we believe that the document $d$ contains $0$ tokens assigned to topic 1, $10$ tokens assigned to topic 2 and $5$ tokens assigned to topic 3, the posterior distribution for its mixing proportions is given by $\text{Dir}(0 + 2, 10 + 1, 5 + 0.5)$. Intuitively, the parameter vector for the prior can then be interpreted as a count vector of past observations, where we can influence the relative importance of terms through $\boldsymbol{\mu}$, and the weight of prior information through $a$.

The use of Dirichlet priors leads to the LDA's tendency to concentrate big part of the probability mass of the mixing proportions in a relatively small number of topics, and big part of the probability mass of the topics' distributions in a relatively small number of terms. This behavior can be best seen when considering the distribution over the topic assignment $z_{d,n}$ for some token $w_{d,n}$, conditional on the topic assignments for the rest of the corpus $\mathcal{W}^{\lnot d,n},\mathcal{Z}^{\lnot d,n}$:
\begin{equation*}
    \mathbb{P}(z_{d,n}=k|w_{d,n} = v_i, \mathcal{W}^{\lnot d,n},\mathcal{Z}^{\lnot d,n}) 
    \propto \underbrace{(N_d^{(k)}+\alpha_k)}_{\text{document}}
    \underbrace{\frac{N^{(k,i)}+\eta_{k,i}} 
        {\sum_{d=1}^D N_d^{(k)}+\sum_{j=1}^{V}\eta_{k,j}}}_{\text{topic}},
\end{equation*}
where $N_d^{(k)} = \sum_{n=1}^{N_d} \mathbbm{1}(z_{d,n} = k)$ is the number of tokens assigned to topic $k$ in document $d$ and $N^{(k,i)} = \sum_{d=1}^D \sum_{n=1}^{N_d} \mathbbm{1} (w_{d,n} = v_i)  \mathbbm{1}(z_{d,n}=k)$ is the number of tokens with the term $v_i$ assigned to topic $k$. This probability consists of two parts:

\begin{itemize}
    \item In the document part, the more tokens within a document are identified as coming from topic $k$, the higher is $N_d^{(k)}$ and the higher the likelihood that the next token is also identified as coming from topic $k$.
    \item In the topic part, the more times the given term $v_i$ is identified as coming from topic $k$ (across the whole corpus), the higher $N^{(k,i)}$ and the higher the likelihood that the next instance of that term is also identified as coming from topic $k$.
\end{itemize}

\subsection{Model specification}

Our approach requires that the following parameters are specified. In the first, unsupervised step the number of topics $K$ is 25; the shape and scale parameters $(c,s)$ for the Gamma priors of the elements of $\bm{\eta}$ and $\bm{\alpha}$ are chosen to be $100^{-1}$ and $100$ respectively. In the second step we split the \textit{tax} topic into \textit{tax increase} and \textit{tax decrease} topic, resulting in $K=26$; the parameters $(c,s)$ for the Gamma priors of the elements of $\bm{\alpha}$ remain at $100^{-1}$ and $100$ respectively, while the vectors $\bm{\eta}_k, \quad k=1,\ldots,K$ are constructed according to \eqref{eq:eta_values_step2}.

In the first step the model consists of the following elements:
\begin{align}
    &\eta_{k,i}=\eta, \quad \eta \sim \text{Gamma}(c,s) & p&(\eta)=\frac{1}{\Gamma(s)c^s} \eta ^{s-1}e^{-\frac{\eta}{c}} \label{eq:dgp_eta}\\
    &\bm{\phi}_k|\bm{\eta}_k \sim \text{Dir}(\bm{\eta}_k) & p&(\boldsymbol{\phi}_k|\boldsymbol{\eta}_k)= \frac{\Gamma\left(\sum_{i=1}^V\eta_{k,i}\right)}
            {\prod_{i=1}^V\Gamma\left(\eta_{k,i}\right)}
            \prod_{i=1}^V {\phi_{k,i}}^{\eta_{k,i}-1} \label{eq:dgp_phi}\\
    &\alpha_k \sim \text{Gamma}(c,s) & p&(\alpha_k)=\frac{1}{\Gamma(s)c^s} \alpha_k ^{s-1}e^{-\frac{\alpha_k}{c}} \label{eq:dgp_alpha}\\
    &\bm{\theta}_d|\bm{\alpha} \sim \text{Dir}(\bm{\alpha}) &  p&(\boldsymbol{\theta}_d|\boldsymbol{\alpha}) = \frac{\Gamma\left(\sum_{k=1}^K\alpha_k\right)}
            {\prod_{k=1}^K\Gamma\left(\alpha_k\right)}
            \prod_{k=1}^K {\theta_{d,k}}^{\alpha_k-1} \label{eq:dgp_theta} \\
    &z_{d,n}|\bm{\theta}_d \sim \text{Cat}(\bm{\theta}_d) & \mathbb{P}&(z_{d,n}=k|\boldsymbol{\theta}_d)=\theta_{d, k}  \nonumber \\
    &w_{d,n}|z_{d,n}=k,\bm{\Phi} \sim \text{Cat}(\bm{\phi}_k) & \mathbb{P}&(w_{d,n}=v_i|z_{d,n}=k) =\phi_{k, i}  \nonumber
\end{align}

In the second step, instead of estimating $\bm{\eta}_k, \quad k=1,\ldots,K$ according to \eqref{eq:dgp_eta} we construct them according to \eqref{eq:eta_values_step2} and treat them as known.

\subsection{Estimation of the topic model} \label{sec:estimation}

Here we describe the Bayesian estimation procedure to obtain the posterior distribution of the document mixing proportions $\boldsymbol{\Theta}$ and topics' term probabilities $\boldsymbol{\Phi}$, conditional on the observed corpus of documents $\mathcal{W}$ and the Dirichlet distribution parameters $\boldsymbol{\alpha}$ and $\boldsymbol{\eta}_k$'s.

The complete data likelihood of the LDA model is
\begin{equation}
\label{eq:completelikelihood}
    p\big(\mathcal{W}, \mathcal{Z}|
    	\boldsymbol{\Theta},\boldsymbol{\Phi}\big) 
    	= \prod_{d=1}^D \prod_{n=1}^{N_d}
    		\prod_{i=1}^{V}\prod_{k=1}^K 
    		\left(
    		\theta_{d,k}\phi_{k,i}
			\right)^{\mathbbm{1}(z_{d,n} = k)\mathbbm{1}(w_{d,n} = v_i)}.
\end{equation}
The posterior of the model is obtained by combining the likelihood in \eqref{eq:completelikelihood} and the priors in \eqref{eq:dgp_phi} and \eqref{eq:dgp_theta}:
\begin{equation*}
\begin{split}
    &p\big(\boldsymbol{\Theta},\boldsymbol{\Phi},
    \boldsymbol{\alpha},\boldsymbol{\eta}_1,\dots,\boldsymbol{\eta}_K,
    \mathcal{Z} |\mathcal{W}\big) \\
    &\qquad  \propto 
    	p\big(\mathcal{W}, \mathcal{Z}|\boldsymbol{\Theta},\boldsymbol{\Phi}\big)   
    	p\big(\boldsymbol{\Theta}|\boldsymbol{\alpha}\big)
	    p\big(\boldsymbol{\Phi}|\boldsymbol{\eta}_1,\dots,\boldsymbol{\eta}_K \big)
	    p\big(\boldsymbol{\alpha}\big)    p\big(\boldsymbol{\eta}_1,\dots,\boldsymbol{\eta}_K \big)	
\\
    &\qquad \propto 
	    \left(\frac{\Gamma\left(\sum_{k=1}^K\alpha_k\right)}
            {\prod_{k=1}^K\Gamma\left(\alpha_k\right)}
        \right)^D 
		\prod_{k=1}^K \left(\frac{\Gamma\left(\sum_{i=1}^{V}\eta_{k,i}\right)}{\prod_{i=1}^{V}\Gamma(\eta_{k,i})}
		\right)
	    p\big(\boldsymbol{\alpha}\big)    p\big(\boldsymbol{\eta}_1,\dots \boldsymbol{\eta}_K \big)
	\\
	& \qquad \quad \times     	
	    \prod_{d=1}^D \prod_{k=1}^K 
	    \theta_{d,k}^{\alpha_k-1
	    		+ \sum_{i=1}^{V}\sum_{n=1}^{N_d} \mathbbm{1}(z_{d,n} = k)\mathbbm{1}(w_{d,n} = v_i)}
	\\
	& \qquad \quad \times 
 		\prod_{k=1}^K \prod_{i=1}^{V} {\phi_{k,i}}^{\eta_{k,i}-1
 		+ \sum_{d=1}^D \sum_{n=1}^{N_d}\mathbbm{1}(z_{d,n} = k)\mathbbm{1}(w_{d,n} = v_i)
 		}.
\end{split}
\label{eq:LDAposterior}
\end{equation*}
Posterior inference of the LDA mdel is computationally demanding due to the size of the corpus. More efficient methods, such as the variational Bayes algorithm of \citet{blei2003latent}, are applicable. We opt for the Gibbs sampler as it has better convergence properties in 
our application. 

Let $\Omega = \{\boldsymbol{\Theta},\boldsymbol{\Phi},
    \boldsymbol{\alpha},\boldsymbol{\eta}_1,\dots,\boldsymbol{\eta}_K \}$ denote the set of all model parameters and $\Omega_{-\kappa} = \Omega \setminus \{ \kappa \} $ for variable $\kappa$ and variable set $\Omega$. The Gibbs sampling steps for the model parameters and latent states are as follows. First,
\begin{equation}
\label{eq:Gibbs_1}
    p\big(\boldsymbol{\theta}_d | \Omega_{-\boldsymbol{\theta}_d}, \mathcal{Z}, \mathcal{W}\big) 
    \propto 
	    \prod_{k=1}^K 
	    \theta_{d,k}^{\alpha_k-1
	    		+ \sum_{i=1}^{V}\sum_{n=1}^{N_d} \mathbbm{1}(z_{d,n} = k)\mathbbm{1}(w_{d,n} = v_i)}, \mbox{ for }d = 1,\ldots,D,
\end{equation}
is proportional to a Dirichlet distribution, as in the conjugate prior in \eqref{eq:dgp_theta}. Next,
\begin{equation}
\label{eq:Gibbs_2}
    p\big(\boldsymbol{\phi}_k | \Omega_{-\boldsymbol{\phi}_k}, \mathcal{Z}, \mathcal{W}\big) 
    \propto 
 		\prod_{i=1}^{V} {\phi_{k,i}}^{\eta_{k,i}-1
 		+ \sum_{d=1}^D \sum_{n=1}^{N_d}\mathbbm{1}(z_{d,n} = k)\mathbbm{1}(w_{d,n} = v_i)
 		}, \mbox{ for }k=1,\ldots,K,
\end{equation}
is proportional to a Dirichlet distribution, as in the conjugate prior in \eqref{eq:dgp_phi}. Similarly,
\begin{equation}
\label{eq:Gibbs_3}
\begin{split}
    & p\big(z_{d,n} = k | \Omega, \mathcal{Z}_{-z_{d,n}}, \mathcal{W}\big) \\
    & \qquad \propto
	    \theta_{d,k}^{\alpha_k-1
	    		+ \sum_{i=1}^{V} \mathbbm{1}(w_{d,n} = v_i)}\prod_{i=1}^{V} {\phi_{k,i}}^{\eta_{k,i}-1
 		+ \mathbbm{1}(w_{d,n} = v_i)
 		}, \mbox{ for }k=1,\ldots,K
\end{split}
\end{equation}
which is a categorical distribution for $d = 1,\ldots, D$ and $n = 1,\ldots, N_d$ with probabilities proportional to the right hand side of \eqref{eq:Gibbs_3}.

For the remaining model parameters, $(\boldsymbol{\alpha},\boldsymbol{\eta}_1,\dots,\boldsymbol{\eta}_K)$, we use the fixed-point iteration (MAP estimator) of \citet{minka2000estimating,wallach2008structured} at each Gibbs iteration. We optimise $\boldsymbol{\alpha}=a\boldsymbol{\mu}$ as \begin{equation}
[a\mu_k]^*=a\mu_k 
            \frac{\sum_{d=1}^D\left[ \Psi \left(N_d^{(k)} + a\mu_k \right) - \Psi \left(a\mu_k \right) \right] + c}
            {\sum_{d=1}^D \left[ \Psi \left(N_d + a \right) - \Psi \left(a \right)\right] - \frac{1}{s}},
\label{eq:Gibbs_3_alpha}
\end{equation}
and, only in the first step, we optimise $\eta$ using
\begin{equation}
\eta^* = \frac{\eta\sum_{k=1}^K\sum_{i=1}^V
            \left[ \Psi \left(N^{(k,i)} + \eta \right) - \Psi \left(\eta \right) \right] + c}
           {V\sum_{k=1}^K \left[ \Psi \left(\sum_{i=1}^V N^{(k,i)} + V\eta \right) - \Psi \left(V \eta \right)\right] - \frac{1}{s}},
\label{eq:Gibbs_3_eta}
\end{equation}
where $N_d^{(k)} = \sum_{n=1}^{N_d} \mathbbm{1}(z_{d,n} = k)$ is the number of tokens assigned to topic $k$ in document $d$, $N^{(k,i)} = \sum_{d=1}^D \sum_{n=1}^{N_d} \mathbbm{1} (w_{d,n} = v_i)  \mathbbm{1}(z_{d,n}=k)$ is the number of tokens with the term $v_i$ assigned to topic $k$, and $(c,s)$ are the shape and scale parameters, respectively, for the Gamma priors exemplified in \eqref{eq:dgp_alpha} and \eqref{eq:dgp_eta}. The steps in \ref{eq:Gibbs_3_alpha} and \ref{eq:Gibbs_3_eta} are repeated until convergence\footnote{Until $\max_k\frac{[a\mu_k]^*-a\mu_k}{a\mu_k}<0.01$ and $\max_k\frac{\eta^*-\eta}{\eta}<0.01$}. The Gibbs sampler for the proposed 2-step LDA model is outlined in Algorithm~\ref{alg:Gibbs} below, for a corpus $\mathcal{W}$ obtained from pre-processed data, and $K$ topics.

\begin{algorithm}
\caption{Gibbs sampling algorithm} \label{alg:Gibbs}
LDA step 1:
\begin{small}
\begin{algorithmic}[1]
\State Set $m=0$, and initialise $\alpha_k^{(m)} = 1$ for $k = 1,\ldots K$ and $\eta =1$.  
\State Initialise $\boldsymbol{\Theta}^{(m)}$ using the Dirichlet distribution in \eqref{eq:dgp_theta}; $z_{d,n}^{(m)}$ for $d = 1,\ldots, D$ and $n=1,\ldots N_d$ using the categorical distribution in \eqref{eq_gen_z}; $\boldsymbol{\phi}_k^{(m)}$ for all $k$ using the Dirichlet distribution in \eqref{eq:dgp_phi}.
\State Set $\Omega^{(m)} = \{\boldsymbol{\Theta}^{(m)},\boldsymbol{\Phi}^{(m)},
    \boldsymbol{\alpha}^{(m)},\boldsymbol{\eta}^{(m)}_1,\dots,\boldsymbol{\eta}^{(m)}_K \}$
\While{$m \leq 15000$} \label{alg:gibbsstart}
	\State Set $m = m+1$
	\State Draw $\boldsymbol{\Theta}^{(m)} \mid \Omega^{(m-1)}_{-\boldsymbol{\Theta}}, \mathcal{Z}^{(m-1)}, \mathcal{W}^{(m-1)}$ using the Dirichlet distribution in \eqref{eq:Gibbs_1}
	\State Update $\Omega^{(m)} = \{\boldsymbol{\Theta}^{(m)},\boldsymbol{\Phi}^{(m-1)},
    \boldsymbol{\alpha}^{(m-1)},\boldsymbol{\eta}^{(m-1)}_1,\dots,\boldsymbol{\eta}^{(m-1)}_K \}$
	\For{$k=1,\ldots,K$} 
	\State Draw $\boldsymbol{\phi}^{(m)}_k \mid \Omega^{(m)}_{\boldsymbol{\phi}_k}, \mathcal{Z}^{(m-1)}, \mathcal{W}$ for all $k$ using the Dirichlet distribution in \eqref{eq:Gibbs_2}
	\State Update $\boldsymbol{\Phi}^{(m)} = \{
	\boldsymbol{\phi}^{(m)}_`,\ldots,
	\boldsymbol{\phi}^{(m)}_k,
	\boldsymbol{\phi}^{(m-1)}_{k+1},\ldots,
	\boldsymbol{\phi}^{(m-1)}_K\}$
	\State Update $\Omega^{(m)} = \{\boldsymbol{\Theta}^{(m)},\boldsymbol{\Phi}^{(m)},
    \boldsymbol{\alpha}^{(m-1)},\boldsymbol{\eta}^{(m-1)}_1, \dots,\boldsymbol{\eta}^{(m-1)}_K \}$
	\EndFor
	\State Set $\mathcal{Z}^{(m)} = \{
	z_{1,1}^{(m-1)}, \ldots, z_{1,N_1}^{(m-1)}, \ldots, z_{D,1}^{(m-1)},\ldots, z_{D,N_D}^{(m-1)}\}$
	\For{$d=1,\ldots,D$}
		\For{$n=1,\ldots,N_d$}
		\State Draw $z_{d,n}^{(m)} \mid \Omega^{(m)}, \mathcal{Z}^{(m)}_{-z_{d,n}}, \mathcal{W}$ using the categorical distribution in \eqref{eq:Gibbs_3}
		\State Update $\mathcal{Z}^{(m)} = \{
	z_{1,1}^{(m)}, \ldots, z_{d,n}^{(m)},
	z_{d,n+1}^{(m-1)},\ldots, z_{d+1,1}^{(m-1)} \ldots, z_{D,N_D}^{(m-1)}\}$
		\EndFor
	\EndFor
    \State Update $\boldsymbol{\alpha}^{(m)}=a\boldsymbol{\mu}$
    using the fixed-point iteration in \eqref{eq:Gibbs_3_alpha}.
    \State Set $\bm{\eta}_{k}^{(m)} = \bm{\eta}^{(m)} = a_{\eta}\bm{\iota}_V$ where $\bm{\iota}_V$ is a $V$-dimensional vector of $1/V$.  
    Update $\bm{\eta}^{(m)}$ using the fixed-point iteration in \eqref{eq:Gibbs_3_alpha}. \label{alg:eta}

\EndWhile \label{alg:gibbsend}
\end{algorithmic}    
LDA step 2
\begin{algorithmic}[1]
\State Set $K = K+1$, set $m=0$, and initialise $\alpha_k^{(m)} \sim \mbox{Gamma}^{(m)}(100^{-1}, 100)$ for $k = 1,\ldots K$ 
\item Set $\bm{\eta}_k, \quad k=1,\ldots,K$ from step 1, using equation~\eqref{eq:eta_values_step2}.
\State Initialise $\boldsymbol{\Theta}^{(m)}$ using the Dirichlet distribution in \eqref{eq:dgp_theta}; $z_{d,n}^{(m)}$ for $d = 1,\ldots, D$ and $n=1,\ldots N_d$ using the categorical distribution in \eqref{eq_gen_z}; $\boldsymbol{\phi}_k^{(m)}$ for all $k$ using the Dirichlet distribution in \eqref{eq:dgp_phi}.
\State Set $\Omega^{(m)} = \{\boldsymbol{\Theta}^{(m)},\boldsymbol{\Phi}^{(m)},
    \boldsymbol{\alpha}^{(m)},\boldsymbol{\eta}^{(m)}_1,\dots,\boldsymbol{\eta}^{(m)}_K \}$
\State Repeat LDA step 1 lines  \ref{alg:gibbsstart}--\ref{alg:gibbsend}, except for \ref{alg:eta}.
\end{algorithmic}
\end{small}
\end{algorithm}

\subsection{Lexicons} \label{sec:lexicons}
Our procedure for constructing priors for the \textit{tax increase} and \textit{tax decrease} topics require specifying the terms whose usage differentiates the two topics. We do that based on reading presidential speeches which are known to provide information about future tax changes. We select a total of 69 speeches used by \citet{romer2009narrative} and \citet{yang2007chronology} in their analyses of tax legislations. The list of the speeches is provided in Appendix \ref{sec:speeches}. In each speech we note the terms used to announce and motivate changes in particular direction. We then aggregate those results and choose terms whose usage is indicative of discussing a tax change with the particular direction.

Importantly, we do not assume that, for example, a term in the \textit{tax increase} lexicon cannot be used when discussing tax cuts. We use those lexicons only to modify the prior probabilities of those terms in the two topics of interest.

\textbf{Tax increase:} ``additional cost'', ``additional revenue'', ``additional tax'', ``balance budget'', ``budget deficit'', ``cut deficit'', ``defense spend'', ``deficit'', ``deficit reduction'', ``fair balance'', ``fair share'', ``fairness'', ``federal revenue'', ``fiscal responsibility'', ``fiscally responsible'', ``government revenue'', ``higher tax'', ``increase revenue'', ``increase tax'', ``military spend'', ``new spend'', ``new tax'', ``propose increase'', ``propose tax'', ``raise revenue'', ``raise tax'', ``reduce debt'', ``reduce deficit'', ``revenue increase'', ``rise cost'', ``sound fiscal'', ``tax impose'', ``tax increase'', ``tax revenue'', ``tax rich''

\noindent \textbf{Tax decrease:} ``boost economy'', ``business incentive'', ``create incentive'', ``create job'', ``cut'', ``cut propose'', ``cut tax'', ``ease burden'', ``economic growth'', ``economic incentive'', ``excessive'', ``grow economy'', ``high rate'', ``incentive'', ``incentive invest'', ``incentivize'', ``increase employment'', ``increase investment'', ``increase production'', ``increase productivity'', ``increase prosperity'', ``investment'', ``investment tax'', ``lower'', ``lower tax'', ``new job'', ``propose cut'', ``provide incentive'', ``rate drop'', ``rate reduction'', ``rebuild economy'', ``recession'', ``reduce burden'', ``reduce rate'', ``reduce unemployment'', ``reduction'', ``relief'', ``relief package'', ``relief program'', ``slow growth'', ``stimulate economic'', ``stimulate economy'', ``stimulate growth'', ``stimulate investment'', ``strengthen economic'', ``strengthen economy'', ``tax break'', ``tax burden'', ``tax credit'', ``tax cut'', ``tax incentive'', ``tax policy'', ``tax rate'', ``tax rebate'', ``tax reduce'', ``tax reduction'', ``tax relief''

\section{Data}\label{sec:data}

\subsection{Text data} 

We analyse the texts of all the documents contained in \textit{The Public Papers of the Presidents}. This includes speeches, but also other public communications (e.g. interviews, letters). We obtained the raw texts from \citeauthor{WoolleyPetersTextData}'s \textit{American Presidency Project} on 2019-03-25.
Our dataset consists of 59,214 texts spanning from 1949-01-20 to 2017-01-19. 

The pre-processing starts with breaking the texts into individual paragraphs, resulting in a total of 1,119,200 documents. The text is then broken along non-alphanumeric characters into tokens (individual words). We then remove stopwords - function words such as prepositions and pronouns, since they are not informative in the \textit{bags-of-words} approach; and rare words which appear in less than 20 documents - those are often spelling mistakes, since they are too rare to be informative about the term co-occurrence pattern. The remaining tokens are then lemmatised - all the nouns are turned into singular form and the verbs are turned into present tense first person form. Lastly, we identify meaningful collocations of two words - bigrams. If a combination of two consecutive words (e.g. ``tax'' and ``cut'') appears more often than predicted by their individual frequency (as measured by a $\chi^2$ score) it is considered a bigram (i.e. ``tax\_cut''). A bigram is treated as a separate term in the vocabulary and each instance of a bigram in the texts is treated as a single token. The final vocabulary consists of 50,851 terms, including 29,815 bigrams.

\subsection{Additional data} 

We use four series of changes in tax liabilities developed by RR \citep{RomerRomer10}. The data-set is part of the paper's supplementary materials \citep{RomerRomer10Data}. The construction of their measures is described in the companion background paper \citep{romer2009narrative}. To create the measures they first identify all the major changes in tax liabilities. For each change they analyse the narrative record, including presidential speeches, to determine the motivation behind it. This can be either exogenous to the macroeconomic conditions (deficit-driven or long-run growth) or endogenous (spending-driven or counter-cyclical).  The size of the changes is the estimated change in tax liabilities at the time of implementation based on contemporary sources. The changes are timed either at the implementation - when they go into effect, or at enactment - when the legislation is signed by the president, in which case the present value of the changes is given.

We also use the speeches identified by \citet{romer2009narrative}, which provide the motivations behind specific tax policy changes, to verify the accuracy of our model in correctly classifying tax-related content. The list of those speeches is provided in Appendix \ref{sec:speeches}.

Throughout our analysis we use additional macroeconomic, financial, and tax data. Further information on these data is provided in Table \ref{tab:variables}. 

\newpage

\begin{scriptsize}
\renewcommand{\arraystretch}{1.5}
\begin{tabularx}{\textwidth}{ 
   >{\raggedright\arraybackslash}p{0.15\textwidth} 
   >{\raggedright\arraybackslash}p{0.25\textwidth} 
   >{\raggedright\arraybackslash}p{0.3\textwidth}
      >{\raggedright\arraybackslash}p{0.3\textwidth}
}
\caption{Overview of additional variables}\label{tab:variables}
\endfirsthead
\toprule
\textbf{Variable} & \textbf{Description} & \textbf{Details} & \textbf{Source} \\ 
\midrule
\\
\multicolumn{4}{l}{\textit{\small{Tax measures:}}}\\
\\

Tax Receipts$^{a}$   & Federal government current tax receipts  & In billions of Dollars, quarterly frequency, seasonally adjusted at annual rate. & \citet{NIPAGovSpend}, account code: W006RC \\ 

Income Taxes$^{a}$ & Federal government personal income tax receipts & In billions of Dollars, quarterly frequency, seasonally adjusted at annual rate. & \citet{NIPAGovSpend}, account code: A074RC \\ 

Corporate Taxes$^{a}$ & Federal government corporate income tax receipts & In billions of Dollars, Quarterly Frequency, seasonally adjusted at annual rate. & \citet{NIPAGovSpend}, account code: B075RC \\

Payroll Taxes$^{a}$ & Federal government payroll tax receipts & Contributions for government social insurance in billions of Dollars, quarterly frequency, seasonally adjusted at annual rate. & \citet{NIPAGovSpend},  account code: W780RC \\

Cyclically Adjusted Revenue & Change in real cyclically adjusted federal government revenues as percent of real GDP &  & \citet{RomerRomer10Data}, variable ``DRCARA''\\

Implicit Tax Rate & Risk-adjusted implicit tax rate with one year maturity as constructed in \citet{leeper2012quantitative} &  \citet{leeper2012quantitative} define the implicit tax rate as $1-Y_m/Y_t$ where $Y_m$ is the yield of a municipal bond and $Y_t$ is the yield of a Treasury bond.
We use municipal and treasury bond yields with one year maturity at monthly frequency. We use their provided Matlab code to construct rolling quarterly averages of monthly data. & \citet{Leeper2012Data} \\ 

RR All Tax Changes & RR's identified endogenous and exogenous tax changes divided by nominal GDP & & \citet{RomerRomer10Data}, variable ``SUMMANRRATIO''\\

RR Exogenous Tax Changes & RR's identified deficit-driven and long-run tax changes divided by nominal GDP &  &\citet{RomerRomer10Data}, variable ``EXOGENRRATIO''\\

RR News All Tax Changes& Present value (at time of enactment) of RR's identified endogenous and exogenous tax changes divided by nominal GDP &  & \citet{RomerRomer10Data}, variable  ``SUMMAPDVRATIO''\\

RR News Exogenous Tax Changes & Present value (at time of enactment) of RR's identified deficit-driven and long-run tax changes divided by nominal GDP &  & \citet{RomerRomer10Data}, variable  ``EXOGEPDVRATIO''\\

MR Unanticipated Tax Changes& \citeauthor{mertens2014reconciliation}'s identified exogenous tax changes for which the legislation and implementation date are less than one quarter apart &  &  \citet{MertensRavn2014Data}, variable ``Tax Narrative''. \\
\\

\multicolumn{4}{l}{\textit{\small{Macroeconomic and financial variables:}}}\\
\\

GDP$^{a}$  & Gross domestic product & In millions of Dollars, quarterly frequency, seasonally adjusted at annual rate. & \citet{NIPAGDP}, account code: A191RC\\

Consumption$^{a}$  & Personal consumption expenditures & In millions of Dollars, quarterly frequency, seasonally adjusted at annual rate. & \citet{NIPAGDP}, account code: DPCERC\\

Investment$^{a}$ & Gross private investment & In millions of Dollars, quarterly frequency, seasonally adjusted at annual rate. & \citet{NIPAGDP}, account code: A006RC\\

Government Spending$^{a}$ & Federal Government current Expenditures & In billions of Dollars, quarterly frequency, seasonally adjusted at annual rate. & \citet{NIPAGovSpend}, account code: W013RC\\

Government Debt$^{a,b}$  & Par value of privately held gross federal debt & In millions of Dollars, monthly frequency, not seasonally adjusted, values reported are for the last business day of the month. & \citet{FedGovDebt} \\

3-month T-Bill$^{b}$ & Secondary market rate 3-month Treasury bill & Daily frequency (downloaded as monthly averages). & \citet{InterestRatesData} \\

Federal Funds Rate$^{b}$ & Federal funds effective rate & Daily frequency (downloaded as monthly averages). & \citet{InterestRatesData} \\

Unemployment Rate$^{b}$ & Unemployed as percentage of the labor force & Monthly frequency, seasonally adjusted. & \citet{UnemploymentData} \\

CPI$^{b}$ & Consumer Price Index for All Urban Consumers: All Items in U.S. City Average &   Index 1982-1984=100, monthly frequency, seasonally adjusted. & \citet{CPIData} \\

S\&P500$^{a,b}$  & S\&P composite prices & Composite prices are daily-close, monthly averages. & \citet{Shiller2020Data} \\

Government Spending News & Nominal present value of \citeauthor{Ramey11}'s government spending news variable divided by nominal GDP of previous quarter &  & \citet{Ramey11Data}, variable ``pdvmily''\\
\\

\multicolumn{4}{l}{\textit{\small{Other variables:}}}\\
\\

Job Approval of the President$^{b}$  & Gallup Approval Poll & Gallup approval poll of the president. Irregular frequency. & \citet{WoolleyPetersTextData}\\

Congress Concurrence & Presidential Victories on Votes in Congress & Percentages indicate the number of congressional votes supporting the president divided by the total number of votes on which the president had taken a clear position. 
The percentages are normalized to eliminate the effects of absences as follows: support = (support)/(support + opposition). Yearly data from 1953 to 2007. & \citet{CongressData}\\
\midrule
\end{tabularx}

\textit{a}: Variable is expressed in real per capita terms. We use a population measure and the GDP implicit price deflator to transform nominal quantities in real per capita terms. To construct the real per capita data we have used the following two series: (i) Population: Civilian noninstitutional population is defined as persons 16 years of age and older residing in the 50 states and the District of Columbia [thousands of persons], monthly frequency (transformed into quarterly averages), not seasonally adjusted. Source:  \citet{PopulationData}, and (ii) Deflator: Gross Domestic Product: Implicit Price Deflator, Index 2012=100, quarterly frequency, seasonally adjusted. Source: \citet{NIPADeflator}, account code: A191RD

\textit{b}: Series is transformed into quarterly averages.

\end{scriptsize}

\clearpage
\section{Topic model results} 

\subsection{Selected tax-related speeches} \label{sec:speeches}

\citet{romer2009narrative} and \citet{yang2007chronology} use presidential speeches to analyse to motivation behind particular tax legislations. We combine their selection of speeches and use them for two purposes. Firstly, the speeches are used to develop the lexicons of terms related to tax increase and tax decrease, as described in Appendix \ref{sec:lexicons}. Secondly, we use them to verify the ability of our model to properly classify a text based on which direction of tax changes it is discussing.

Below we present a list of speeches announcing legislations whose overall effect was a tax increase (Table \ref{tab:sel_speeches_inc}) and tax decrease (Table \ref{tab:sel_speeches_dec}). Importantly, in our estimation we treat individual paragraphs as documents, each having separate mixing proportions. Especially in case of long speeches, such as the \textit{State of the Union Addresses}, many of those paragraphs discuss issues other than taxation. Additionally, certain legislations consisted of measures that increased some taxes, but cut others. In the tables those are indicated as ``mixed''. For speeches related to those legislations we expect some paragraphs to have high \textit{tax increase} proportions, and other to have high \textit{tax decrease} proportions. For each speech we therefore show the minimum, the average and the maximum mixing proportion across all of its paragraphs, for both \textit{tax increase} and \textit{tax decrease} topics, as estimated in the second step. Overall the results fit our expectations, supporting the claim that our model is able to distinguish between the direction of the discussed changes.

\begin{landscape}
\scriptsize
\begin{longtable}[c]{llllllllllll}

\caption{List of speeches announcing legislation whose net effect was an increase in tax revenue.} \\

\multicolumn{12}{c}{\textbf{Title of the speech}} \\
\multirow{2}{*}{\textbf{Date}} & \multicolumn{1}{c}{\multirow{2}{*}{\textbf{Announced legislation}}} &
\multicolumn{1}{c}{\multirow{2}{*}{\textbf{Mixed}}} & \multicolumn{1}{c}{\multirow{2}{*}{\textbf{\begin{tabular}[c]{@{}c@{}}Number of\\      paragraphs\end{tabular}}}} & \multicolumn{1}{c}{\textbf{}} & \multicolumn{3}{c}{\textbf{\textit{Tax increase} proportion}} & \multicolumn{1}{c}{\textbf{}} & \multicolumn{3}{c}{\textbf{\textit{Tax decrease} proportion}} \\
 & \multicolumn{1}{c}{} & \multicolumn{1}{c}{} & \multicolumn{1}{c}{} &  & \textbf{Min.} & \textbf{Av.} & \textbf{Max.} & \textbf{} & \textbf{Min.} & \textbf{Av.} & \textbf{Max.} \\
 \endhead
\multicolumn{12}{l}{\textit{In a special   message to Congress on tax policy}} \\
1950/01/23 & Revenue Act of 1950 & Yes & 58 &  & 0.010 & 0.530 & 0.855 &  & 0.004 & 0.015 & 0.197 \\
\multicolumn{12}{l}{\textit{Midyear Economic   Report of the President}} \\
1950/07/26 & Revenue Act of 1950 & Yes & 92 &  & 0.004 & 0.383 & 0.848 &  & 0.003 & 0.059 & 0.387 \\
\multicolumn{12}{l}{\textit{Letter to   Committee Chairmen on Taxation of Excess Profits}} \\
1950/11/14 & Revenue Act of 1950 & Yes & 9 &  & 0.016 & 0.373 & 0.788 &  & 0.004 & 0.012 & 0.027 \\
\multicolumn{12}{l}{\textit{Annual Budget   Message to Congress}} \\
1951/01/15 & Revenue Act of 1951 & Yes & 399 &  & 0.002 & 0.262 & 0.907 &  & 0.002 & 0.017 & 0.352 \\
\multicolumn{12}{l}{\textit{Special Message to   the Congress Recommending a “Pay as We Go” Tax Program}} \\
1951/02/02 & Revenue Act of 1951 & Yes & 48 &  & 0.019 & 0.500 & 0.845 &  & 0.004 & 0.020 & 0.223 \\
\multicolumn{12}{l}{\textit{Radio address}} \\
1953/05/19 & \begin{tabular}[c]{@{}l@{}}Extending the Excess Profits\\ Tax Act of 1950\end{tabular} & No & 76 &  & 0.005 & 0.217 & 0.759 &  & 0.004 & 0.017 & 0.060 \\
\multicolumn{12}{l}{\textit{Annual Message to   the Congress on the State of the Union}} \\
1956/01/05 & Federal-Aid Highway Act of 1956 & No & 128 &  & 0.004 & 0.126 & 0.822 &  & 0.002 & 0.021 & 0.274 \\
\multicolumn{12}{l}{\textit{Annual Budget   Message to the Congress}} \\
1965/01/25 & Social Security Amendments of 1965 & No & 251 &  & 0.006 & 0.203 & 0.785 &  & 0.003 & 0.028 & 0.474 \\
\multicolumn{12}{l}{\textit{Annual Budget   Message to the Congress}} \\
1966/01/24 & Tax Adjustment Act of 1966 & No & 304 &  & 0.004 & 0.177 & 0.802 &  & 0.003 & 0.024 & 0.357 \\
\multicolumn{12}{l}{\textit{Special Message to   the Congress on Fiscal Policy}} \\
1966/09/08 & Public Law 89-800 & No & 97 &  & 0.007 & 0.277 & 0.809 &  & 0.003 & 0.064 & 0.608 \\
\multicolumn{12}{l}{\textit{State of Union   address}} \\
1967/01/10 & \begin{tabular}[c]{@{}l@{}}Revenue and Expenditure\\ Control Act of 1968\end{tabular} & Yes & 178 &  & 0.004 & 0.075 & 0.851 &  & 0.004 & 0.028 & 0.498 \\
\multicolumn{12}{l}{\textit{Annual Budget   Message to the Congress}} \\
1967/01/24 & \begin{tabular}[c]{@{}l@{}}Revenue and Expenditure\\ Control Act of 1968\end{tabular} & Yes & 285 &  & 0.005 & 0.176 & 0.871 &  & 0.003 & 0.024 & 0.558 \\
\multicolumn{12}{l}{\textit{Special Message to   the Congress: The State of the Budget and the Economy}} \\
1967/08/03 & \begin{tabular}[c]{@{}l@{}}Revenue and Expenditure\\ Control Act of 1968\end{tabular} & Yes & 130 &  & 0.006 & 0.303 & 0.776 &  & 0.005 & 0.033 & 0.498 \\
\multicolumn{12}{l}{\textit{Annual Budget   Message to the Congress}} \\
1968/01/29 & \begin{tabular}[c]{@{}l@{}}Revenue and Expenditure\\ Control Act of 1968\end{tabular} & Yes & 419 &  & 0.004 & 0.165 & 0.824 &  & 0.003 & 0.029 & 0.581 \\
\multicolumn{12}{l}{\textit{Special Message to   Congress on Fiscal Policy}} \\
1969/03/26 & Extending the Ten Percent Surtax & No & 18 &  & 0.124 & 0.435 & 0.805 &  & 0.005 & 0.035 & 0.248 \\
\multicolumn{12}{l}{\textit{Windfall Profits   Tax and Energy Security Trust Fund Message to the Congress}} \\
1979/04/26 & Crude Oil Windfall Profit Tax Act of 1980 & No & 77 &  & 0.011 & 0.304 & 0.782 &  & 0.003 & 0.019 & 0.150 \\
\multicolumn{12}{l}{\textit{Statement   Announcing the Establishment of the National Commission on Social Security   Reform}} \\
1981/12/16 & Social Security Amendments of 1983 & No & 7 &  & 0.040 & 0.302 & 0.547 &  & 0.005 & 0.009 & 0.015 \\
\multicolumn{12}{l}{\textit{State of the Union   address}} \\
1982/01/26 & \begin{tabular}[c]{@{}l@{}}Tax Equity and Fiscal\\ Responsibility Act of 1982\end{tabular} & No & 77 &  & 0.004 & 0.205 & 0.826 &  & 0.003 & 0.068 & 0.576 \\
\multicolumn{12}{l}{\textit{Address Before a   Joint Session of the Congress on the State of the Union}} \\
1984/01/25 & Deficit Reduction Act of 1984 & No & 80 &  & 0.003 & 0.116 & 0.753 &  & 0.003 & 0.030 & 0.516 \\
\multicolumn{12}{l}{\textit{Message to the   Congress Transmitting the Fiscal Year 1985 Budget}} \\
1984/02/01 & Deficit Reduction Act of 1984 & No & 105 &  & 0.003 & 0.208 & 0.807 &  & 0.002 & 0.037 & 0.440 \\
\multicolumn{12}{l}{\textit{Remarks to   Reporters Announcing a Deficit Reduction Plan}} \\
1984/03/15 & Deficit Reduction Act of 1984 & No & 26 &  & 0.012 & 0.184 & 0.772 &  & 0.004 & 0.016 & 0.028 \\
\multicolumn{12}{l}{\textit{Joint Session of   Congress on the State of the Union}} \\
1987/01/27 & Omnibus Budget Reconciliation Act of 1987 & No & 34 &  & 0.003 & 0.073 & 0.631 &  & 0.003 & 0.040 & 0.682 \\
\multicolumn{12}{l}{\textit{Statement on   Proposed Tax Increases}} \\
1987/10/15 & Omnibus Budget Reconciliation Act of 1987 & No & 4 &  & 0.350 & 0.461 & 0.618 &  & 0.004 & 0.006 & 0.010 \\
\multicolumn{12}{l}{\textit{Statement on the   federal budget negotiations}} \\
1990/06/26 & Omnibus Budget Reconciliation Act of 1990 & No & 3 &  & 0.010 & 0.268 & 0.777 &  & 0.008 & 0.011 & 0.017 \\
\multicolumn{12}{l}{\textit{Address to the   Nation on the Federal Budget Agreement}} \\
1990/10/02 & Omnibus Budget Reconciliation Act of 1990 & No & 15 &  & 0.009 & 0.261 & 0.704 &  & 0.004 & 0.057 & 0.334 \\
\multicolumn{12}{l}{\textit{Address to the   nation on the economic program}} \\
1993/02/15 & Omnibus Budget Reconciliation Act of 1993 & Yes & 20 &  & 0.007 & 0.243 & 0.740 &  & 0.005 & 0.109 & 0.666 \\
\multicolumn{12}{l}{\textit{Address Before a   Joint Session of Congress on Administration Goals}} \\
1993/02/17 & Omnibus Budget Reconciliation Act of 1993 & Yes & 75 &  & 0.003 & 0.262 & 0.789 &  & 0.002 & 0.054 & 0.634 \\
\multicolumn{12}{l}{\textit{Radio Address}} \\
1993/05/15 & Omnibus Budget Reconciliation Act of 1993 & Yes & 17 &  & 0.006 & 0.346 & 0.819 &  & 0.005 & 0.098 & 0.486 \\

\label{tab:sel_speeches_inc}
\end{longtable}
\end{landscape}
\begin{landscape}
\scriptsize
\begin{longtable}[c]{llllllllllll}

\caption{List of speeches announcing legislation whose net effect was a decrease in tax revenue.} \\

\multicolumn{12}{c}{\textbf{Title of the speech}}                                                                                                                                                                                                                                                                                                                                                                                                                                       \\
\multicolumn{1}{c}{\multirow{2}{*}{\textbf{Date}}} & \multicolumn{1}{c}{\multirow{2}{*}{\textbf{Announced legislation}}}                                   & \multicolumn{1}{c}{\multirow{2}{*}{\textbf{Mixed}}} & \multicolumn{1}{c}{\multirow{2}{*}{\textbf{\begin{tabular}[c]{@{}c@{}}No.\\ paragraphs\end{tabular}}}} & \multicolumn{1}{c}{\textbf{}} & \multicolumn{3}{c}{\textbf{\textit{Tax increase} proportion}}       & \multicolumn{1}{c}{\textbf{}} & \multicolumn{3}{c}{\textbf{\textit{Tax decrease} proportion}}       \\
\multicolumn{1}{c}{}                               & \multicolumn{1}{c}{}                                                                                  & \multicolumn{1}{c}{}                                & \multicolumn{1}{c}{}                                                                                   &                               & \textbf{Min.} & \textbf{Av.} & \textbf{Max.} & \textbf{}                     & \textbf{Min.} & \textbf{Av.} & \textbf{Max.} \\
 \endhead
\multicolumn{12}{l}{\textit{Letter to the   President of the Senate and to the Speaker of the House of Representatives   Urging Continuation of Corporation Tax Rates}}                                                                                                                                                                                                                                                                                                                 \\
1958/05/26                                         & Tax Rate Extension Act of 1958                                                                        & No                                                  & 3                                                                                                      &                               & 0.008         & 0.013        & 0.016         &                               & 0.011         & 0.205        & 0.472         \\
\multicolumn{12}{l}{\textit{Radio and   Television Report to the American People on the State of the National Economy}}                                                                                                                                                                                                                                                                                                                                                                 \\
1962/08/13                                         & \begin{tabular}[c]{@{}l@{}}Changes in Depreciation Guidelines\\ and Revenue Act of 1962\end{tabular}  & Yes                                                 & 68                                                                                                     &                               & 0.004         & 0.150        & 0.750         &                               & 0.005         & 0.218        & 0.744         \\
\multicolumn{12}{l}{\textit{Annual Message to   the Congress on the State of the Union}}                                                                                                                                                                                                                                                                                                                                                                                                \\
1963/01/14                                         & Revenue Act of 1964                                                                                   & Yes                                                 & 86                                                                                                     &                               & 0.002         & 0.037        & 0.525         &                               & 0.003         & 0.120        & 0.895         \\
\multicolumn{12}{l}{\textit{Special Message to   the Congress on Tax Reduction and Reform}}                                                                                                                                                                                                                                                                                                                                                                                             \\
1963/01/24                                         & Revenue Act of 1964                                                                                   & Yes                                                 & 145                                                                                                    &                               & 0.002         & 0.025        & 0.331         &                               & 0.011         & 0.572        & 0.899         \\
\multicolumn{12}{l}{\textit{Remarks at the   National Conference of the Business Committee for Tax Reduction}}                                                                                                                                                                                                                                                                                                                                                                          \\
1963/09/10                                         & \begin{tabular}[c]{@{}l@{}}Changes in Depreciation Guidelines\\ and Revenue Act of 1962\end{tabular}  & Yes                                                 & 36                                                                                                     &                               & 0.004         & 0.078        & 0.590         &                               & 0.012         & 0.452        & 0.847         \\
\multicolumn{12}{l}{\textit{Remarks Upon   Announcing Plans to Recommend a Reduction in Excise Taxes}}                                                                                                                                                                                                                                                                                                                                                                                  \\
1965/05/15                                         & Excise Tax Reduction Act of 1965                                                                      & No                                                  & 14                                                                                                     &                               & 0.007         & 0.132        & 0.571         &                               & 0.010         & 0.252        & 0.647         \\
\multicolumn{12}{l}{\textit{Special Message to   the Congress Recommending Reduction of Excise Taxes and Increases in User   Charges}}                                                                                                                                                                                                                                                                                                                                                  \\
1965/05/17                                         & Excise Tax Reduction Act of 1965                                                                      & No                                                  & 112                                                                                                    &                               & 0.003         & 0.048        & 0.579         &                               & 0.011         & 0.311        & 0.729         \\
\multicolumn{12}{l}{\textit{Special Message to   the Congress Recommending Reinstatement of the Investment Tax Credit and   Accelerated Depreciation Investment Incentives}}                                                                                                                                                                                                                                                                                                            \\
1967/03/09                                         & Public Law 90-26                                                                                      & Yes                                                 & 33                                                                                                     &                               & 0.008         & 0.090        & 0.546         &                               & 0.007         & 0.202        & 0.668         \\
\multicolumn{12}{l}{\textit{Special Message to   the Congress on Reform of the Federal Tax System}}                                                                                                                                                                                                                                                                                                                                                                                     \\
1969/04/21                                         & Tax Reform Act of 1969                                                                                & No                                                  & 38                                                                                                     &                               & 0.004         & 0.044        & 0.434         &                               & 0.022         & 0.400        & 0.786         \\
\multicolumn{12}{l}{\textit{Statement   Announcing Changes in Administration of the Depreciation Provisions of the   Tax Laws}}                                                                                                                                                                                                                                                                                                                                                         \\
1971/01/11                                         & Reform of Depreciation Rules                                                                          & No                                                  & 17                                                                                                     &                               & 0.005         & 0.067        & 0.337         &                               & 0.012         & 0.300        & 0.801         \\
\multicolumn{12}{l}{\textit{Annual Budget   Message to the Congress}}                                                                                                                                                                                                                                                                                                                                                                                                                   \\
1971/01/29                                         & Reform of Depreciation Rules                                                                          & No                                                  & 224                                                                                                    &                               & 0.003         & 0.023        & 0.524         &                               & 0.005         & 0.140        & 0.794         \\
\multicolumn{12}{l}{\textit{Challenge of Peace}}                                                                                                                                                                                                                                                                                                                                                                                                                                        \\
1971/08/15                                         & Revenue Act of 1971                                                                                   & No                                                  & 56                                                                                                     &                               & 0.006         & 0.090        & 0.395         &                               & 0.004         & 0.125        & 0.652         \\
\multicolumn{12}{l}{\textit{Address to the   Nation on Energy and Economic Programs}}                                                                                                                                                                                                                                                                                                                                                                                                   \\
1975/01/13                                         & Tax Reduction Act of 1975                                                                             & Yes                                                 & 41                                                                                                     &                               & 0.005         & 0.062        & 0.499         &                               & 0.009         & 0.203        & 0.818         \\
\multicolumn{12}{l}{\textit{Address Before a   Joint Session of the Congress Reporting on the State of the Union}}                                                                                                                                                                                                                                                                                                                                                                      \\
1975/01/15                                         & Tax Reduction Act of 1975                                                                             & Yes                                                 & 91                                                                                                     &                               & 0.003         & 0.040        & 0.622         &                               & 0.005         & 0.188        & 0.748         \\
\multicolumn{12}{l}{\textit{Annual Budget   Message to the Congress}}                                                                                                                                                                                                                                                                                                                                                                                                                   \\
1975/02/03                                         & Tax Reduction Act of 1975                                                                             & Yes                                                 & 89                                                                                                     &                               & 0.002         & 0.033        & 0.449         &                               & 0.003         & 0.319        & 0.875         \\
\multicolumn{12}{l}{\textit{Address to the   Nation on Federal Tax and Spending Reduction}}                                                                                                                                                                                                                                                                                                                                                                                             \\
1975/10/06                                         & Revenue Adjustment Act of 1975                                                                        & No                                                  & 28                                                                                                     &                               & 0.003         & 0.081        & 0.439         &                               & 0.013         & 0.337        & 0.766         \\
\multicolumn{12}{l}{\textit{Annual Message to   Congress}}                                                                                                                                                                                                                                                                                                                                                                                                                              \\
1976/01/26                                         & Tax Reform Act of 1976                                                                                & No                                                  & 24                                                                                                     &                               & 0.002         & 0.130        & 0.737         &                               & 0.002         & 0.168        & 0.687         \\
\multicolumn{12}{l}{\textit{Economic Recovery   Program—Message to the Congress}}                                                                                                                                                                                                                                                                                                                                                                                                       \\
1977/01/31                                         & Tax Reduction and Simplification Act of 1977                                                          & Yes                                                 & 34                                                                                                     &                               & 0.007         & 0.056        & 0.458         &                               & 0.022         & 0.334        & 0.711         \\
\multicolumn{12}{l}{\textit{Fiscal Year 1978   Budget Revisions Message to the Congress Transmitting the Revisions}}                                                                                                                                                                                                                                                                                                                                                                    \\
1977/02/22                                         & Tax Reduction and Simplification Act of 1977                                                          & Yes                                                 & 13                                                                                                     &                               & 0.004         & 0.027        & 0.126         &                               & 0.013         & 0.286        & 0.696         \\
\multicolumn{12}{l}{\textit{State of the Union   address}}                                                                                                                                                                                                                                                                                                                                                                                                                              \\
1978/01/19                                         & Revenue Act of 1978                                                                                   & No                                                  & 305                                                                                                    &                               & 0.003         & 0.028        & 0.616         &                               & 0.002         & 0.085        & 0.695         \\
\multicolumn{12}{l}{\textit{Tax Reduction and   Reform Message to the Congress}}                                                                                                                                                                                                                                                                                                                                                                                                        \\
1978/01/20                                         & Revenue Act of 1978                                                                                   & No                                                  & 204                                                                                                    &                               & 0.003         & 0.042        & 0.533         &                               & 0.018         & 0.432        & 0.870         \\
\multicolumn{12}{l}{\textit{Budget Message to   the Congress Transmitting the Fiscal Year 1979 Budget}}                                                                                                                                                                                                                                                                                                                                                                                 \\
1978/01/20                                         & Revenue Act of 1978                                                                                   & No                                                  & 35                                                                                                     &                               & 0.003         & 0.025        & 0.168         &                               & 0.004         & 0.187        & 0.751         \\
\multicolumn{12}{l}{\textit{Inaugural Address}}                                                                                                                                                                                                                                                                                                                                                                                                                                         \\
1981/01/20                                         & Economic Recovery Tax Act of 1981                                                                     & Yes                                                 & 35                                                                                                     &                               & 0.004         & 0.022        & 0.336         &                               & 0.004         & 0.054        & 0.471         \\
\multicolumn{12}{l}{\textit{Address to the   Nation on the Economy}}                                                                                                                                                                                                                                                                                                                                                                                                                    \\
1981/02/05                                         & Economic Recovery Tax Act of 1981                                                                     & Yes                                                 & 50                                                                                                     &                               & 0.004         & 0.080        & 0.624         &                               & 0.008         & 0.349        & 0.775         \\
\multicolumn{12}{l}{\textit{Address before a   Joint Session of the Congress on the Program for Economic Recovery}}                                                                                                                                                                                                                                                                                                                                                                     \\
1981/02/18                                         & Economic Recovery Tax Act of 1981                                                                     & Yes                                                 & 63                                                                                                     &                               & 0.004         & 0.082        & 0.645         &                               & 0.005         & 0.329        & 0.778         \\
\multicolumn{12}{l}{\textit{Address Before a   Joint Session of the Congress on the Program for Economic Recovery}}                                                                                                                                                                                                                                                                                                                                                                     \\
1981/04/28                                         & Economic Recovery Tax Act of 1981                                                                     & Yes                                                 & 40                                                                                                     &                               & 0.004         & 0.053        & 0.762         &                               & 0.005         & 0.201        & 0.820         \\
\multicolumn{12}{l}{\textit{Address to the   Nation on Tax Reform}}                                                                                                                                                                                                                                                                                                                                                                                                                     \\
1985/05/28                                         & Tax Reform Act of 1986                                                                                & Yes                                                 & 53                                                                                                     &                               & 0.003         & 0.057        & 0.402         &                               & 0.005         & 0.360        & 0.818         \\
\multicolumn{12}{l}{\textit{Address Before a   Joint Session of Congress on the State of the Union}}                                                                                                                                                                                                                                                                                                                                                                                    \\
1986/02/04                                         & Tax Reform Act of 1986                                                                                & Yes                                                 & 34                                                                                                     &                               & 0.002         & 0.028        & 0.451         &                               & 0.003         & 0.115        & 0.750         \\
\multicolumn{12}{l}{\textit{Radio Address to   the Nation on Tax Reform}}                                                                                                                                                                                                                                                                                                                                                                                                               \\
1986/05/10                                         & Tax Reform Act of 1986                                                                                & Yes                                                 & 7                                                                                                      &                               & 0.006         & 0.051        & 0.205         &                               & 0.015         & 0.376        & 0.738         \\
\multicolumn{12}{l}{\textit{The President’s   Radio Address}}                                                                                                                                                                                                                                                                                                                                                                                                                           \\
1997/02/22                                         & \begin{tabular}[c]{@{}l@{}}Taxpayer Relief Act of 1997\\ and Balanced Budget Act of 1997\end{tabular} & Yes                                                 & 8                                                                                                      &                               & 0.003         & 0.119        & 0.530         &                               & 0.009         & 0.354        & 0.865         \\
\multicolumn{12}{l}{\textit{Remarks on   Departure for Boston}}                                                                                                                                                                                                                                                                                                                                                                                                                         \\
1997/06/30                                         & \begin{tabular}[c]{@{}l@{}}Taxpayer Relief Act of 1997\\ and Balanced Budget Act of 1997\end{tabular} & Yes                                                 & 40                                                                                                     &                               & 0.005         & 0.067        & 0.487         &                               & 0.003         & 0.236        & 0.760         \\
\multicolumn{12}{l}{\textit{A press conference}}                                                                                                                                                                                                                                                                                                                                                                                                                                        \\
2001/02/05                                         & \begin{tabular}[c]{@{}l@{}}Economic Growth and Tax Relief\\ Reconciliation Act of 2001\end{tabular}   & No                                                  & 20                                                                                                     &                               & 0.009         & 0.043        & 0.188         &                               & 0.008         & 0.397        & 0.740         \\
\multicolumn{12}{l}{\textit{The President’s   Agenda for Tax Relief}}                                                                                                                                                                                                                                                                                                                                                                                                                   \\
2001/02/08                                         & \begin{tabular}[c]{@{}l@{}}Economic Growth and Tax Relief\\ Reconciliation Act of 2001\end{tabular}   & No                                                  & 1                                                                                                      &                               & 0.129         & 0.129        & 0.129         &                               & 0.207         & 0.207        & 0.207         \\
\multicolumn{12}{l}{\textit{The President’s   Radio Address}}                                                                                                                                                                                                                                                                                                                                                                                                                           \\
2001/03/17                                         & \begin{tabular}[c]{@{}l@{}}Economic Growth and Tax Relief\\ Reconciliation Act of 2001\end{tabular}   & No                                                  & 10                                                                                                     &                               & 0.010         & 0.169        & 0.615         &                               & 0.083         & 0.303        & 0.738         \\
\multicolumn{12}{l}{\textit{Remarks to   Business}}                                                                                                                                                                                                                                                                                                                                                                                                                                     \\
2001/10/26                                         & Job Creation and Worker Assistance Act of 2002                                                        & Yes                                                 & 43                                                                                                     &                               & 0.005         & 0.054        & 0.426         &                               & 0.003         & 0.084        & 0.745         \\
\multicolumn{12}{l}{\textit{State of the Union   address}}                                                                                                                                                                                                                                                                                                                                                                                                                              \\
2002/01/29                                         & Job Creation and Worker Assistance Act of 2002                                                        & Yes                                                 & 64                                                                                                     &                               & 0.003         & 0.034        & 0.689         &                               & 0.003         & 0.051        & 0.680         \\
\multicolumn{12}{l}{\textit{Remarks to the   Economic Club of Chicago in Chicago}}                                                                                                                                                                                                                                                                                                                                                                                                      \\
2003/01/07                                         & \begin{tabular}[c]{@{}l@{}}Jobs and Growth Tax Relief\\ Reconciliation Act of 2003\end{tabular}       & Yes                                                 & 52                                                                                                     &                               & 0.003         & 0.114        & 0.702         &                               & 0.004         & 0.223        & 0.815         \\
\multicolumn{12}{l}{\textit{The President’s   Radio Address}}                                                                                                                                                                                                                                                                                                                                                                                                                           \\
2003/01/11                                         & \begin{tabular}[c]{@{}l@{}}Jobs and Growth Tax Relief\\ Reconciliation Act of 2003\end{tabular}       & Yes                                                 & 11                                                                                                     &                               & 0.008         & 0.086        & 0.396         &                               & 0.011         & 0.226        & 0.795         \\
\multicolumn{12}{l}{\textit{Remarks to the Tax   Relief Coalition}}                                                                                                                                                                                                                                                                                                                                                                                                                     \\
2003/05/06                                         & \begin{tabular}[c]{@{}l@{}}Jobs and Growth Tax Relief\\ Reconciliation Act of 2003\end{tabular}       & Yes                                                 & 47                                                                                                     &                               & 0.004         & 0.090        & 0.748         &                               & 0.004         & 0.194        & 0.839         \\
\multicolumn{12}{l}{\textit{State of the Union   address}}                                                                                                                                                                                                                                                                                                                                                                                                                              \\
2004/01/20                                         & Working Families Tax Relief Act of 2004                                                               & No                                                  & 68                                                                                                     &                               & 0.004         & 0.050        & 0.728         &                               & 0.004         & 0.058        & 0.747         \\
\multicolumn{12}{l}{\textit{Remarks on the   national economy}}                                                                                                                                                                                                                                                                                                                                                                                                                         \\
2008/01/18                                         & Economic Stimulus Act of 2008                                                                         & No                                                  & 14                                                                                                     &                               & 0.004         & 0.094        & 0.563         &                               & 0.007         & 0.182        & 0.824       \\ 

\label{tab:sel_speeches_dec}
\end{longtable}
\end{landscape}

\subsection{Wordclouds - composition of the topics} \label{sec:wordclouds}

Figure \ref{wc:all} shows the 26 topic distributions estimated in the second step of our LDA approach in the form of wordclouds. For each topic 150 terms with the highest probability of being used are shown. The size of a term indicates its relative probability. For each topic we provide an interpretation which we believe best captures its likely usage. As expected the majority of estimated topics relate to particular issues in U.S. politics. It is important to note however that this is not the case for all of them. Topics are based on terms that tend to co-occur in the documents, and that co-occurrence can be caused not only by the what issues are discussed. For example, our data-set includes interviews with the president, and the particular language used in those seems to ``picked up'' by topic shown in Fig. \ref{wc:interviews}. Another example is the topic which we call \textit{Informal Speech} shown in Fig. \ref{wc:informal_speech}. The fact that it heavily features the term ``laughter'' used in the transcripts to indicate when the president or the audience is laughing - suggests that it concerns the informal part of presidential speeches. Other terms however do not seem to be connected by any particular theme, and as such, it might be an artifact of our pre-processing approach. In particular, terms that show up frequently regardless of the actual content might co-occur ``naturally'' and be grouped into a common topic. 

\begin{figure}[ht]
\includegraphics[width=0.95\linewidth]{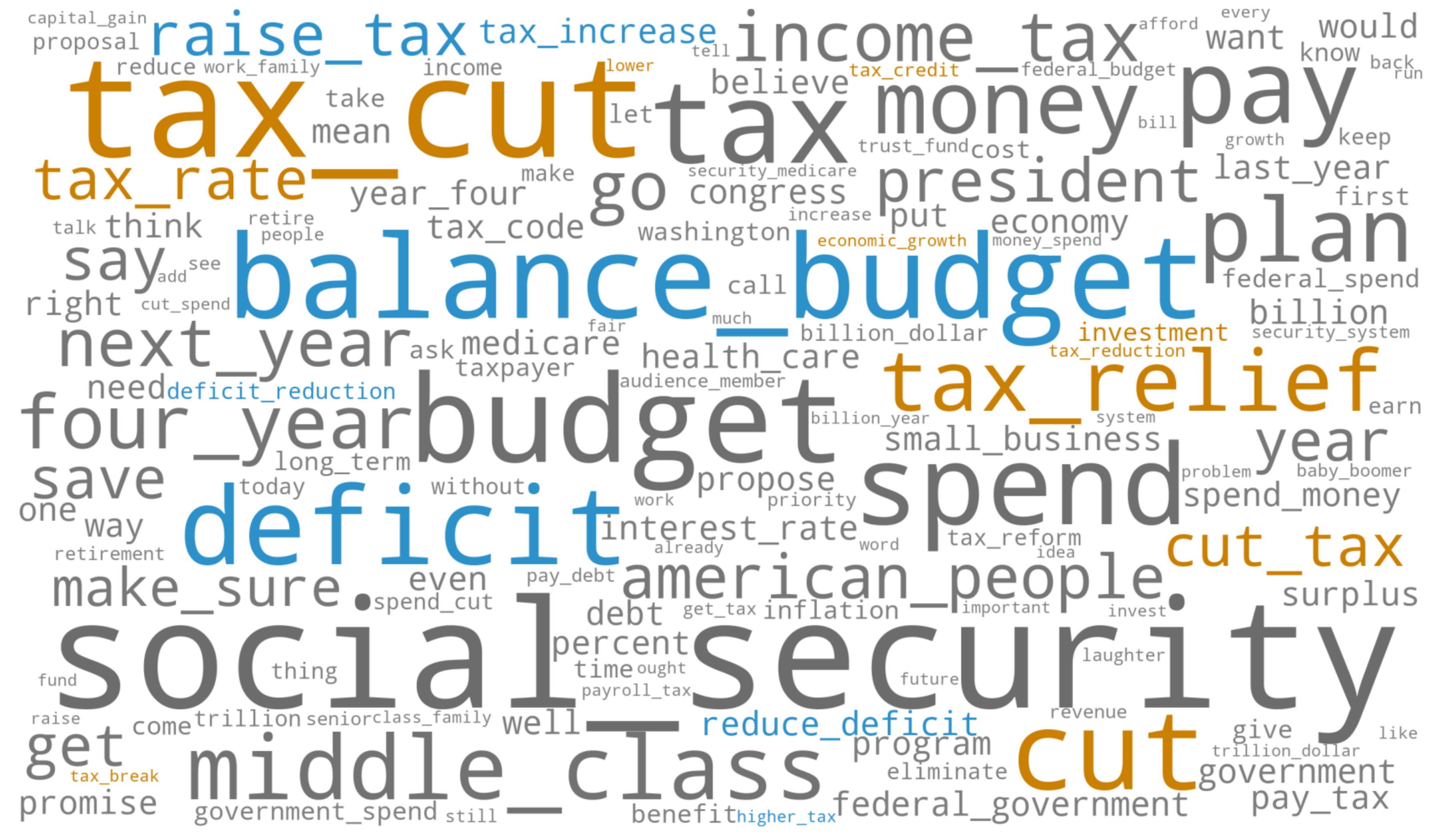} 
 \caption{General tax topic from the (unsupervised) first step LDA.}
 \label{fig:wc_gen}
 \end{figure}

\clearpage

\begin{figure}

\begin{subfigure}{0.49\columnwidth}
\includegraphics[width=0.95\linewidth]{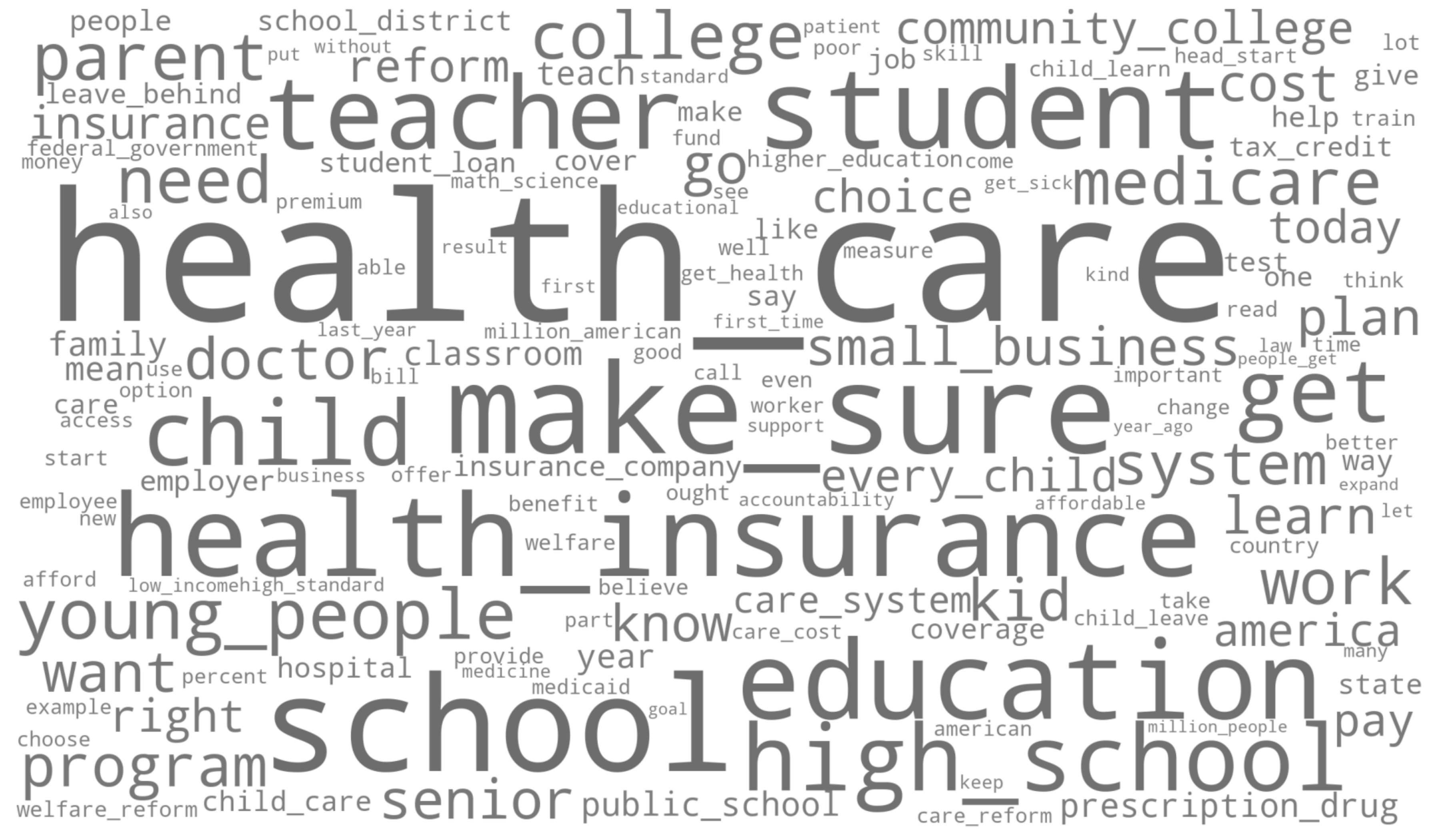}
\caption{Social Services}
\end{subfigure}
\begin{subfigure}{0.49\columnwidth}
\includegraphics[width=0.95\linewidth]{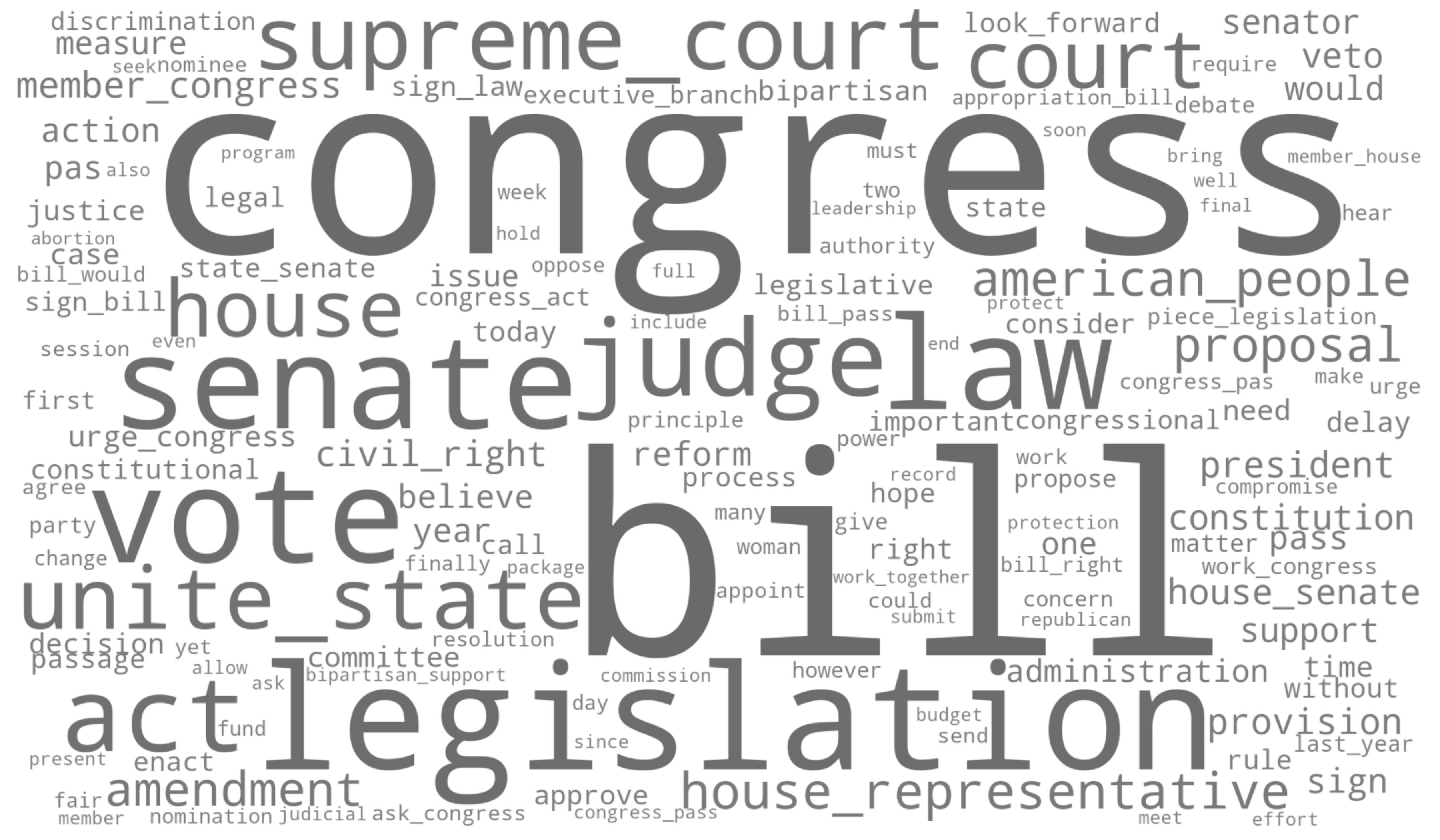}
\caption{Legislative Process}
\end{subfigure}
\begin{subfigure}{0.49\columnwidth}
\includegraphics[width=0.95\linewidth]{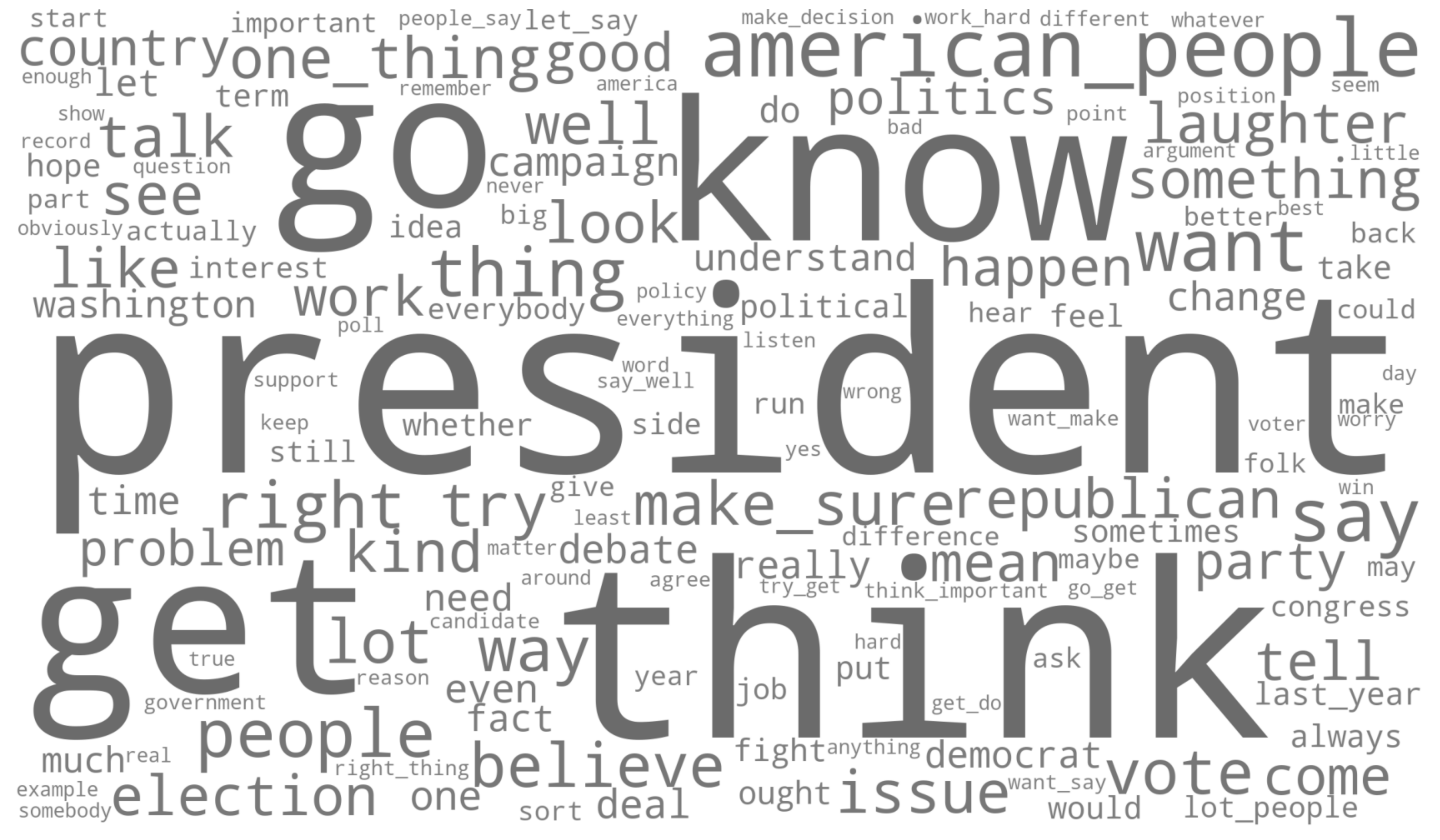}
\caption{Persuasive Rhetoric}
\end{subfigure}
\begin{subfigure}{0.49\columnwidth}
\includegraphics[width=0.95\linewidth]{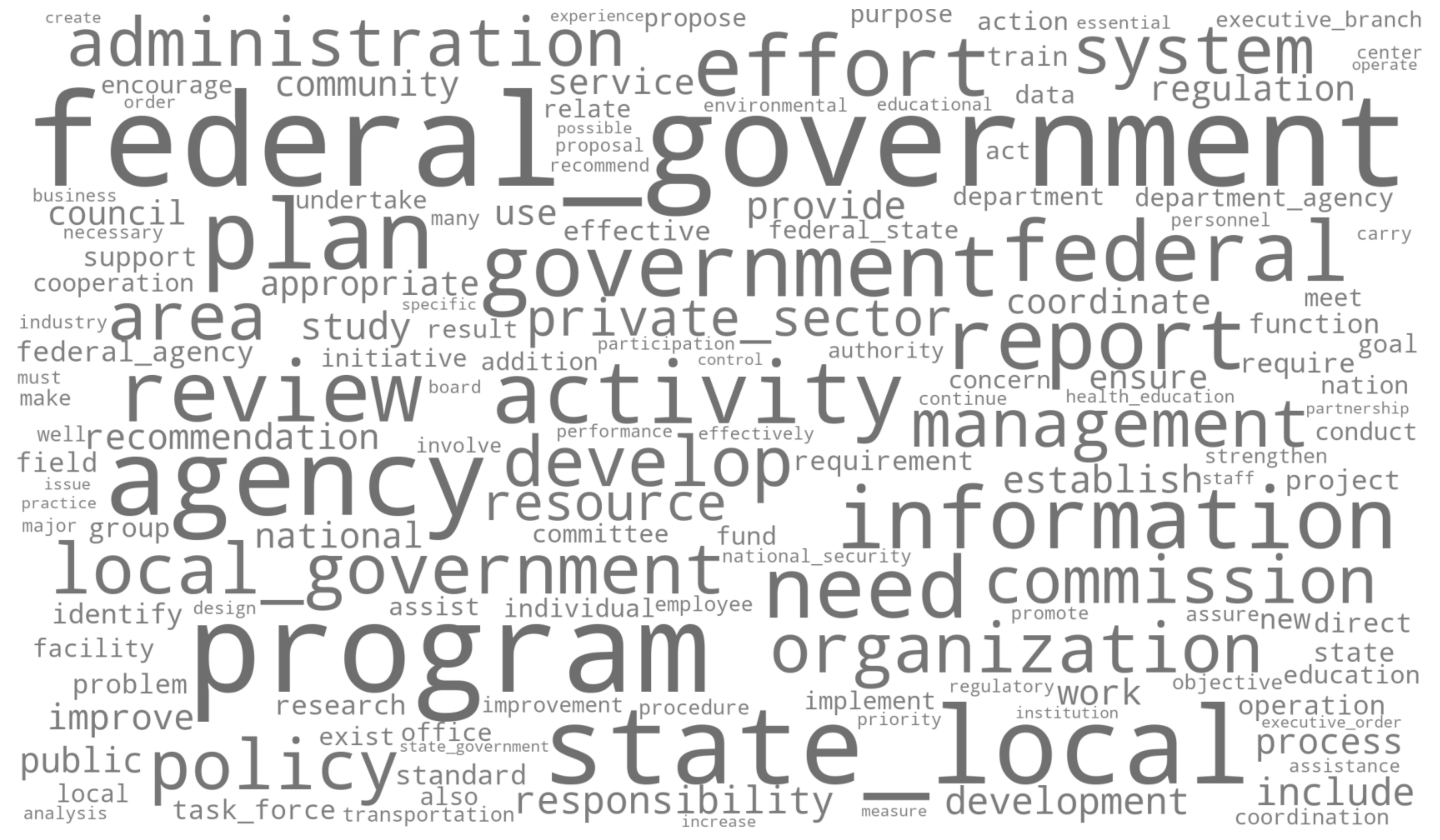}
\caption{Public Administration}
\end{subfigure}
\begin{subfigure}{0.49\columnwidth}
\includegraphics[width=0.95\linewidth]{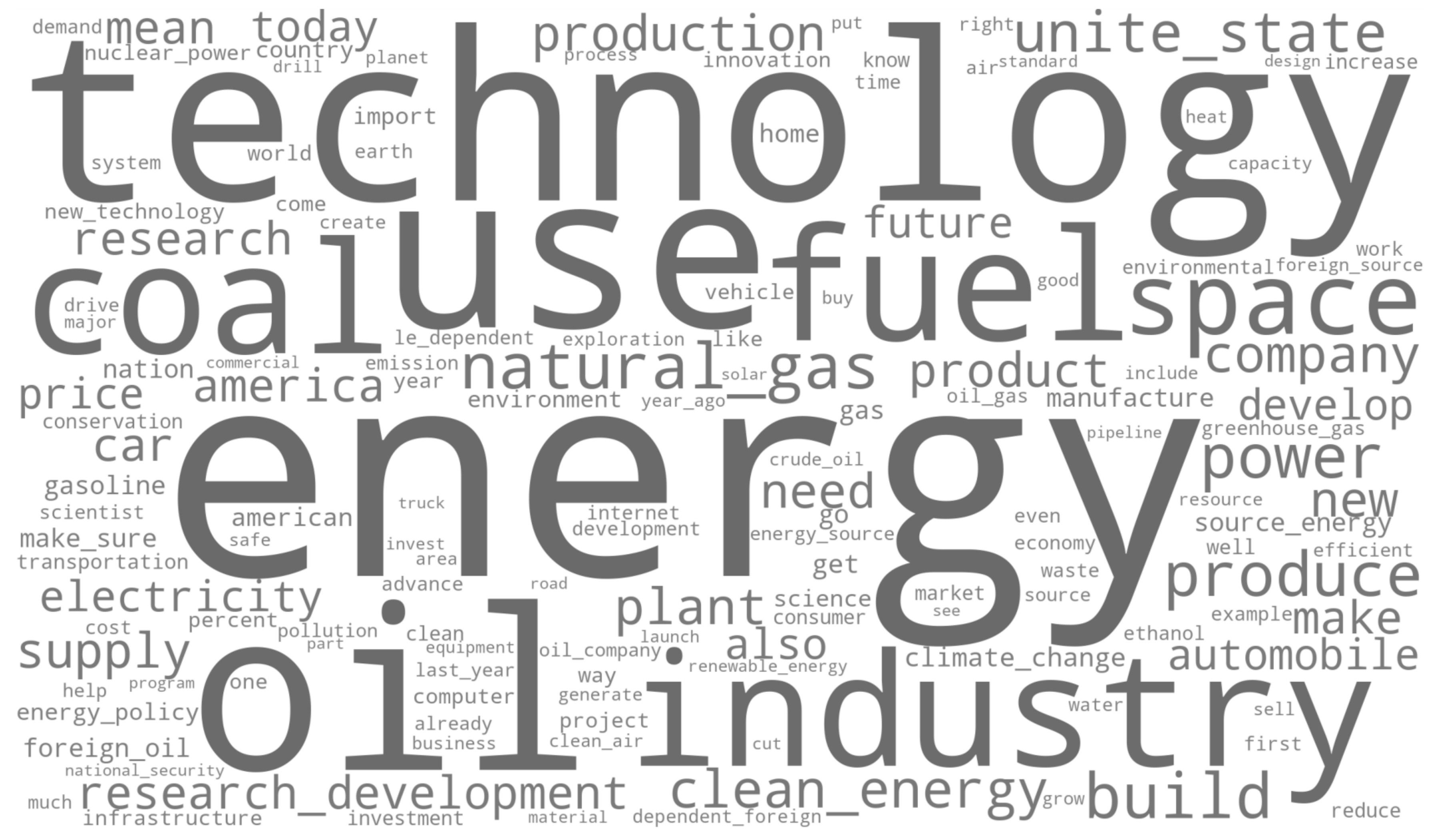}
\caption{Natural Resources, Energy \& Technology}
\end{subfigure}
\begin{subfigure}{0.49\columnwidth}
\includegraphics[width=0.95\linewidth]{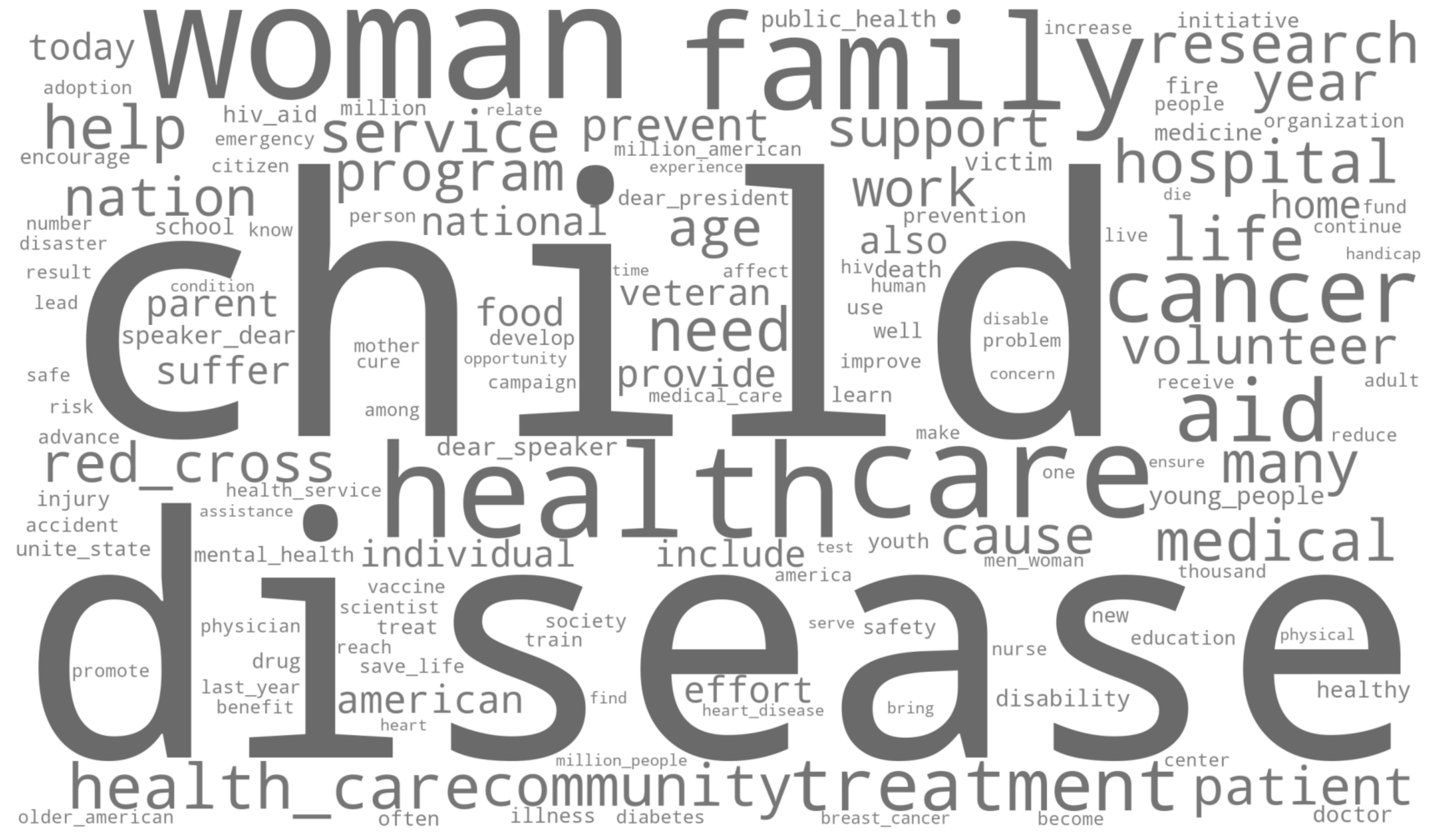} 
\caption{Health Care}
\end{subfigure}
\begin{subfigure}{0.49\columnwidth}
\includegraphics[width=0.95\linewidth]{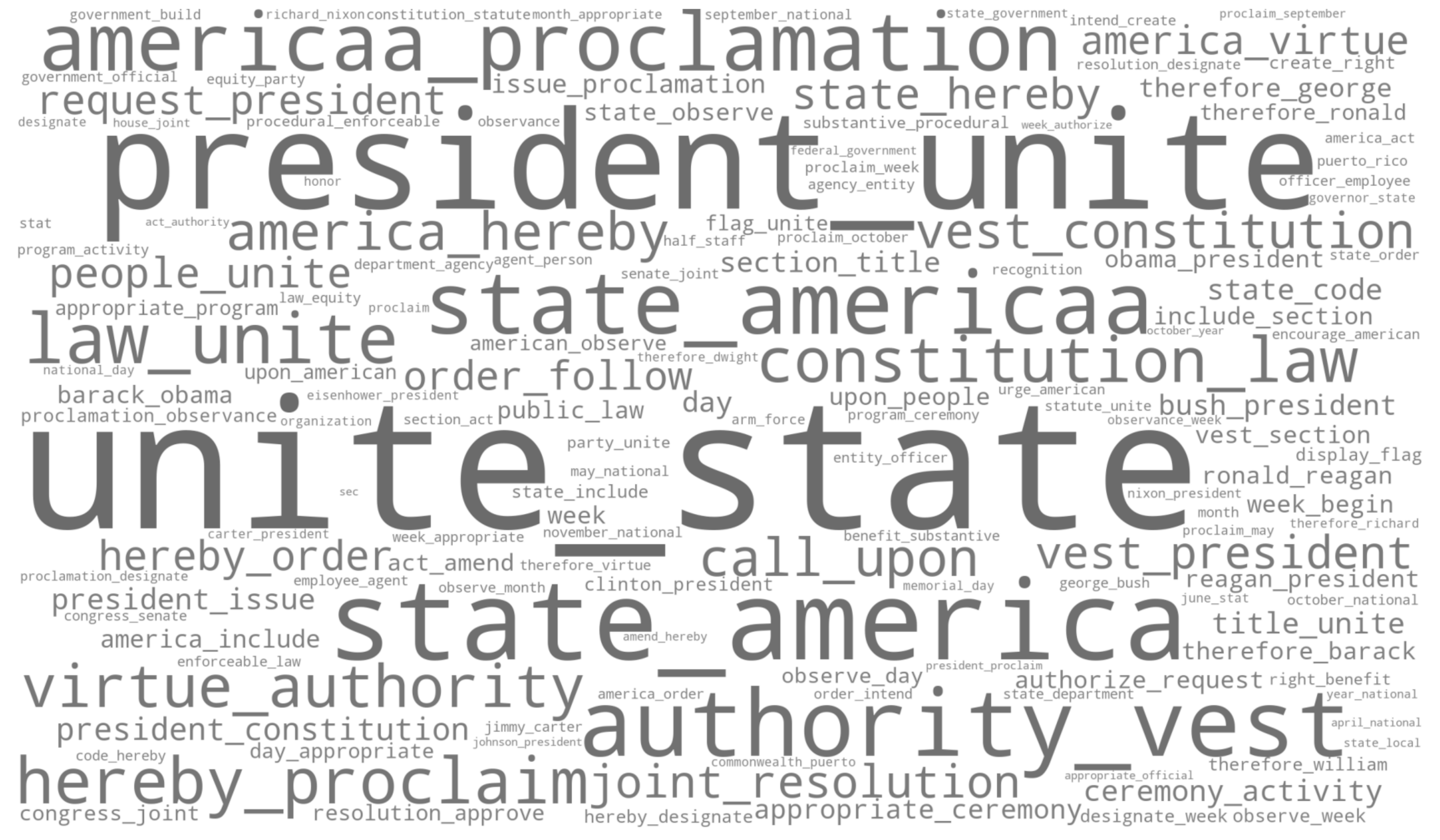}
\caption{Proclamations of Unity \& Patriotism}
\end{subfigure}
\begin{subfigure}{0.49\columnwidth}
\includegraphics[width=0.95\linewidth]{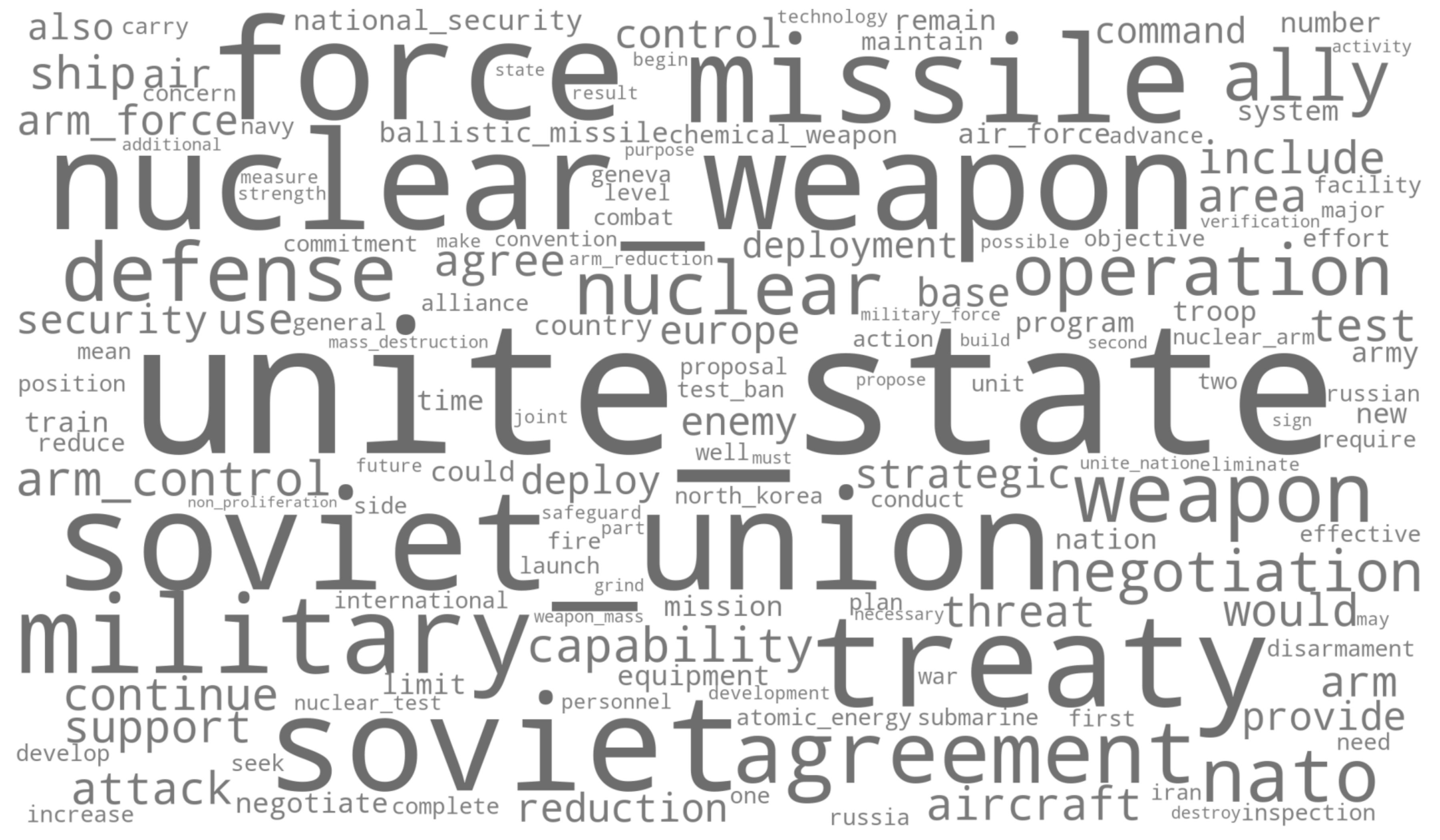}
\caption{Cold War}
\end{subfigure}
\caption{Wordclouds of the non-tax topics}
\end{figure}


\begin{figure}
\ContinuedFloat
\begin{subfigure}{0.49\columnwidth}
\includegraphics[width=0.95\linewidth]{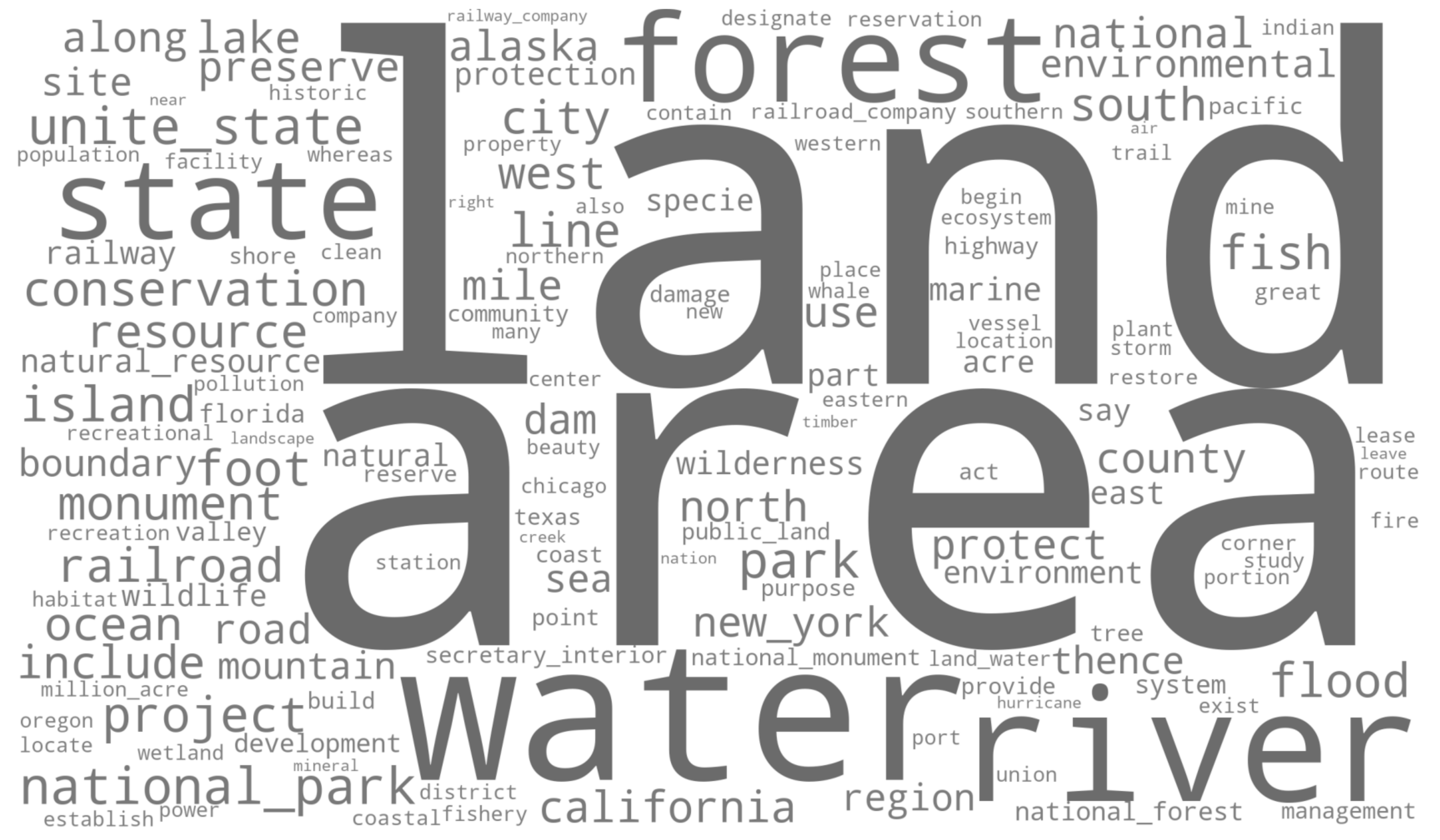}
\caption{Nature \& Environment}
\end{subfigure}
\begin{subfigure}{0.49\columnwidth}
\includegraphics[width=0.95\linewidth]{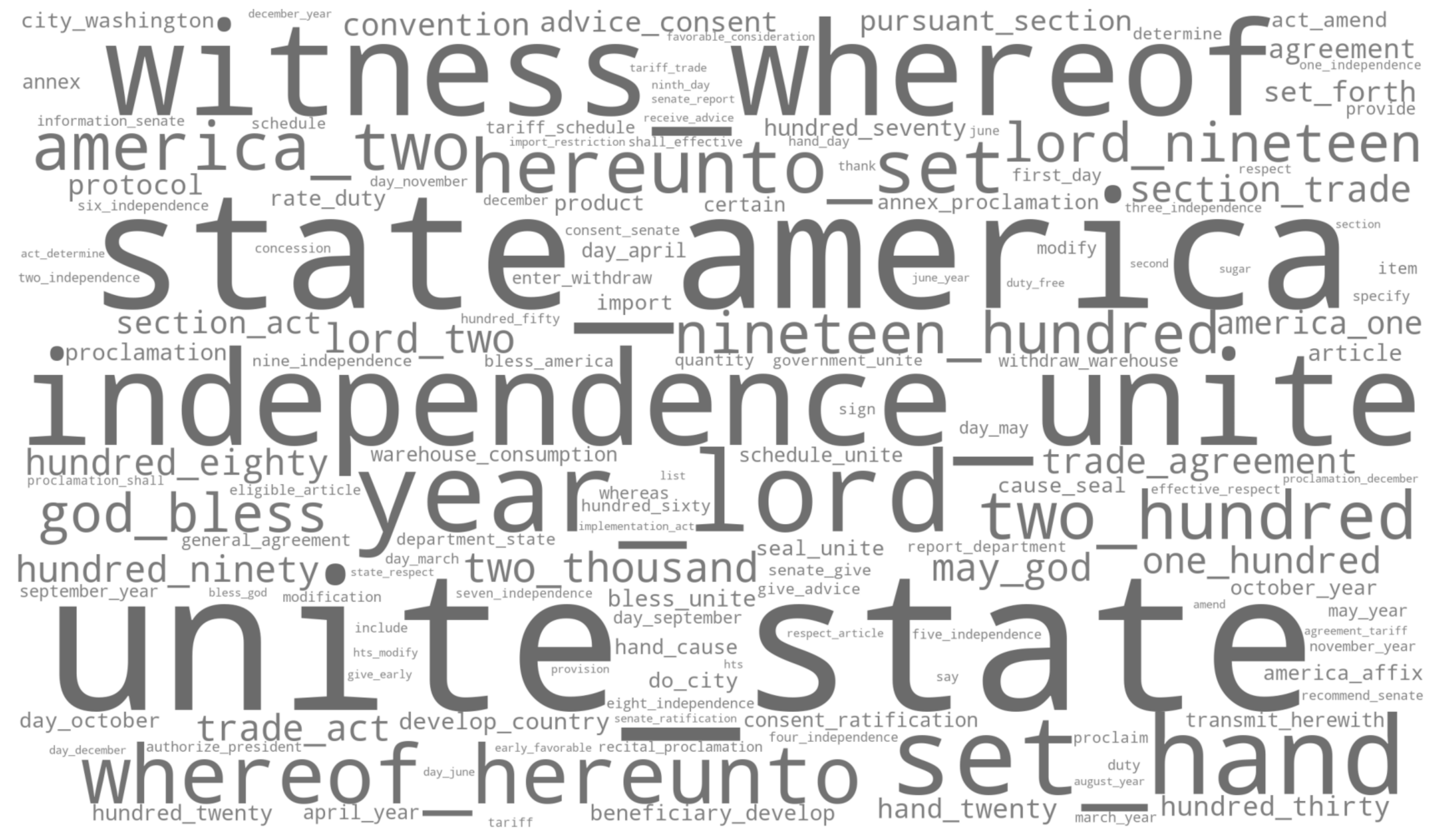}
\caption{Tradition and Conservative Values}
\end{subfigure}
\begin{subfigure}{0.49\columnwidth}
\includegraphics[width=0.95\linewidth]{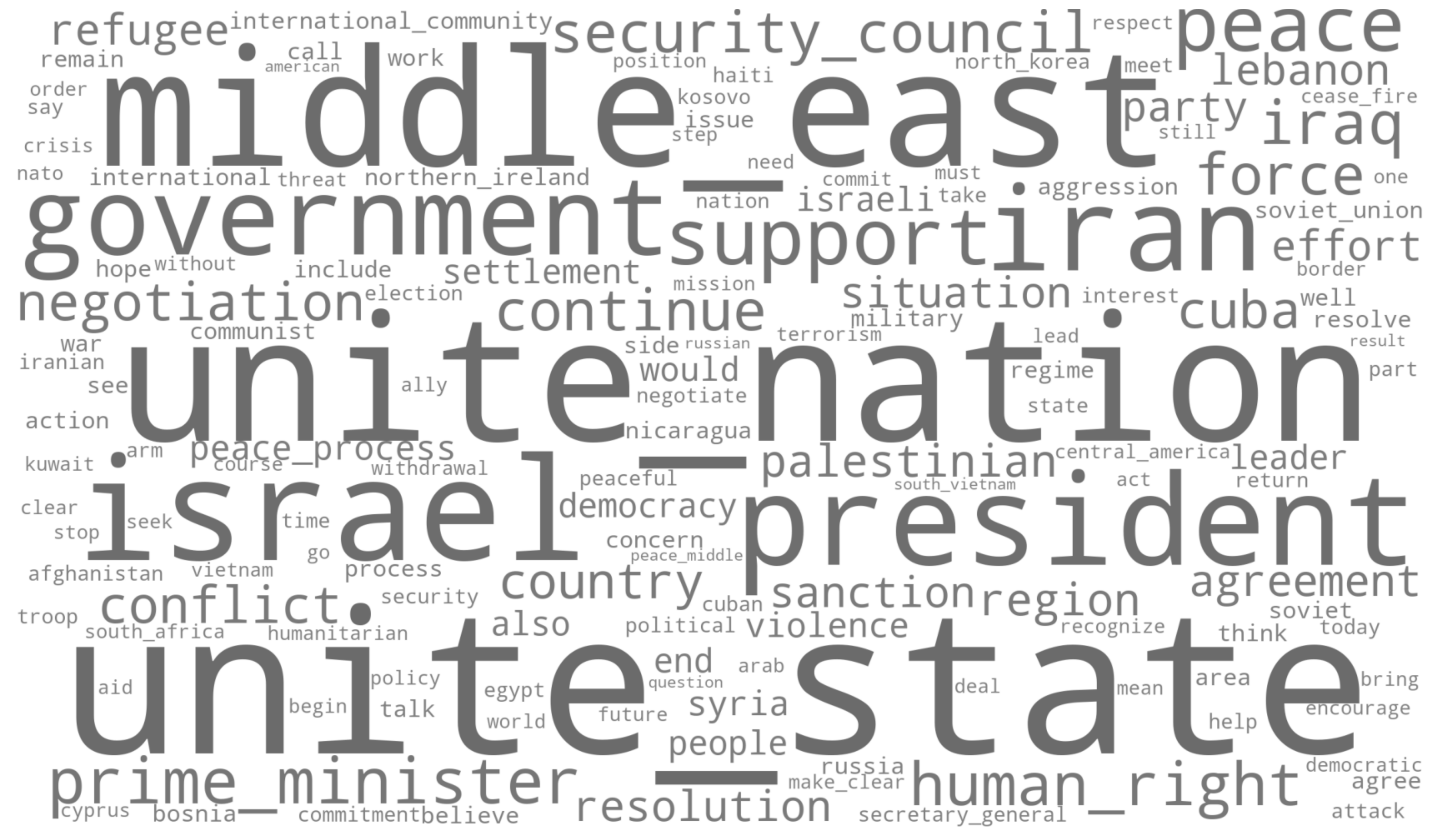}
\caption{Middle East}
\end{subfigure}
\begin{subfigure}{0.49\columnwidth}
\includegraphics[width=0.95\linewidth]{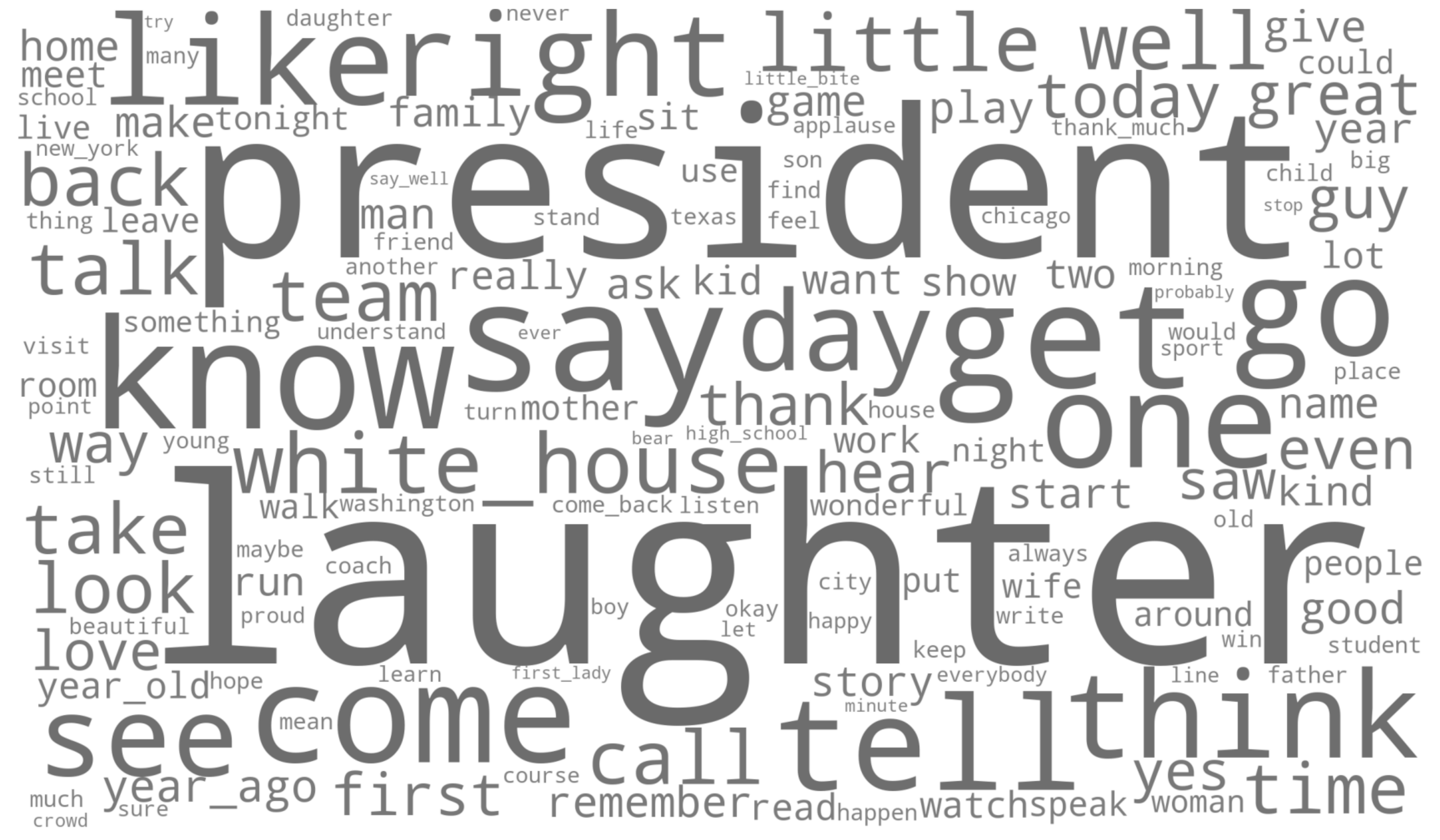}
\caption{Informal Speech}
\label{wc:informal_speech}
\end{subfigure}
\begin{subfigure}{0.49\columnwidth}
\includegraphics[width=0.95\linewidth]{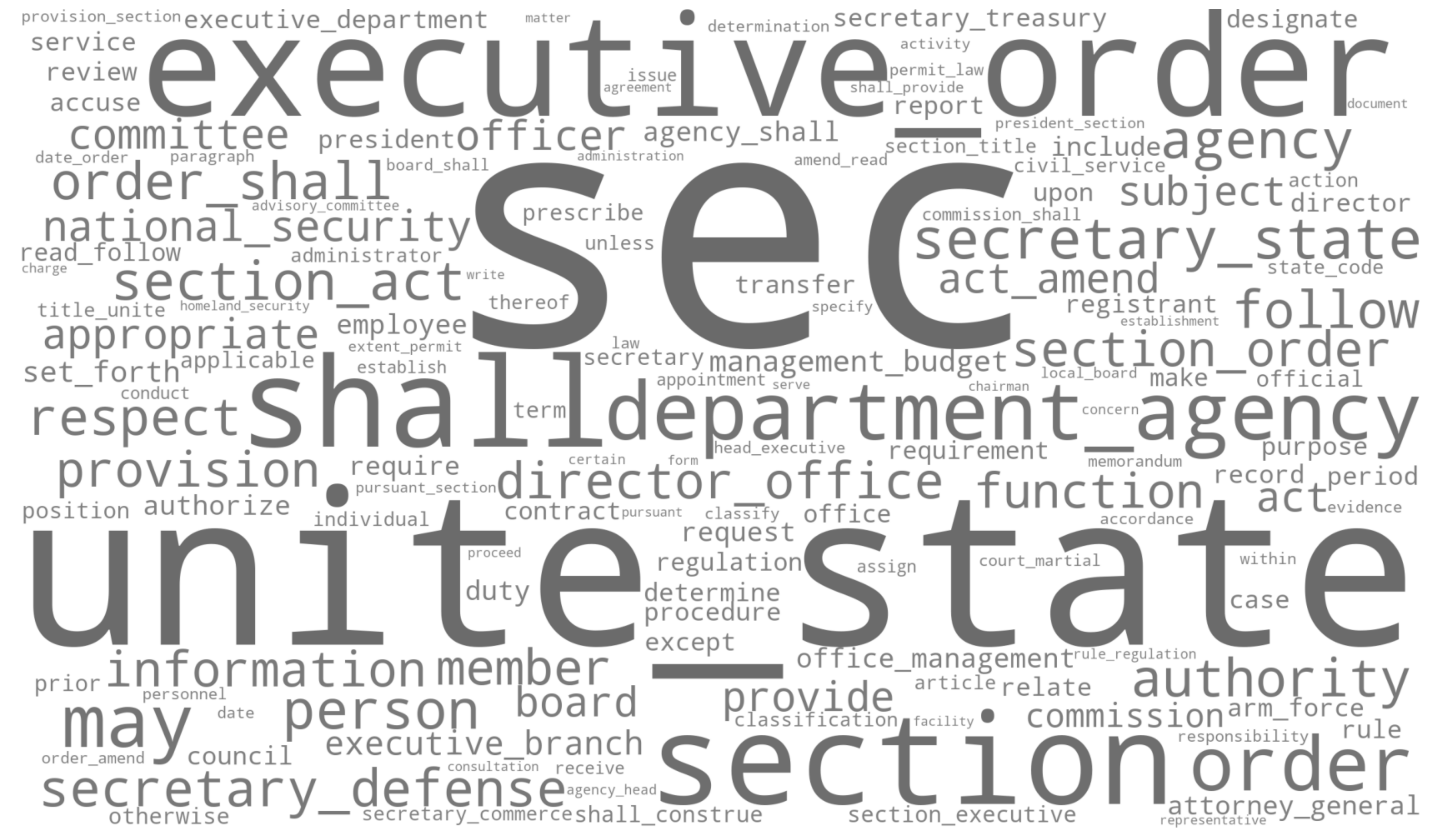}
\caption{Executive Orders}
\end{subfigure}
\begin{subfigure}{0.49\columnwidth}
\includegraphics[width=0.95\linewidth]{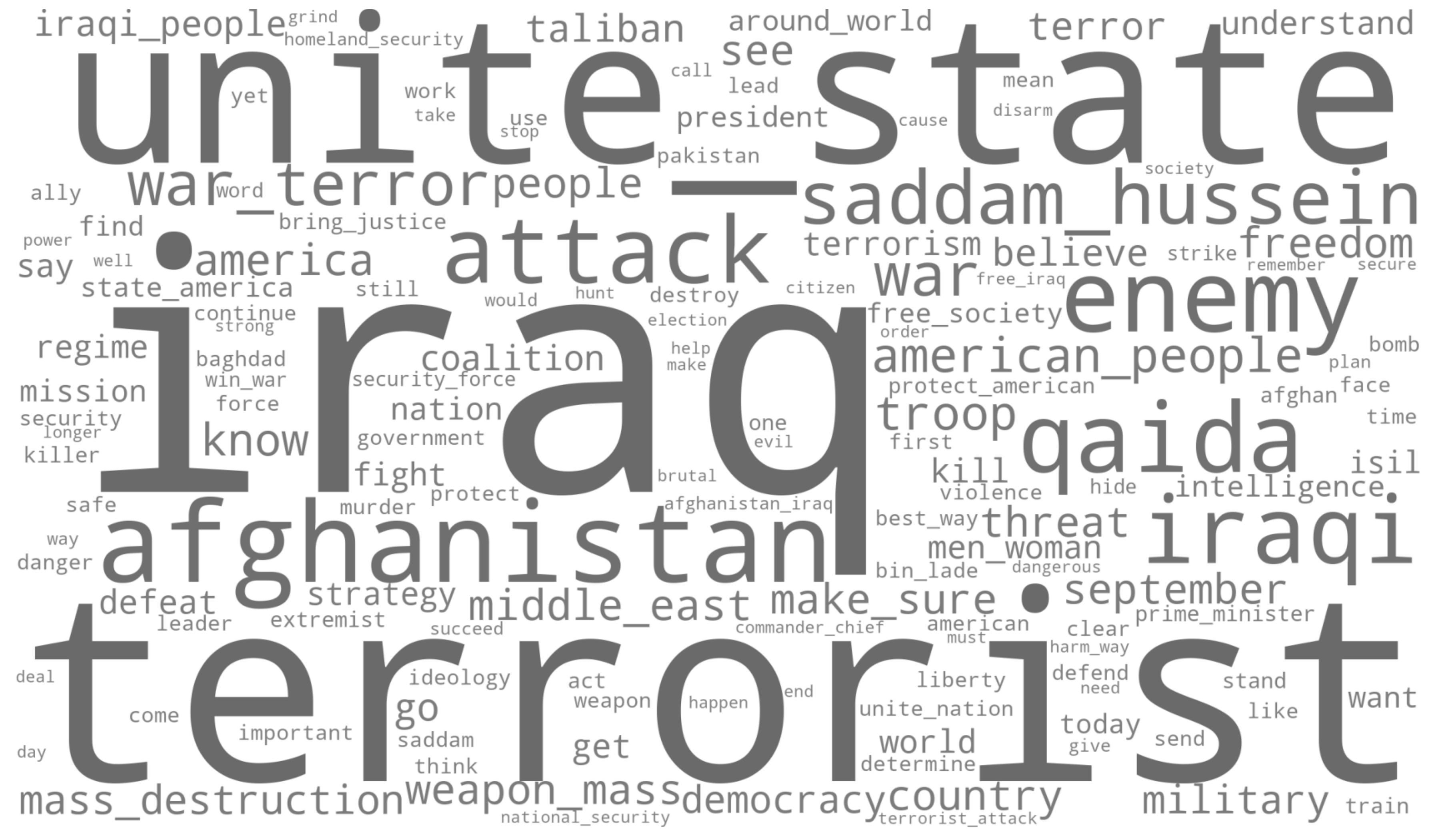} 
\caption{War on Terror}
\end{subfigure}
\begin{subfigure}{0.49\columnwidth}
\includegraphics[width=0.95\linewidth]{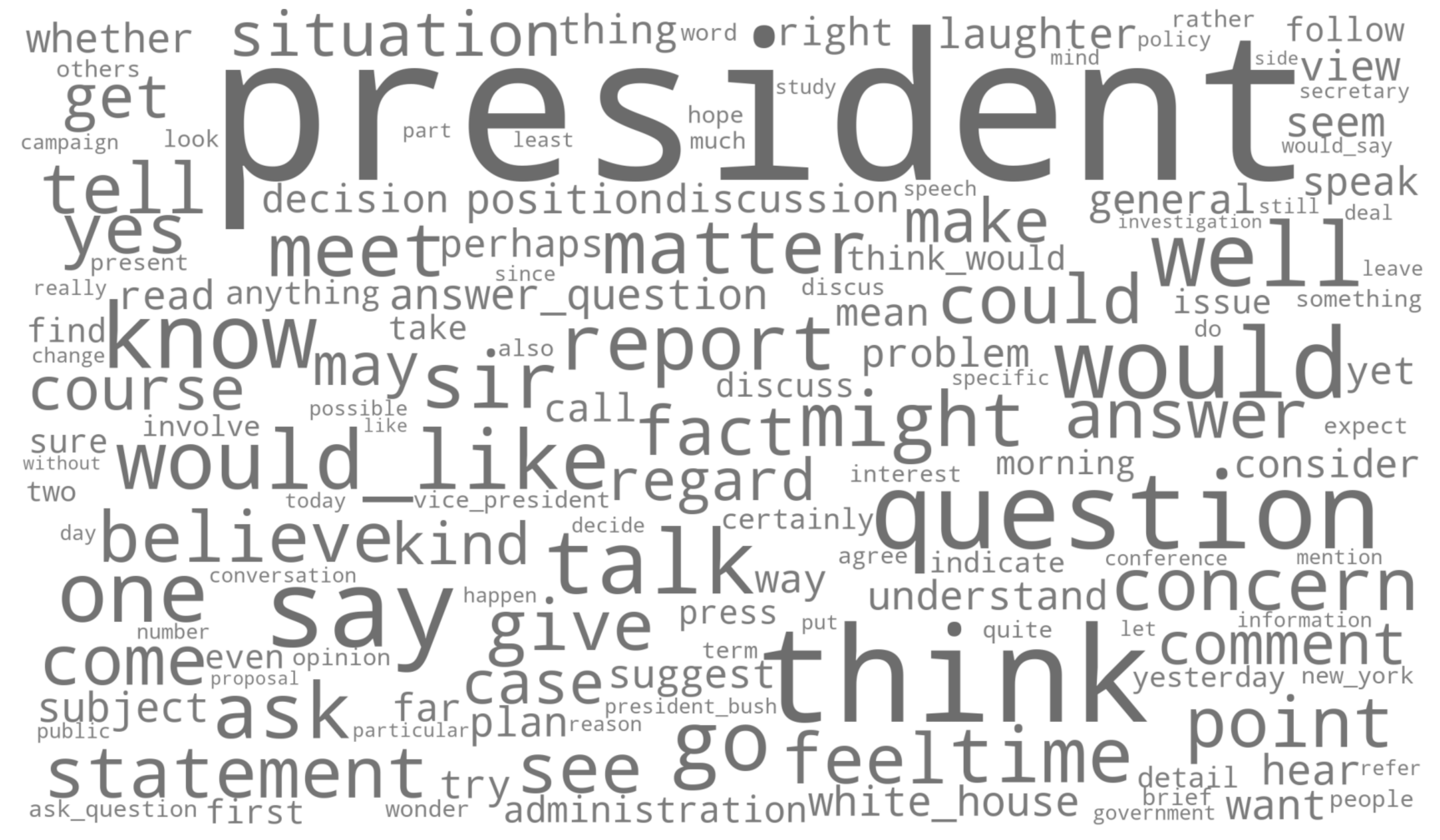}
\caption{Interviews}
\label{wc:interviews}
\end{subfigure}
\begin{subfigure}{0.49\columnwidth}
\includegraphics[width=0.95\linewidth]{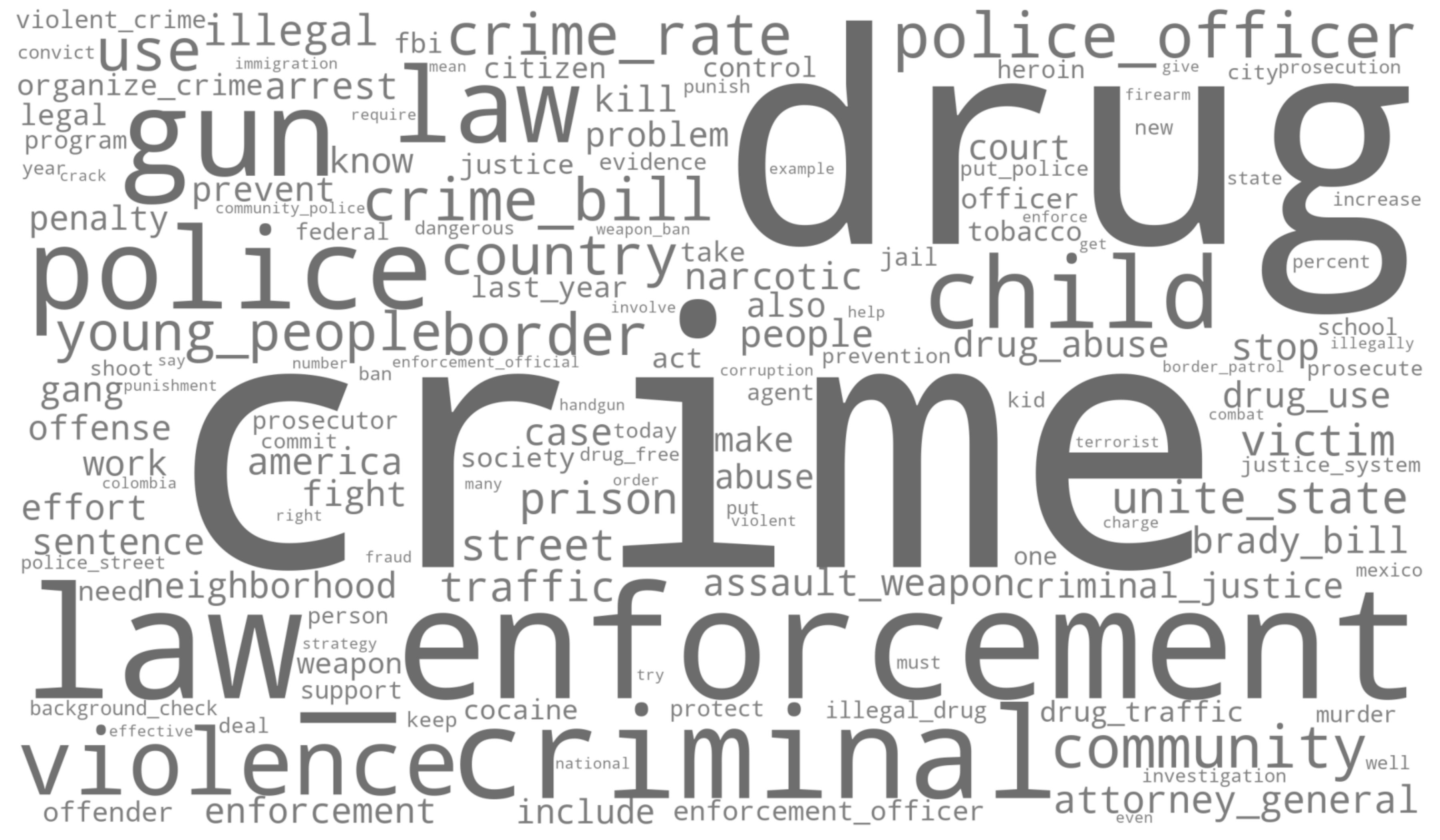}
\caption{Crime}
\end{subfigure}
\caption{Wordclouds of the non-tax topics}
\end{figure}


\begin{figure}
\ContinuedFloat
\begin{subfigure}{0.49\columnwidth}
\includegraphics[width=0.95\linewidth]{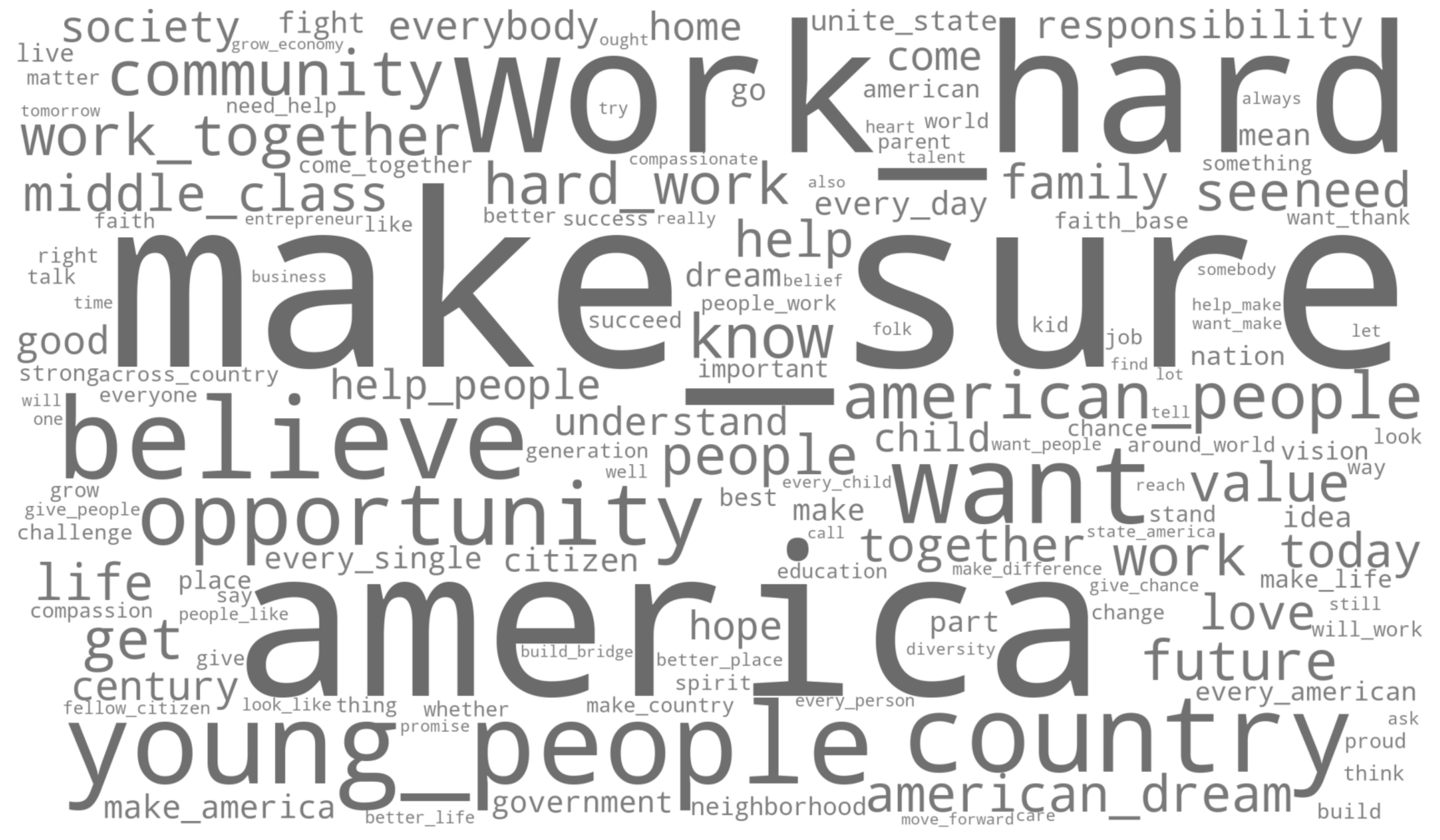}
\caption{American Values}
\end{subfigure}
\begin{subfigure}{0.49\columnwidth}
\includegraphics[width=0.95\linewidth]{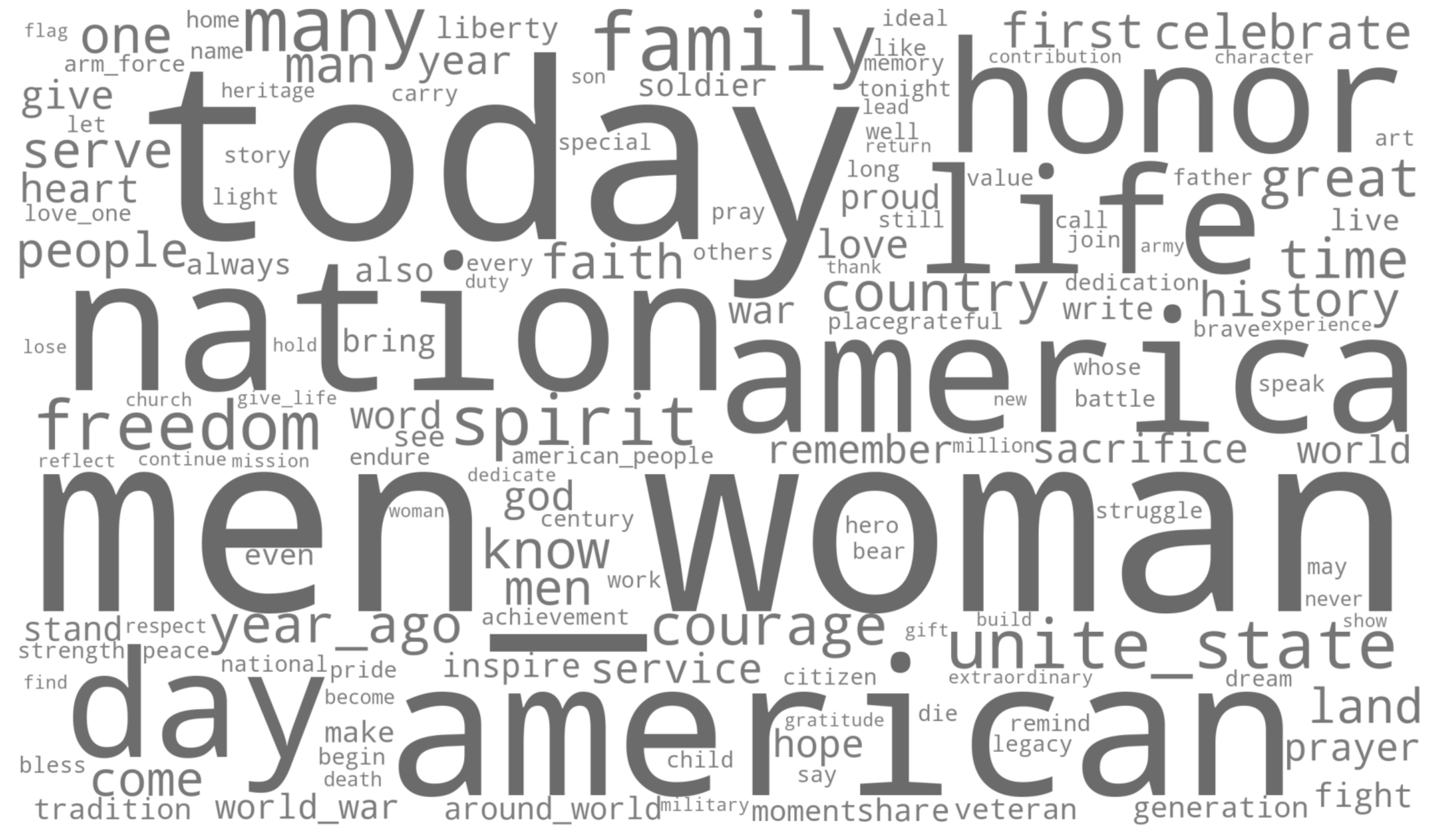}
\caption{Paying Respect to Armed Forces}
\end{subfigure}
\begin{subfigure}{0.49\columnwidth}
\includegraphics[width=0.95\linewidth]{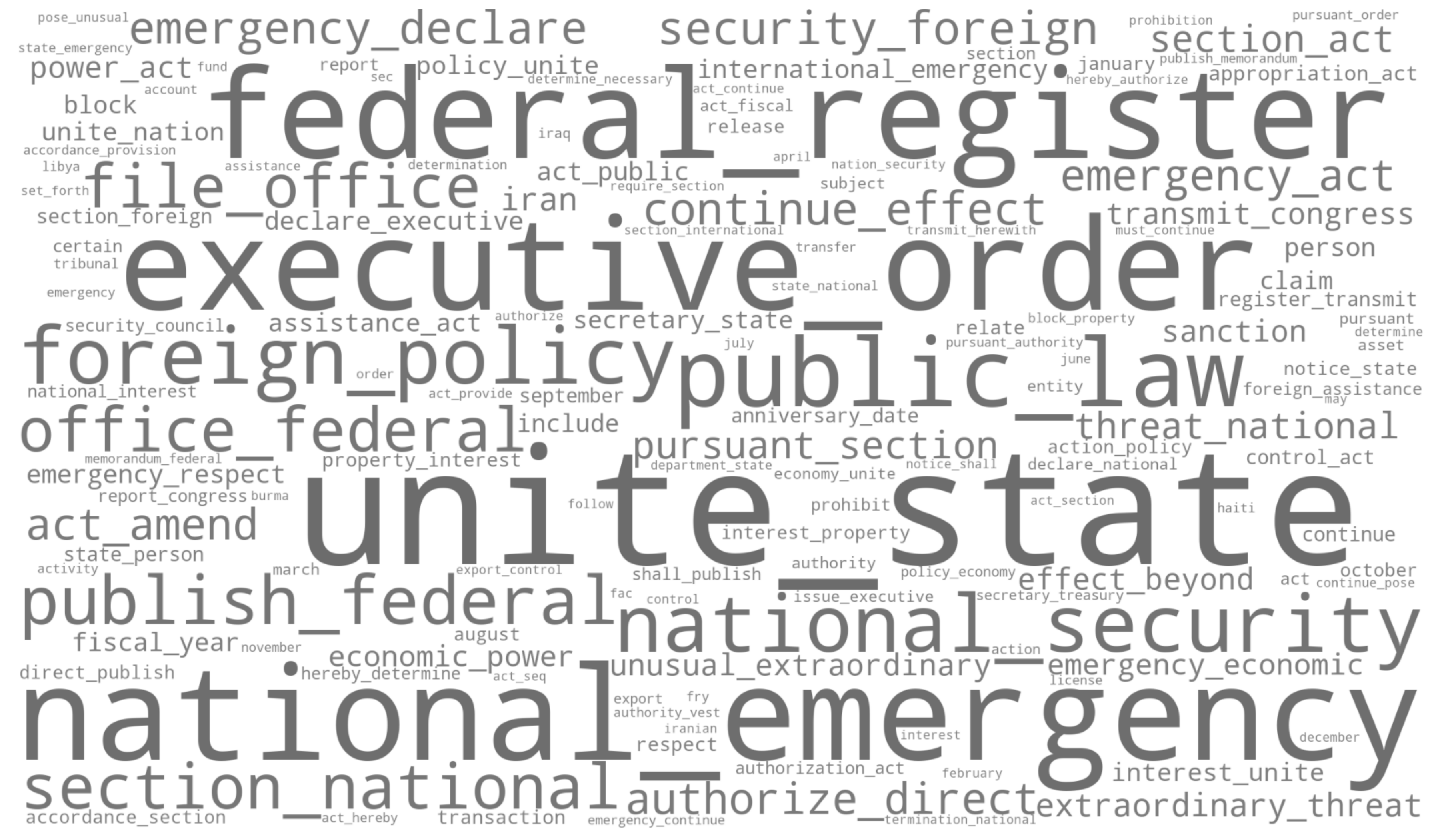}
\caption{National Security \& Emergencies}
\end{subfigure}
\begin{subfigure}{0.49\columnwidth}
\includegraphics[width=0.95\linewidth]{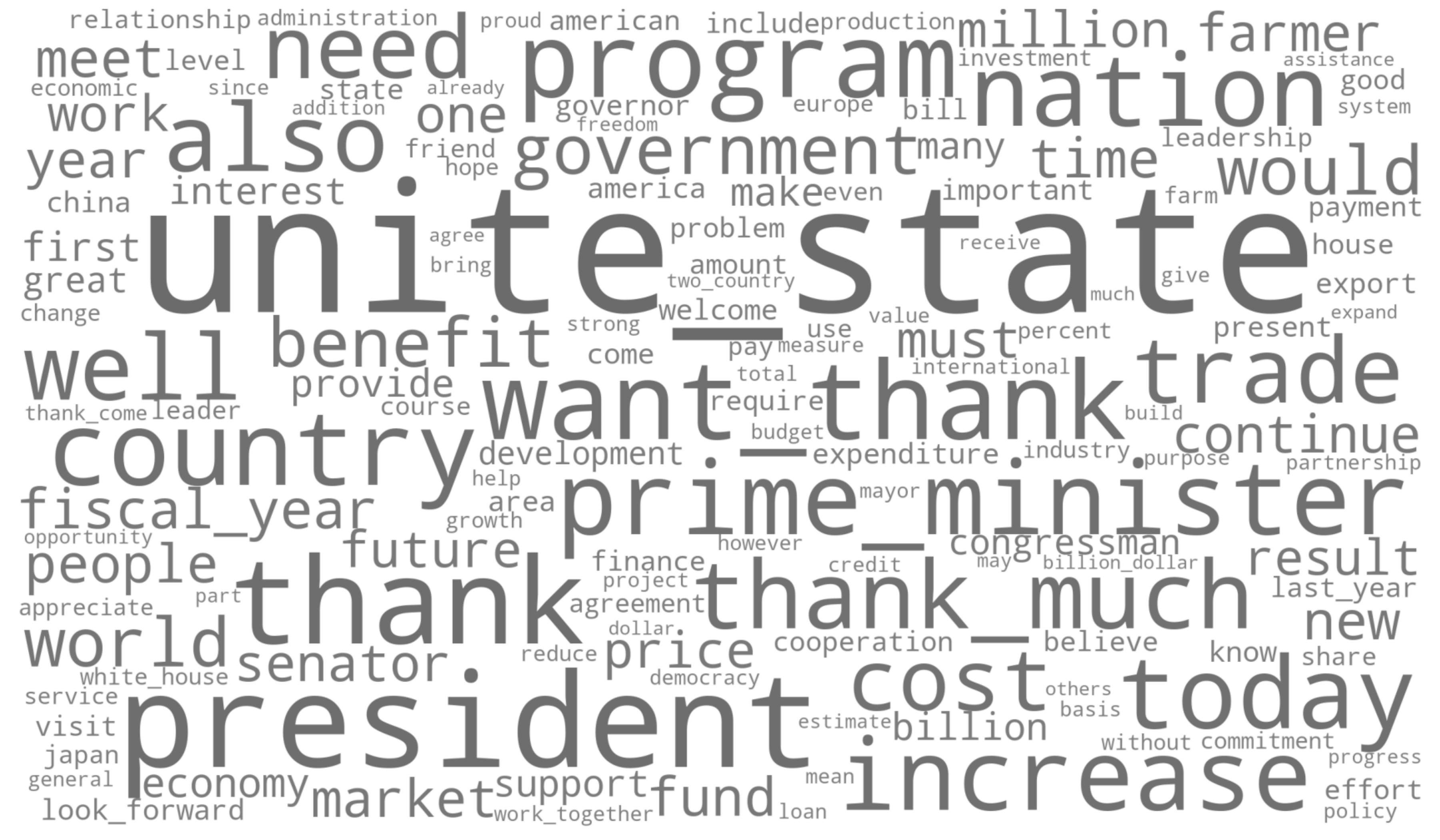}
\caption{Economic Development}
\end{subfigure}
\begin{subfigure}{0.49\columnwidth}
\includegraphics[width=0.95\linewidth]{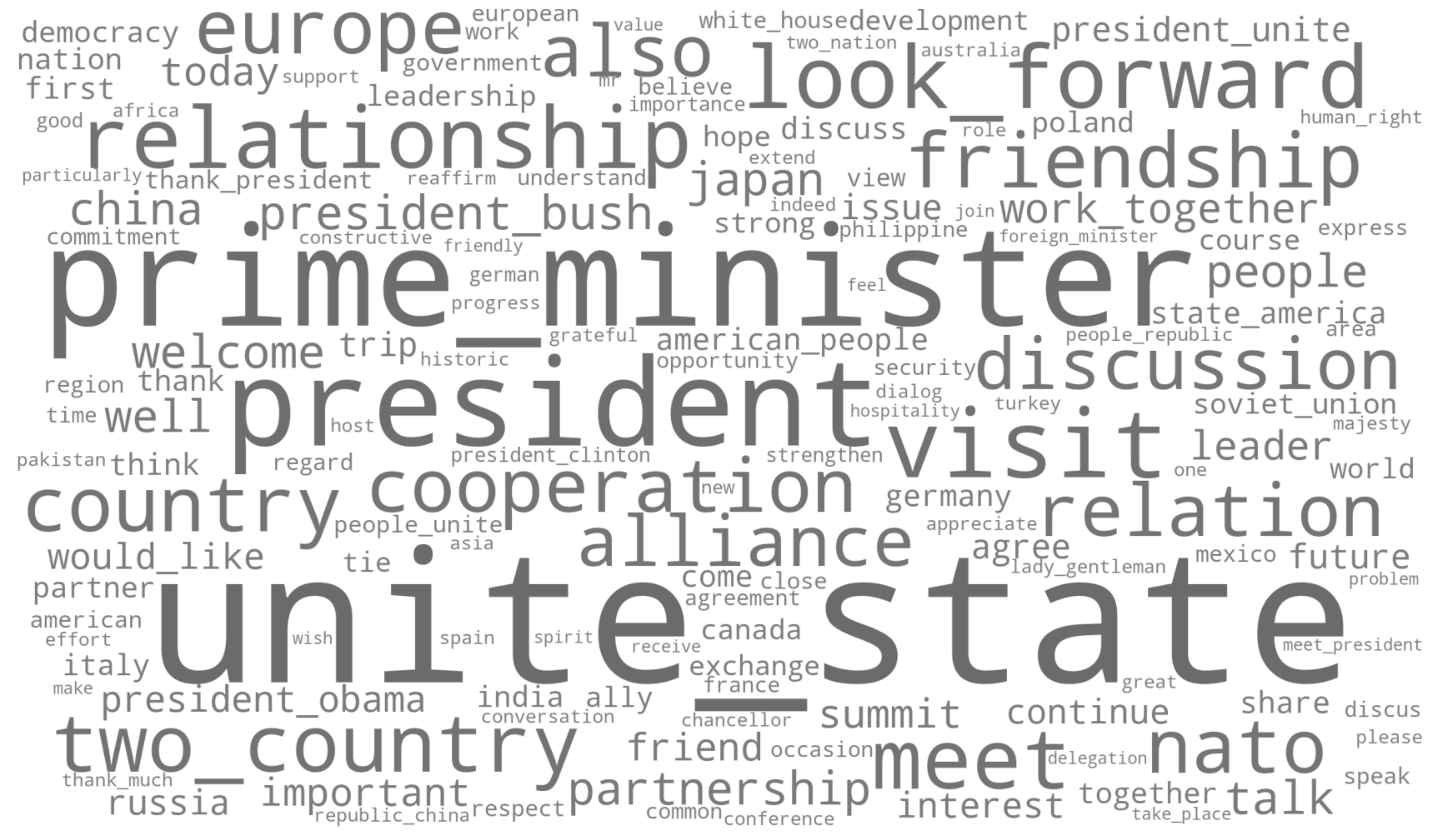}
\caption{International Relations/Diplomacy}
\end{subfigure}
\begin{subfigure}{0.49\columnwidth}
\includegraphics[width=0.95\linewidth]{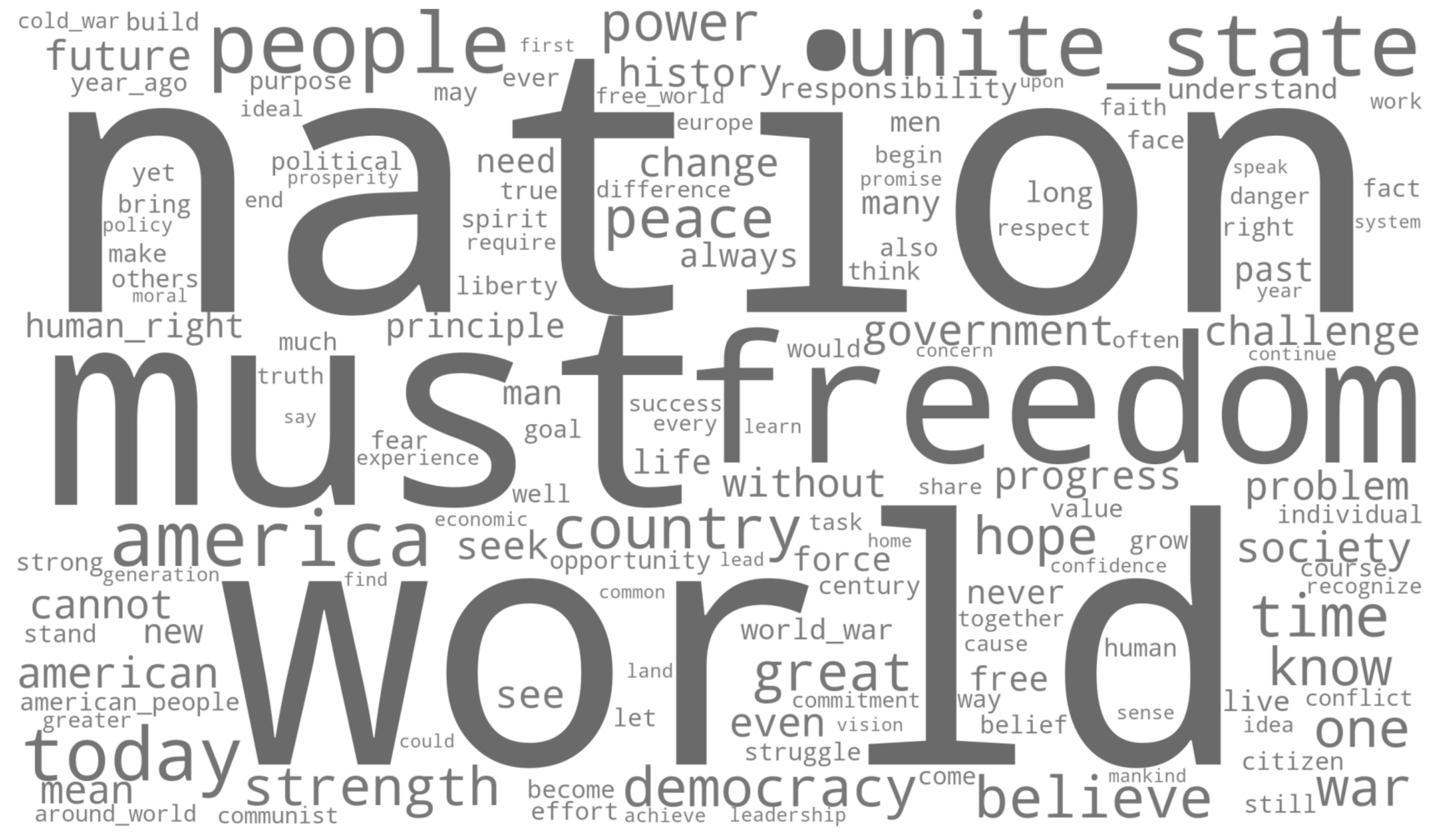}
\caption{Freedom \& Patriotism}
\end{subfigure}
\begin{subfigure}{0.49\columnwidth}
\includegraphics[width=0.95\linewidth]{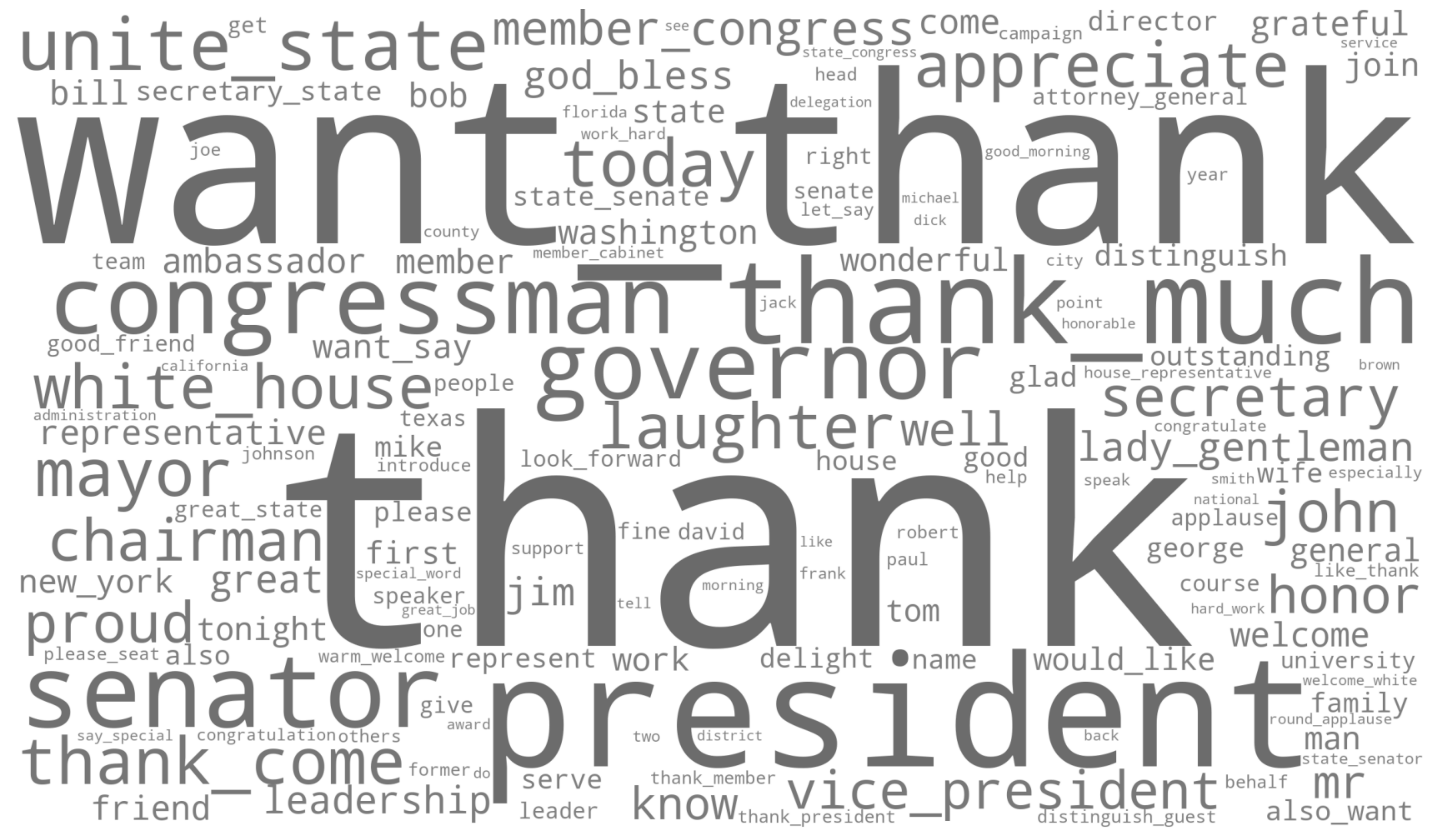}
\caption{Expressing Gratitude in Speeches}
\end{subfigure}
\begin{subfigure}{0.49\columnwidth}
\includegraphics[width=0.95\linewidth]{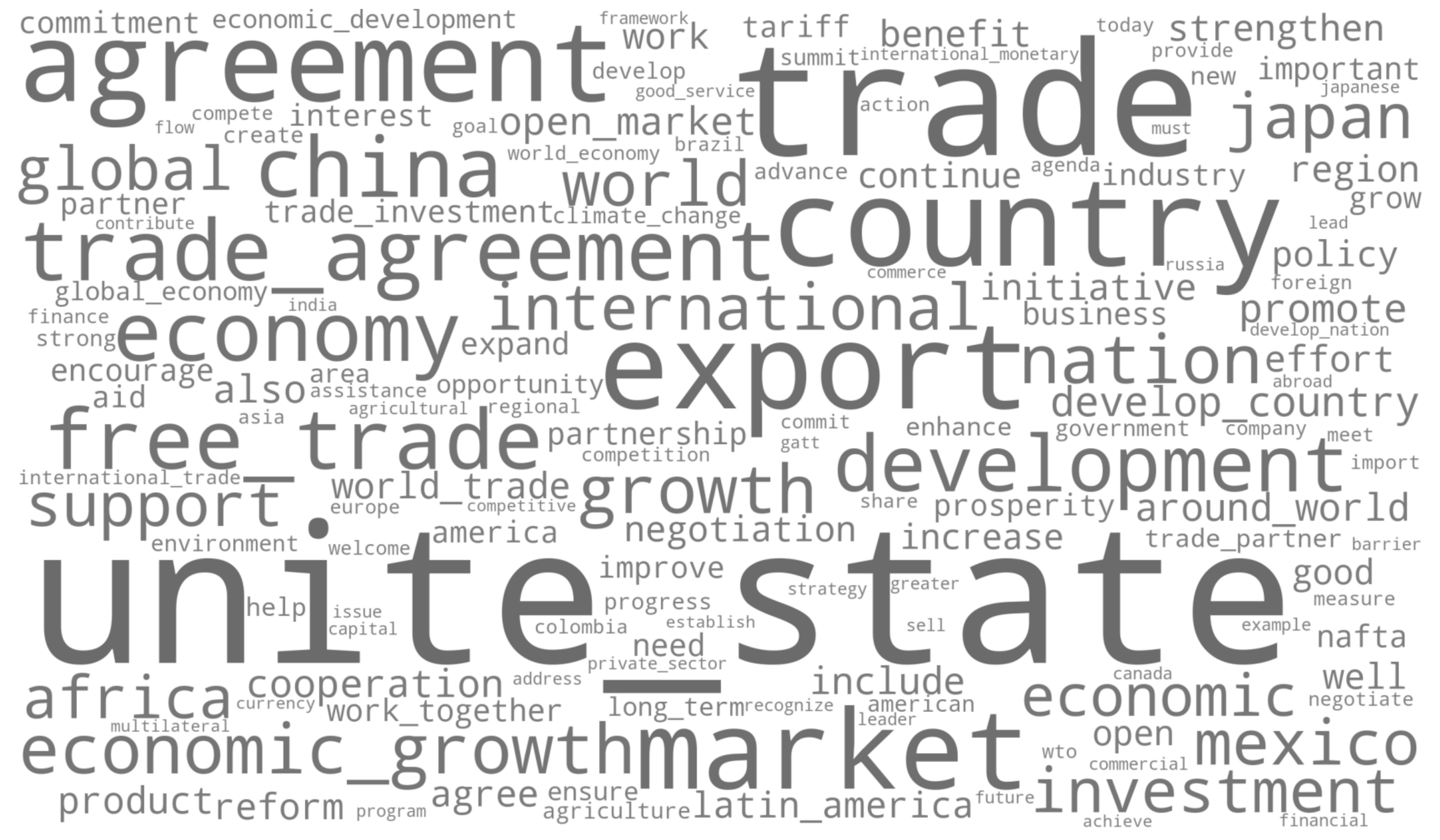}
\caption{International Economics \& Trade}
\end{subfigure}

\caption{Wordclouds of the non-tax topics}
\label{wc:all}

\end{figure}

\end{appendices}

\end{document}